\documentclass[rmp,reprint,floatfix,longbibliography,footinbib]{revtex4-1} 
\usepackage{amsmath}  
\usepackage{amsfonts} 
\usepackage{graphicx} 
\usepackage{bm}
\usepackage[usenames,dvipsnames]{color}
\usepackage{verbatim}
\usepackage[T1]{fontenc}
\usepackage{MnSymbol}
\usepackage{wasysym}
\usepackage{soul}

\def\Thad#1{{#1}} 
\def\Tom#1{{#1}}

\def\PJ#1{{#1}}

\def\be{\begin{eqnarray}}
\def\ee{\end{eqnarray}}

\newcommand{\dens}[1]{{\rm [#1]}}
\def\expect#1{{\left\langle#1\right\rangle}}

\def\He3{$^3$He}
\def\Het{$^3$He}
\def\Hef{$^4$He}
\def\Hemet{$\mathrm{He}^{\mathrm{\ast}}$}
\def\He3{$^3$He}
\def\densrat{{\cal D}}
\def\esS{2$^3$S}
\def\esP{2$^3$P}

\RequirePackage{lineno}

\begin{document}

\setpagewiselinenumbers

\modulolinenumbers[5]

\title{Optically polarized \He3}

\author{T. R. Gentile}
\affiliation{National Institute of Standards and Technology (NIST), Gaithersburg, Maryland 20899, USA}
\author{P. J. Nacher}
\affiliation{Laboratoire Kastler Brossel, ENS-PSL Research University, CNRS, UPMC-Sorbonne Universit\'es, Coll\`ege de France, Paris, France}
\author{B. Saam}
\affiliation{Department of Physics and Astronomy, Washington State University, Pullman, WA 99164, USA}
\author{T. G. Walker}
\email{tgwalker@wisc.edu} 
\affiliation{Department of Physics, University of Wisconsin-Madison, Madison, Wisconsin 53706, USA}


\date{\today}

\begin{abstract}{
This article reviews the physics and technology of producing  large quantities of highly spin-polarized \He3 nuclei using
spin-exchange (SEOP) and metastability-exchange (MEOP) optical pumping.  Both technical developments and deeper understanding of the physical processes involved have led to substantial improvements in the capabilities of both methods.  For SEOP, the use of spectrally narrowed lasers and K-Rb mixtures has substantially
increased the achievable polarization and polarizing rate.  For MEOP nearly lossless compression allows for rapid production of polarized \He3\,
and operation in high magnetic fields has likewise significantly increased the pressure at which this method
can be performed, and revealed new phenomena. Both methods have benefitted from development of storage methods that allow for spin-relaxation times of hundreds of hours, and specialized precision methods for polarimetry.  SEOP and MEOP are now widely applied for spin-polarized targets,  neutron spin filters, magnetic resonance imaging, and precision measurements.
}\end{abstract}


\maketitle

\tableofcontents



\section{Introduction}
\subsection{Overview and roadmap}

Gases of nuclear spin-polarized \He3, highly or "hyper"-polarized { to close to 100\%} in quantities on the order of a standard liter, have extensive  {scientific} applications. As a target for nuclear and particle physics with charged particle and photon beams (see Sec.~\ref{targets}), 
polarized \He3\ provides a reasonable approximation to a polarized free neutron target.  Neutron spin filters (Sec.~\ref{NSFs})
can polarize neutron beams because of the
large spin dependence of the cross section for the absorption of neutrons by \He3.  For hyperpolarized magnetic resonance imaging (MRI, Sec.\ref{MRI}), long relaxation times {\it in vivo} due to the chemical inertness of He and the absence of an electric quadrupole moment, along with a large magnetic moment, yield the highest resolution imaging of human air spaces.
Relaxation times of hundreds of hours (Sec.~\ref{Relax}) make polarized \He3 extremely stable and sensitive for precision measurements (Sec.~\ref{spect}).

These applications are enabled by specialized optical pumping methods that polarize {large volumes of \He3\ nuclei to polarizations approaching unity.}
{This review focuses on the } significant developments in the theory, practice and applications of these methods in the last two to three decades.

{Due} to the small hyperfine splitting in the \He3 $1s2p$ state and resulting long hyperfine mixing time
relative to the short \Thad{excited-state} lifetime, it is not possible to directly optically pump \He3 {to a useful polarization level} using  the 1$s^2-1s2p$ transition,
even if the required 58 nm radiation were conveniently available.  Hence two indirect optical pumping methods, spin-exchange (SEOP)
and metastability-exchange (MEOP), are employed to hyperpolarize \He3 \ nuclei.  

 In the SEOP method (Sec. \ref{SEOP}), electronic polarization produced
in alkali metal atoms by optical pumping \Thad{at bar-scale pressures} is slowly transferred to \He3\ nuclei during collisions via the Fermi-contact hyperfine interaction
between the alkali electron and the \He3\ nucleus.
The MEOP method (Sec.~\ref{Sec_MEOP}) \Thad{rapidly} produces nuclear polarization
in metastable \He3\ atoms \Thad{at mbar-scale pressures} by a combination of optical pumping and hyperfine mixing.  The \Thad{metastable nuclear spin polarization}
is then rapidly transferred to the ground state population via metastability exchange collisions.  \Thad{Typically the gas is then compressed for use in applications.}

Brief overviews of the two techniques are presented in Secs.~\ref{SEOPover} and \ref{MEOPover}, with a comparison of their relative merits  in Sec.~\ref{compare}.
Although limited by different issues, the polarizations achievable by each method for applications have remained roughly comparable,
with the current achievable {time-averaged} values between 55\% and  {85\%}, depending on conditions.
The polarizations and polarizing rates and thus the capabilities for applications have dramatically
improved with the advent of new lasers and their continual {progress} in capability and convenience.
In both methods, increased capabilities have led to deeper examinations of their respective {physical} limits.

For SEOP (Sec. \ref{SEOP}) the discovery of an unexpected \Thad{\He3} relaxation mechanism proportional to the alkali-metal density  
has modified our view of the maximum {attainable} polarization, 
but the limit is still not completely clear. 
 Motivated by the potential of lower alkali-metal spin relaxation rates, studies of optical pumping
with \Thad{ K and "hybrid" K-Rb mixtures have been performed 
with hybrid K-Rb emerging as the favored approach at this time.} 
With the elucidation of the physical processes that increase the laser power demand, 
the sophistication of SEOP models has increased 
but they are still not fully mature and verified.

\Thad{Major thrusts for MEOP (Sec.~\ref{Sec_MEOP}) have been the parallel  development of different compression approaches and extensions to high-pressure ($\sim 0.1 $bar), high-field ($\sim 1$ T) operation.  
  Large scale piston compressors are now used with mbar-scale MEOP to rapidly produce highly polarized gas at pressures of up to several bar.
For MRI applications,  compact inexpensive peristaltic pumps are being employed that exploit the reduced compression ratio requirements allowed by high pressure, high field MEOP.}
For MEOP itself, there have been advances in the theory of this complex process. 
Recent studies of the \Thad{ maximum attainable polarization have revealed a light-induced relaxation mechanism that is particularly important
 at high pressures.}

\Thad{Spin-relaxation} remains a key issue for practical application of polarized \He3 (Sec.~\ref{Relax}).  Despite much technical progress, with relaxation times in the hundreds of hours now quite common, new unexplained aspects of wall relaxation
have been observed and the subject remains poorly understood from a fundamental perspective.   
Extensive work has been done on storage of polarized \He3 in glass containers of various types, and with a variety of coatings. 
 Magnetostatic cavities allow long \He3 relaxation times to be maintained even in the large magnetic field gradients  of neutron scattering experiments.
 At cryogenic temperatures cesium has been employed
to inhibit wall relaxation and superfluid film flow.

Accurate absolute polarization metrology (Sec.~\ref{metrology}) methods are crucial for target and neutron spin filter applications.
Techniques now include water-calibrated NMR, 
neutron transmission, 
and magnetometry. 
Method-specific approaches include alkali-metal electron paramagnetic resonance (EPR) polarimetry 
for SEOP, and calibrated fluorescence 
and light absorption polarimetry 
for MEOP optical pumping cells.


All of these improvements have greatly \Thad{enhanced} applications of polarized \He3.
\Thad{As an example, order-of-magnitude increases in the luminosity of polarized \He3\ targets }for nuclear and particle physics with charged particle 
and photon 
beams (Sec.\ref{targets}) have enabled high precision studies of topics such as 
 nucleon electric and magnetic form factors,
spin structure functions, 
and three-body nuclear physics.

The large spin dependence of the cross section for absorption of neutrons by \He3\
allows polarization of neutron beams (Sec.~\ref{NSFs}).
Such "neutron spin filters" (NSFs)
are being developed worldwide
and applications to neutron scattering are growing rapidly. 
Examples of topics and materials under study include magnetic ordering, 
  magnetic multilayers 
and magnetic nanoparticles.
 Applications of
NSFs to fundamental neutron physics include studies of neutron beta
decay,  measurements of parity-violating asymmetries, and measurements
of the spin dependence of the neutron-\He3 scattering length.

A variety of polarized \He3 \ apparatus are being employed for medical studies of human lung airways (Sec.~\ref{MRI}). 
  \Thad{  Although} 
the range of imaging studies with $^{129}$Xe  is larger because it dissolves in water and fat,
the \Thad{ larger magnetic moment and generally higher attainable polarizations for \He3 generally} makes it the preferred choice
for the \Thad{highest resolution} lung images. 


Precision spectroscopy on \He3\ (Sec.~\ref{spect}) has resulted in, for example,
new magnetometry methods, 
searches for violations of fundamental symmetries, 
and searches for interactions mediated by axion-like particles. 

\Thad{\He3 has been used for other applications that we will not consider further in this paper.  \citet{Newbury_prl91} studied polarized muonic He by capturing muons on polarized \He3.  
The macroscopic behavior of spin-polarized \He3 fluids 
is modified at low temperature by polarization of the \He3        
nuclei \cite{OwersBradley97,Lhuillier79,Lhuillier82,Castaing79}.
Nonlinear spin dynamics \cite{Akimoto00,Desvaux13}, and NMR time reversal
 \cite{Baudin08} have been studied in \Thad{\He3-$^4$He} mixtures. }

We conclude the review by discussing future trends in Sec. \ref{future}.

\subsection{Overview of SEOP physics and apparatus} \label{SEOPover}

In the SEOP method, electronic polarization produced in alkali metal atoms by optical pumping is slowly transferred to \He3\ nuclei
during collisions via the Fermi-contact hyperfine interaction between the alkali electron and the \He3 \ nucleus \cite{Bouchiat60,Walker97}.
\He3\ gas \Thad{at typical} pressures between 1~bar and 10~bar\footnote{All pressures refer to room temperature unless otherwise noted.} is contained in sealed glass cells along with on the order
of 0.1 g of alkali-metal, usually rubidium or a rubidium-potassium mixture.  Pure Rb \Thad{or} K-Rb cells are typically heated to
170 $^\circ$C or 220 $^\circ$C, respectively, to establish an alkali-metal density \Thad{around} $3\times 10^{14}$ cm$^{-3}$.  
 Due to the short  diffusion length of the alkali-metal atoms,  the entire volume of the cell must be immersed in circularly polarized laser light,
typically provided by diode laser arrays with output power on the order of 100~W, an air wavelength of 794.7 nm and a typical linewidth of 0.25~nm.  {A number of factors such as absorption by optically thick alkali-metal vapor, focusing and/or distortion of the laser
light  by non-uniform blown glass cells, and in some cases poor spatial mode quality make it difficult to realize the goal of uniform optical pumping rate
at all points in the cell, \Thad{which can} be partially addressed by pumping from opposing directions  \cite{Chann03,Chen14}}.
The absorption width of the Rb vapor is determined by pressure broadening \Thad{of} 0.040 nm/amg \cite{Romalis97,Kluttz13}, so the atomic 
pressure broadened width is comparable to the laser linewidth for 
\Thad{high density (10 amg}\footnote{1 amagat (amg)=2.69$\times$10$^{19}$ cm$^{-3}$ is the density of an ideal gas at standard temperature and pressure. }) targets but substantially smaller than the laser linewidth for typical neutron spin filters (1.5 amg).
{The degree of the circular polarization (99\% or better is usually attained with commercial wave plates) is not highly critical because the relatively high alkali-metal density strongly absorbs the undesired \Thad{photon} spin state \citep{Bhaskar79,Chann_skew02}.}
To suppress radiation trapping from radiative decay of the excited alkali-metal atoms, on the order of 0.1 amg of
nitrogen \Thad{ gas} is added to provide rapid collisional de-excitation \cite{Walker97,Lancor10}.

The alkali-metal polarization $P_{\mathrm A}$ is determined by the ratio of the electronic spin relaxation rate to the optical pumping rate.
Rubidium spin relaxation rates are typically a few hundred~s$^{-1}$ whereas
optical pumping rates are much higher, hence $P_{\mathrm A}$ near unity is established on a very short time scale.
The temperature must be maintained at a low enough value so that absorption of the laser light by the optically thick \Thad{alkali-metal vapor} does
not yield too low an optical pumping rate \Thad{in the interior} of the cell.     
Alkali-metal spin relaxation results primarily from alkali-alkali collisions (dominant at low \He3\ gas pressures) and
alkali-\He3 \ collisions (dominant for high \He3\ gas pressures), with some contribution from alkali-nitrogen collisions \cite{Baranga98}.
For the same spin-exchange rate, the spin relaxation for potassium is typically about five times lower than that of rubidium,
hence the use of K-Rb mixtures or pure K increases the efficiency of SEOP \cite{Babcock03}.
In practice K-Rb mixtures are typically employed because of greater laser availability at 795 nm for Rb pumping
as compared to K pumping at 770~nm \cite{Chen07hyb}.

The \He3\ polarization is determined by the ratio of \He3 \ nuclear spin relaxation to the spin-exchange rate.
The spin-exchange rate is typically on the order of 0.1~h$^{-1}$, hence a day is required to {approach the} maximum polarization.
This slow time scale makes long \He3\ relaxation times critical.   Sealed cells made from fully blown aluminosilicate glass
are typically employed for charged particle and photon scattering and neutron spin filters, whereas
borosilicate glass and/or open systems are more common for polarized gas MRI and other applications.
\Tom{Aluminosilicate glass has low  \He3 \ permeability and is alkali-metal resistant, and for neutron applications GE180\footnote{GE Lighting Component Sales, 
Cleveland, OH 
Certain trade names and company products are mentioned in the text or identified in an illustration in order to adequately specify the experimental procedure and equipment used. In no case does such identification imply recommendation or endorsement by the National Institute of Standards and Technology, nor does it imply that the products are necessarily the best available for the purpose.
}  is particularly desirable because it is boron-free.}
The use of fully blown glass \cite{Newbury93,Chen11} and the presence of the alkali-metal \cite{Heil95} are both important for achieving the longest relaxation times.
\He3 \ cells with room temperature relaxation times on the order of 100~h
or longer \cite{Chen11} ostensibly make wall relaxation a minor contributor to limiting \He3 polarization, but 
a strongly temperature-dependent relaxation mechanism has been found to limit the achievable \He3\ polarization \cite{Babcock06}.
{For poorer cells,  short wall relaxation times can in principle be overcome by increasing the alkali density (and using hybrid pumping) and thus increasing the spin-exchange rate, but this is only possible to the extent that sufficient laser power is available.}

\begin{figure}
\includegraphics[width=3.3 in]{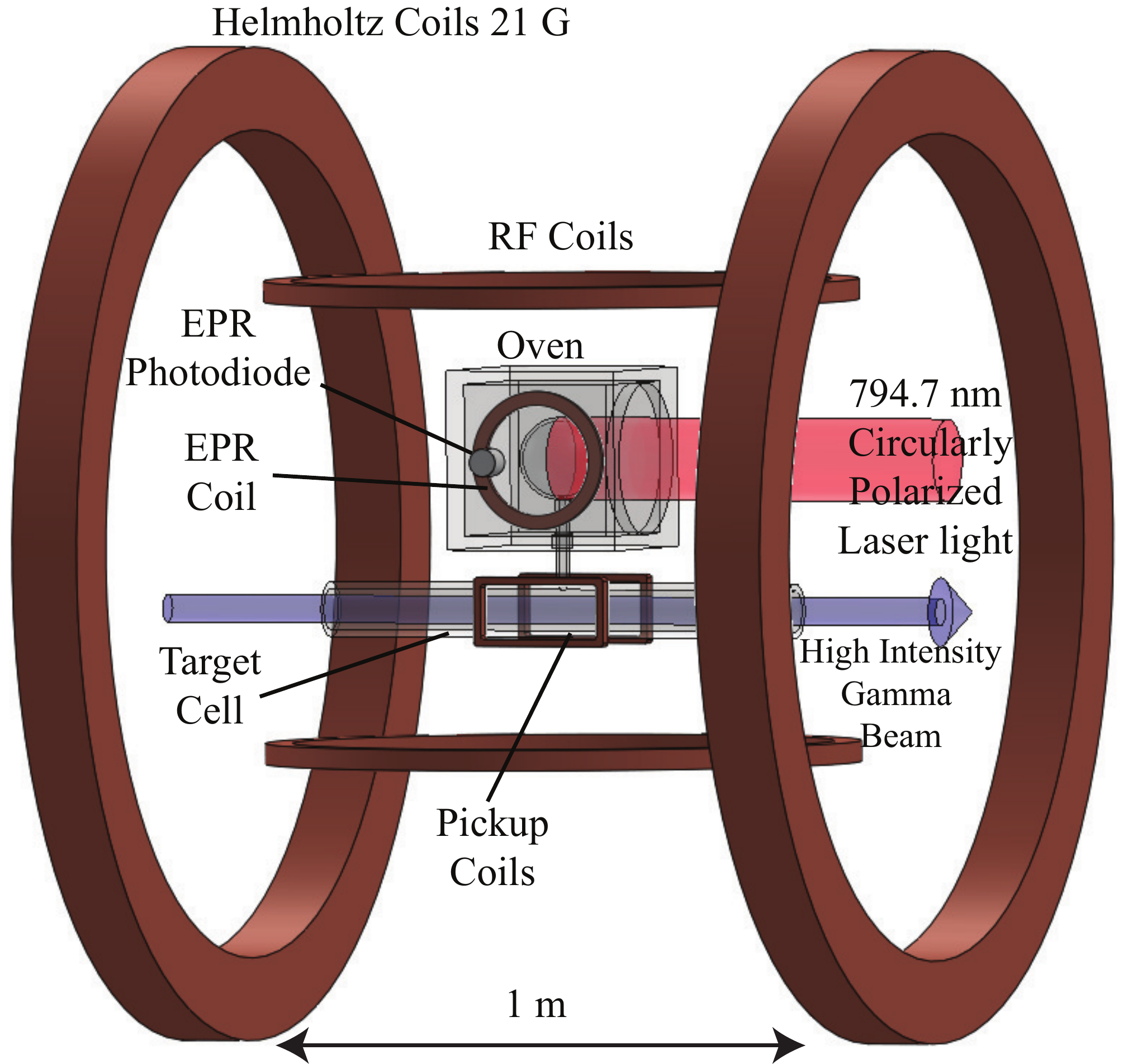}
\caption{An example of an apparatus for spin-exchange optical pumping (SEOP).
This apparatus has been employed for photon scattering experiments.
Whereas SEOP is typically performed in single cells for neutron spin-filter and magnetic resonance imaging applications,
both electron and photon scattering applications employ a double cell configuration in
which a target cell (TC) is linked to an optical pumping cell (OPC) through a connecting tube.
Electron paramagnetic resonance and nuclear magnetic resonance are employed to measure the \He3\ polarization in the OPC and TC, respectively. Adapted from  \citet{Ye10}}
\label{TUNLfigure}
\end{figure}

A typical SEOP apparatus (Fig.~\ref{TUNLfigure}) consists of a non-magnetic oven to heat a \He 3\ cell, a uniform magnetic field
provided by Helmholtz coils, and a spectrally-narrowed high-power diode array laser with suitable optics for
producing circular polarization and focusing and steering the laser beam.
Hot air is typically used for heating but non-magnetic electrical heating has also been employed \citep{Tong10,Babcock16}.
Although Helmholtz coils are the most common, four-coil systems, compensated solenoids, and
magnetostatic cavities have also been used.  Diode laser bars are spectrally narrowed
with diffraction gratings \cite{Babcock05laser} or volume holographic gratings (VHGs) \cite{Nikolau13,Chenjap14,Liu15}.  \He3\ polarization is monitored
using either adiabatic fast passage (AFP) or free induction decay (FID) NMR. AFP typically
requires a pair of drive coils large enough to immerse the cell in a reasonably uniform
radiofrequency (RF) field and pickup coil(s) to detect the precessing magnetization.  FID can also be performed
in this scheme but more often a small surface coil that both transmits a short RF pulse and detects
the small precessing transverse magnetization is employed.  SEOP apparatus may also
employ other diagnostics  such as EPR \cite{Romalis98}, Faraday rotation \cite{Vliegen01}, {and transmission spectrum monitoring with a small diffraction grating spectrometer.}
Other characteristics of SEOP apparatus are discussed under individual applications.

\subsection{Overview of MEOP physics and apparatus} \label{MEOPover}

In the MEOP method, nuclear polarization is produced in metastable \He3\ atoms by a combination of optical pumping and hyperfine mixing,
and then rapidly transferred to the ground state population via metastability exchange collisions \cite{Colegrove63,Batz11}.
Traditional low-field MEOP is performed in pure \He3\ or \He3 - $^4$He\ mixtures \cite{Stoltz96} at pressures on the order of 1 mbar, whereas high-field MEOP
has been performed at pressures up to a few hundred millibar \cite{Nikiel13}.   An electrical discharge produced by external electrodes is employed  to produce metastable densities on the order of 10$^{10}$~cm$^{-3}$.  Low gas-phase impurity levels are required because metastable atoms have 20 eV of energy, which is sufficient
to ionize most common impurities, \Thad{ resulting in destruction} of the metastable atoms.
In practice, highly pure source gas and baked and discharge-cleaned glass walls are required to achieve
an adequately high metastable density.  A hand-held spectrometer provides a simple and convenient method
to evaluate the purity of the \He3\ gas, \Thad{since  the light emitted from an MEOP cell should not show
broadband background and/or impurity emission lines.}

Due to the relatively low metastable density the gas is optically thin, hence long cells can be employed to
absorb more laser light and thus {yield larger quantities of hyperpolarized gas}.  However, due to the weak absorption, the degree of circular polarization
is more critical than for SEOP.  {The spatial profile of the light somewhat underfills the pumping cell, approximately matching the metastable density profile that 
vanishes at the cell walls.}
 Radiation trapping is not a major issue
\Thad{since} the diameter of the cells are typically less than $\approx$ 7 cm to avoid this consideration.
Optical pumping is typically performed with Yb fiber lasers with output power on the order of 10 W,
an air wavelength of 1083 nm, and a typical linewidth of 2 GHz to match the Doppler-broadened absorption width.
The polarizations of the ground state and metastable populations are strongly coupled \cite{Nacher85} and evolve
together on typical time scales of seconds to minutes, depending on the cell size and laser power \cite{Gentile93}.
The \He3 ground state relaxation time is generally dominated by the discharge and ranges from a fraction of a minute to several minutes.
Stronger discharges yield higher metastable densities and thus higher polarizing rates, but also increased relaxation.
In many applications the polarized gas is compressed to obtain pressures of between 1 bar and 4 bar at polarizing rates
of a few  \Thad{amg-L/h},
with a typical time scale of a few hours or less \Thad{to fill a cell} 
 \cite{Batz05,Lelievre07}.
Storage cells are typically made of fused quartz or aluminosilicate glass and coated with
alkali-metal (usually cesium) to reduce relaxation \cite{Heil95}.

\begin{figure}
\includegraphics[width=3.2 in]{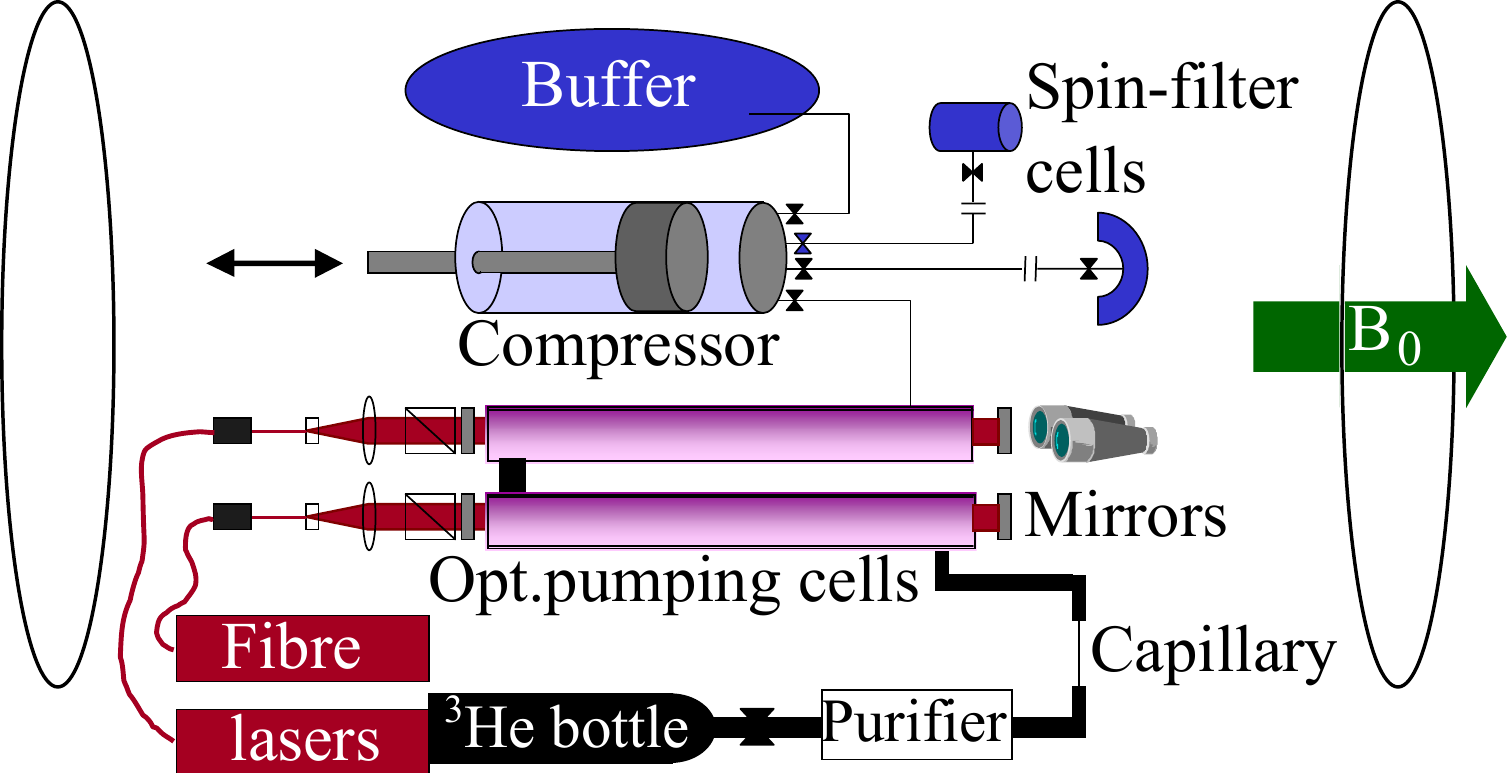}
\caption{An example of an apparatus for metastability exchange optical pumping, adapted from  \citet{Andersen05}.
\Tom{This apparatus has been employed for compressing gas into neutron spin filter cells.
Nine 2 m diameter coils provide the uniform magnetic field for the 2.3 m long optical pumping cells (OPC).
\He3\ gas is purified, polarized in the OPCs, compressed in two stages with an
intermediate buffer cell}, and \Thad{dispensed into} detachable neutron spin filter cells.  The capillary serves
to control the flow rate and restrict diffusion for the typical 1 mbar pressure in the OPCs.
 }
\label{NacherMEOP}
\end{figure}

A typical MEOP apparatus (Fig.~\ref{NacherMEOP}) consists of a radio-frequency  discharge, a uniform magnetic field
provided by a set of coils or solenoid, and a Yb fiber laser with suitable optics for
producing circular polarization and focusing and steering the laser beam \cite{Andersen05}.
Although Helmholtz coils are  common, multiple-coil systems are also  used,
in particular for large compression apparatus.   In the optical pumping cell, \He3\ polarization is monitored
by measuring the degree of circular polarization from the 668 nm \He3 \ emission line or absorption of
a probe laser at the pumping wavelength. Compression apparatus range from large-scale piston
compressors for a range of applications to small-scale peristaltic pumps  typically
employed for polarized gas MRI \cite{Nacher99,Nikiel07}.   In these small-scale systems high-field pumping
in the bore of the MRI magnet can be employed.    After compression, either  AFP or FID \Thad{NMR}
is used.  Other characteristics of MEOP apparatus are discussed under individual applications.


\begin{figure}
\includegraphics[width=3.5 in]{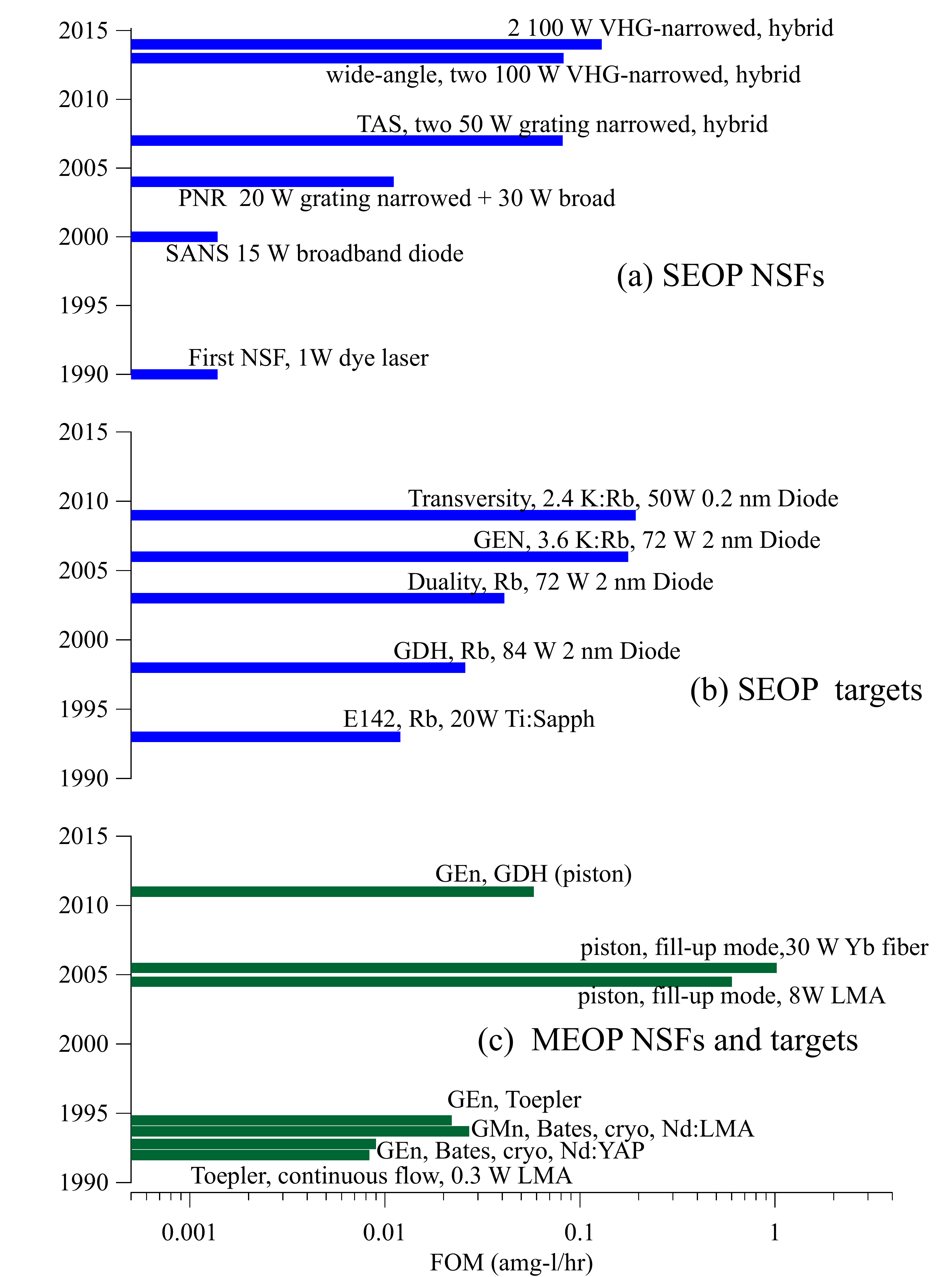}
\caption{\Tom{Development of neutron spin filters (NSFs) and spin-polarized \He3\ targets, illustrated by representative devices and experiments.  In each case, key parameters are listed to show how advances for each method improved performance,
e.g. spectrally narrowed lasers and K-Rb mixtures for spin-exchange optical pumping (SEOP), and filling of cells with piston compressors and high power laser sources for metastability exchange optical pumping (MEOP).  See Secs. \ref{targets} and \ref{NSFs} for definitions of identifiers in this plot.   The FOM is defined to be $P^2_{\rm He} N/\tau$ for SEOP NSFs and targets, and MEOP cryogenic targets, where $P_{\rm He}$ is the \He3 polarization, $N$ is the total number of atoms in the cell in units of amg-L, and $\tau$ is the time constant for polarizing the cell. (Reminder: to reach 90\% of the maximum polarization requires 2.3 time constants.) For compression-based MEOP NSFs and targets, $\tau$ is replaced by $T$, where $T$ is the time to refresh the gas for continuous flow or the time between cell exchange for remotely operated compression. 
 (a) SEOP NSFs:  First NSF \cite{Coulter90}), SANS \cite{Jones00,Gentile_sans00}, PNR \cite{Chen04}, TAS \cite{Chen07}, wide-angle \cite{Ye13}, and VHG-narrowed \cite{Chen14}.
(b) SEOP targets for electron-scattering experiments.  Here the FOM is equivalent to the potential effective luminosity of \citet{Singh15}.  Data provided by J. Singh.   E142 \cite{Anthony93}, GDH \cite{Amarian02},
GEN \cite{Riordan10}, Duality \citep{Solvignon08},  Transversity \cite{Qian11}
(c)  MEOP NSFs and targets.   GEn (Bates, cryogenic, Nd:YAP)~\cite{Jones93}, GMn (Bates, cryogenic, Nd:LMA)~\cite{Gao94}, GEn (Toepler pump)~\cite{Meyerhoff94}, GEn, GDH (piston)~\cite{Krimmer09, Schlimme13, Krimmer11}, (Toepler pump)~\cite{Eckert92}, piston fill-up mode \cite{Batz05}.}}
\label{PolTarget}
\end{figure}

\subsection{SEOP/MEOP compare and contrast} \label{compare}
Although SEOP and MEOP address the problem of polarizing \He3\ nuclei quite differently, the needs of
applications are met by both methods, with comparable practical results.  Figure \ref{PolTarget}  shows representative performance of the two methods as the advances discussed in this review occurred.


%

The key feature of SEOP is the ability to polarize \He3\ directly at a wide range of pressures 
(typically between 0.5 bar and 13 bar), which is required for most applications.  In contrast MEOP is typically performed at pressures of order 1 mbar,
thus requiring polarization preserving compression.  The key feature of MEOP is the ability to 
produce polarized \He3\ at rates of a few \Thad{amg-L/h} \cite{Batz05}, typically an order of magnitude faster
than most SEOP apparatus.  Most applications have been approached by both methods,
but in some cases one may be preferred, e.g. dual species masers, in which both \He3\ and
$^{129}$Xe can be simultaneously polarized by SEOP, \Thad{and internal targets, for} which the low operating pressure and
high polarizing rate are well-matched to MEOP.\\

A comparison of the key parameters of each method reveals how their practical performance is comparable.
MEOP starts off with a rate constant nearly nine orders of magnitude higher than SEOP,
primarily because of the inherent weakness of the spin-exchange cross section between alkali-metal atoms and \He3  nuclei as compared to the nearly gas-kinetic cross section for metastability exchange. However, this advantage is dramatically reduced because the typical
alkali-metal density is about four orders of magnitude higher than the typical metastable density thus resulting
in a difference of nearly five orders of magnitude in the polarizing time constant, e.g. 15~h for pure Rb SEOP 
and 1~s for low-field MEOP.  SEOP is performed at pressures 10$^3$ to 10$^4$ times
higher than low-field MEOP, hence overall the polarizing rate for MEOP is about an order of magnitude faster than SEOP
(see Fig.~\ref{PolTarget}).  The operational approach of each method
reflects exploitation of their respective strengths:  SEOP is slow but compact and can operate
unattended, hence one operates continuously on charged particle beam lines and/or makes
use of overnight operation with multiple polarizing  stations to remotely polarize cells for neutron beam lines.
MEOP is fast but compression requires greater attention and the speed
is maximized with large optical pumping volumes, hence one operates with 
replaceable cells that are rapidly filled with polarized gas by a remote compressor. \\

In recent years the limitations of each method are being addressed by new methods.
More compact piston compression apparatus have been developed \cite{Mrozik2011,Beecham11,Kraft2014}
and for MRI applications
 MEOP has been extended to optical pumping pressures up to hundreds of millibar
at high magnetic fields \cite{Nikiel13}.   The use of K-Rb mixtures has increased the polarizing rate for SEOP \cite{Chen07hyb,Singh15} and large scale production of polarized \He3\ via SEOP is under development \cite{Hersman13}.

\section{Spin-Exchange Optical Pumping}
\label{SEOP}

Spin-exchange optical pumping of \He3 was pioneered by \citet{Bouchiat60} using lamps.  With some exceptions \citep{Gamblin65,Fitzsimmons69,Grover78} the topic was largely ignored until the proposal and initial demonstration of \citet{Chupp87} for producing high density polarized targets using SEOP with tunable lasers.  Subsequent experimental developments are described in detail throughout this review.  Basic descriptions of SEOP physics \citep{Chupp87,Wagshul94,Walker97} have been presented previously, as well as much more detailed discussions \cite{Appelt98,HJW}.  In this section we present the current understanding of SEOP physics, with particular emphasis on new understanding obtained since the RMP Colloquium of 1997 \citep{Walker97}.

\subsection{Spin-exchange collisions} \label{sec:SpinExchange}

\begin{figure}
\includegraphics[width=3.0 in]{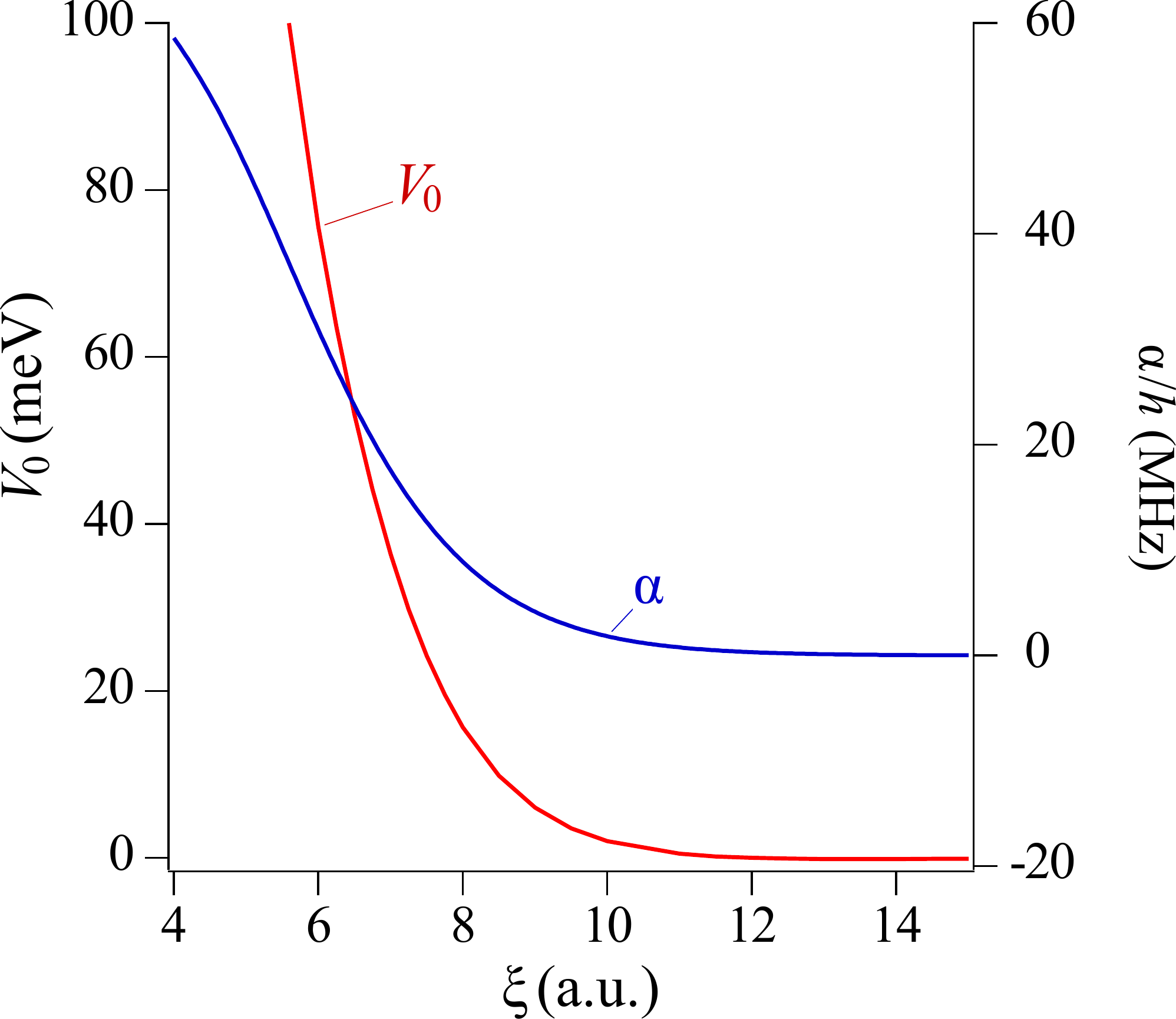}
\caption{Calculated spin-independent ($V_0(\xi)$) \cite{Partridge2001} and Fermi-contact ($\alpha(\xi)$) \cite{Tscherbul11} potentials for K-\He3 molecules, as a function of interatomic separation $\xi$ in atomic units. 
}
\label{potentials}
\end{figure}

Spin-exchange optical pumping of \He3 transfers spin-polarization between alkali-metal electron spins (A) and \He3 nuclear spins during binary collisions:
\be
\mbox{A}\uparrow+\mbox{\He3}\downarrow \longleftrightarrow \mbox{A}\downarrow+\mbox{\He3}\uparrow
\ee
The dominant interactions experienced during  A-He collisions are
\be
V(\xi)=V_0(\xi)+\alpha(\xi)
\Thad{ {\bf S}\cdot {\bf I}_{\rm He}}+\gamma(\xi) {\bf N}\cdot{\bf S}
\label{VAHe}
\ee
where $V_0(\xi)$ is the spin-independent interaction potential between the two atoms \Thad{separated by  a distance ${ \xi}$ }and $\alpha(\xi)$ is the strength of the Fermi-contact hyperfine interaction between the alkali electron spin $\bf S$ and the \He3 \Thad{nuclear} spin \Thad{${\bf I}_{\rm He}$}.  Recent theoretical calculations \citep{Tscherbul11,Partridge2001} of these for K-\He3  are shown in Fig.~\ref{potentials}.  The extremely weak attractive portion of $V_0$, supporting at most one bound state, \cite{Kleinekathofer1999} is imperceptible on the scale of typical thermal collision energies so that the dominant spin-transfer occurs at the inner turning points of binary collisions.  The spin-rotation interaction $\gamma(\xi) {\bf N}\cdot{\bf S}$, which couples the electron spin to the rotational angular momentum $\bf N$ of the alkali-metal-He pair, is a major source of angular momentum loss in the system and governs the maximum possible efficiency with which spin exchange can occur. 

The curves of Fig.~\ref{potentials} can be used with time-dependent perturbation theory to estimate the spin-exchange rate coefficient, $k_{SE}\approx v \sigma (\alpha\tau/\hbar)^2$,  to be within a  factor of two or so of the measured spin-exchange rate coefficients, around $6\times 10^{-20}$ cm$^3$/s.  Here $v\sigma$ is the gas-kinetic rate coefficient and $\tau$ the collision time.  The probability of spin-exchange during a single collision, $ (\alpha\tau/\hbar)^2$, is on the order of $10^{-10}$. With alkali-metal densities $\dens{A}$  in the range of $10^{14}-10^{15}$ cm$^{-3}$, this implies a spin-exchange time constant $(k_{SE}\dens{A})^{-1}$ of 5-50 hours.  These numbers illustrate \Thad{why 
the technical development of storage containers with \He3 lifetimes of hundreds of hours (Sec.~\ref{Relax}) is vital to attaining polarizations that approach unity}.

The \He3 polarization dynamics due to spin exchange and \Thad{other sources of relaxation (rate $\Gamma_w$, usually dominated by wall collisions)} can under most conditions be accurately modeled as
\be
{d\over dt}P_{\rm He}=k_{SE}\dens{A}(P_A-P_{\rm He})-\Gamma_{\rm w} P_{\rm He}
\label{SEdynamics}
\ee
where the alkali-metal electron polarization $P_A$, due to its sub-second relaxation times, is essentially constant compared to the hour-scale variations of $P_{\rm He}$.  Thus the noble gas polarization builds up to a steady-state value
\be
P_{\rm He \infty}=P_A{k_{SE}\dens{A}\over k_{SE}\dens{A}+\Gamma_{\rm w}}
\label{PKinf}
\ee
with a time constant $\tau$ obeying
\be
{1\over \tau}=k_{SE}\dens{A}+\Gamma_{\rm w}
\ee
This latter relation seems to imply a simple method for measuring $k_{SE}$, namely measure the time constant as a function of the alkali density $\dens{A}$.  This method does not work, however, since for reasons not yet understood the wall relaxation rate $\Gamma_{\rm w}$ in spin-exchange cells is observed to increase rapidly with temperature.  This issue is discussed in detail in Sec.~\ref{sec:X}.

\subsubsection{Measurements of spin-exchange collision rates} \label{semeas}

 Several wall-independent methods have been used to determine $k_{SE}$ for different species.  The repolarization method \cite{Baranga98,Chann02c} measures the alkali polarization produced by spin-exchange in the absence of optical pumping, 
 \be
 P_{A}^{\rm re}=k_{SE}\dens{He}P_{\rm He}/\Gamma_A \label{repol}
 \ee
 where $\Gamma_A$ is the measured alkali spin-relaxation rate.  The rate balance method \cite{Chann02c} measures \Thad{$P_{\rm He\infty}$}, $\tau$, $P_A$, and the alkali density to deduce $k_{SE}$ from Eq.~\eqref{PKinf}. A combination of these two methods, measuring the time rate of change of the repolarization signal, was used by \citet{Borel03}.
  For alkali atoms with small $\Gamma_A$, the spin-exchange rate can also be deduced by measuring the difference between $\Gamma_A$ for \He3 and for $^4$He,  with a correction for the reduced-mass scaling of the spin-relaxation contribution \cite{Walker10}. 
 Most recently, \citet{Singh15} used absolute alkali polarimetry and density measurements , combined with the initial slope from Eq.~\eqref{SEdynamics}, to infer $k_{SE}$. Table~\ref{setable} shows the status of wall-independent spin-exchange rate coefficient measurements for the various alkali-metal atoms.  \Thad{For potassium, the recent result of \citet{Singh15} is 30~\% higher than
the weighted average of three prior measurements.  They speculated
that  this difference may be due to their operation at substantially higher temperature,
but the origin of the disagreement has not been established.}  
 
\begin{table} 
\begin{tabular}{||c|c|c|c||}
\hline \hline
 &Na&K&Rb\\ \hline
& &6.1(4)\footnote{\cite{Babcock03}}&\\ 
$k_{SE}$ &6.1(6)\footnote{\cite{Borel03}}&5.5(2)\footnote{\cite{BabcockPhD}} &6.7(6)\footnote{\cite{Baranga98}}\\ 
& &6.1(7)\footnote{\cite{Walker10}}&6.8(2)\footnote{\cite{Chann02c}}\\ 
& &7.5(5)\footnote{\cite{Singh15}}&\\ \hline
$\kappa_0$&4.72(09)\footnote{\cite{Babcock05}, \Thad{200 $^\circ$C}} & 6.01(11)\footnote{\cite{Babcock05}, \Thad{200 $^\circ$C}} & 6.15(09)\footnote{\cite{Romalis98}, 175$^\circ$C}\\ \hline
$d\kappa_0/dT$&  0.00914(56)\footnote{\cite{Babcock05},\Tom{ 210--350$^\circ$C}} &\Thad{0.0086(20)}\footnote{\cite{Babcock05}, \Tom{150--220$^\circ$C} }& 0.00934(14)\footnote{\cite{Romalis98}, \Tom{110-172 $^\circ$C}}  \\
& && 0.00916(26)\footnote{\cite{Babcock05}, \Tom{170--350$^\circ$C}}  \\
\hline \hline
\end{tabular}
\caption{Spin-exchange rate coefficient($k_{SE}$) measurements, in units of $10^{-20}$ cm$^3$/s, using wall-independent methods;  EPR frequency shift enhancement factors $\kappa_0$ and $d\kappa_0/dT$ (\Thad{units K$^{-1}$}; see Sec.~\ref{EPRpol}).  Throughout this paper, numbers in parentheses represent the uncertainty in the last digit(s). }
\label{setable}
\end{table}

\subsubsection{Spin-exchange efficiency} \label{sec:eff}

Despite the very slow time constants associated with spin-exchange pumping of \He3,  the efficiency of transfer of angular momentum from the photons in the optical pumping light to the nuclei can, under ideal conditions, be surprisingly high.  If fully polarized alkali metal atoms are assumed not to scatter optical pumping light (an assumption to be examined in Sec.~\ref{sec:lightprop}), the loss of angular momentum by $\dens{A}V$ polarized alkali metal             atoms in volume $V$ occurs at a rate $\dens{A}V\Gamma_{A}P_A$.  Comparing this to the rate $V\dens{He}dP_{\rm He}/dt$ at which angular momentum is added to the noble-gas nuclei, gives the collisional efficiency
\be
\eta(P_{\rm He})={V\dens{He}dP_{\rm He}/dt\over \dens{A}V\Gamma_{A}P_A}={k_{SE}\dens{He}\over \Gamma_A}\left(1-{P_{\rm He}\over P_{\rm He\infty}}\right)
\label{eff}
\ee
 The collisional efficiency is maximum at low He polarizations, then decreases at higher polarizations as significant amounts of angular momentum are returned to the alkali metal atoms via spin-exchange collisions from the polarized He nuclei.  Direct measurements of the efficiency \cite{Baranga98} were made using  Eq.~\eqref{repol} rewritten as
 $\eta(0)=P_{A}^{\rm re}/P_{\rm He}$.  Variations from atom to atom in $\Gamma_A$ are the primary determining factor in the efficiency.

\begin{figure}[htbp]
\includegraphics[width=3.3 in]{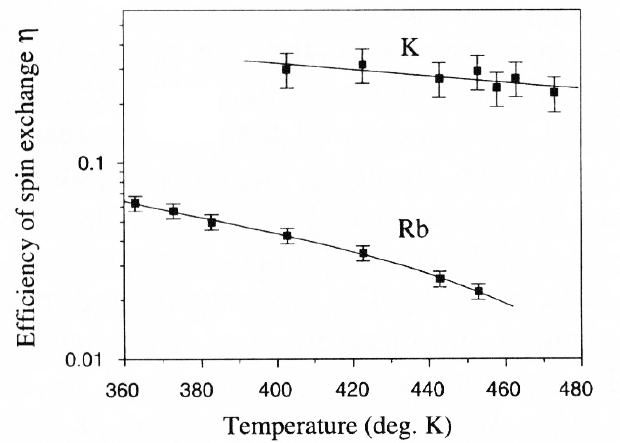}
\caption{Spin-exchange efficiency \Thad{($\eta_{SE}$)} measurement for K and Rb, from \citet{Baranga98}.}
\label{fig:efficiency}
\end{figure}

 There are many contributing processes to the alkali spin-relaxation rate $\Gamma_A$  (Sec.~\ref{sec:alkalispinrelaxation}), but at a given $\dens{He}$ the minimum relaxation rate is $\Gamma_A=(k_{SR}+k_{SE})\dens{He}$, where $k_{SR}$ is the rate coefficient for relaxation due to the spin-rotation interaction.  Thus when relaxation due to He atoms dominates the alkali-metal spin-relaxation, the spin-exchange limited efficiency is
 \be
 \eta_{SE}={k_{SE}\over k_{SE}+k_{SR}}
 \ee
Measurements of the spin-exchange efficiency for Rb and K  \cite{Baranga98}, shown in Fig.~\ref{fig:efficiency}, obtained $ \eta_{SE}\sim 1/50 $ and $1/6$ under typical conditions.  For Na the spin-exchange efficiency is consistent with 1 \cite{Borel03}.  The unmeasured Cs spin-exchange efficiency is predicted to be about 7 times lower than Rb using estimates from \citet{HJW}.
 
\subsection{Optical pumping}

In order to produce substantial quantities of \Thad{highly polarized} \He3 by SEOP, it is necessary to spin polarize large volumes of  high density  alkali-metal vapor interacting with \He3 gas at densities ranging from 0.5 amg for neutron spin filters to 10 amg for targets.  The optical pumping is typically done using 100 W scale lasers whose spectral line widths are comparable to or broader than the atomic lines. These extreme conditions raise a number of issues that are typically not encountered in other optical pumping contexts\footnote{{Recent treatments of optical pumping include \citet{Auzinsh} and \citet{HJW}.  The classic review of the subject is \protect\citet{Happer1972}, and a good introduction is \citet{Happer1987}.}}.  The breakdown of free-atom light selection rules, 
 light propagation and spectral evolution in optically thick conditions, and dissipation of heat are examples of issues that are key to understanding the SEOP process.  These and other effects are discussed in this section.

\subsubsection{\Thad{High pressure optical pumping}} \label{sec:opcycle}

\begin{figure}
\includegraphics[width=3.3 in]{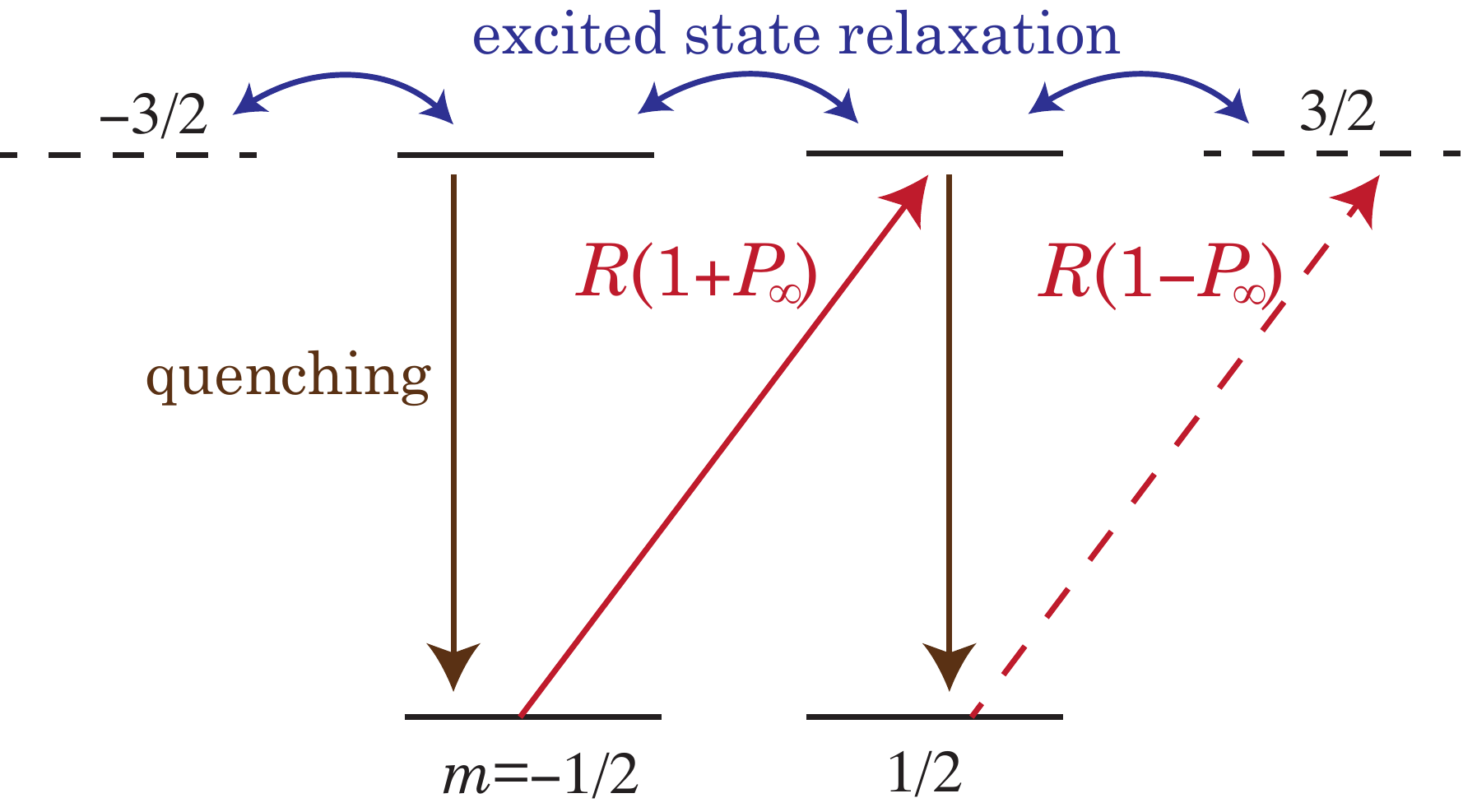}
\caption{Key elements of the optical pumping cycle for light tuned to the alkali metal S$_{1/2}$---P$_{1/2}$ resonance in the presence of high pressure He gas.  Collisions with the He atoms mix the P$_{1/2}$ and P$_{3/2}$ levels, altering the selection rules for light absorption.  Atoms in the $m_s=\pm1/2$ Zeeman ground-state sublevels absorb photons with relative probabilities $1\mp P_\infty$.  Once excited, rapid collisional spin-relaxation occurs.  Quenching collisions with N$_2$ molecules randomly repopulates the ground-state sublevels.  The atoms  accumulate in the $m_S=1/2$ sublevel, acquiring a steady-state spin-polarization of $P_\infty$ \Thad{in the absence of ground-state spin-relaxation}. }
\label{fig:optpump}
\end{figure}

Figure \ref{fig:optpump} shows the energy levels and photon absorption rates relevant for optical pumping of alkali-metal atoms in the S$_{1/2}$ ground state using circularly polarized light tuned to the P$_{1/2}$ "D1" resonance. { In the presence of He atoms, there is  collisional mixing of the P$_{1/2}$ and P$_{3/2}$ levels, so that the P$_{1/2}$ level acquires some P$_{3/2}$ character, indicated by the dashed Zeeman sublevels in the figure.}  
 The relative absorption probability, parameterized by $P_\infty$, for atoms in the $m_S=-1/2$ Zeeman sublevel is $1+P_\infty$, much greater for atoms than the relative absorption probability  $1-P_\infty$ for those in the  $m_S=1/2$  sublevel.  Thus atoms in the $-1/2$ sublevel are selectively excited by the light.  The excited atoms experience rapid spin-relaxation in collisions with the \Thad{ He and N$_2$} buffer gases, randomizing populations among the $P_{1/2}$ and, to a significant extent, P$_{3/2}$ sublevels.   \Thad{Quenching collisions with N$_2$ molecules then resonantly transfer the P-state energy to 
excited N$_2$ vibrational levels, returning the alkali atoms to the ground S$_{1/2}$ state.} 
Atoms that return to the $m_S=1/2$ state only rarely absorb the polarized light, while those that return to $m_S=-1/2$ will be efficiently re-excited.  In this manner atoms will preferentially populate the $m_S=1/2$ state, reaching a steady-state population of $P_\infty$ \citep{Lancor10prl}.


Rapid quenching of the excited atoms by N$_2$ molecules plays an essential role in this process \citep{Lancor10}.  Since SEOP typically takes place in extremely optically thick cells, the reabsorption probability for photons emitted by the excited alkali atoms can be considered to be nearly unity.  Since such photons constitute an essentially unpolarized background of resonant light, they act as a relaxation mechanism and decrease the efficiency of optical pumping.

 In the limit of large pressure broadening and short excited-state quenching times, the alkali-metal nuclear spin $I_A$ can be considered to be conserved in the optical pumping cycle, an excellent approximation for SEOP in the 5 to 10 amg He density range.  At lower densities, however,  as the electrons are rapidly relaxed in the excited state, hyperfine coupling with the nuclei causes some nuclear spin-relaxation,  reducing the optical pumping efficiency and requiring more photons per atom to reach full polarization \citep{Lancor10}.

 Under the high alkali density conditions of SEOP, spin-exchange collisions between the alkali metal atoms are by far the dominant spin-dependent collision process. Since alkali-metal spin-exchange collisions conserve total angular momentum, they mainly serve, in concert with the hyperfine coupling between the nuclei and electrons, to reach a spin-temperature equilibrium where the fraction of atoms in any total angular momentum state $F,m_F$ is \Thad{ $\rho(F,m_F)\propto e^{\beta m_F}$} \cite{Anderson59,Happer1972,Appelt98}.  The spin-temperature parameter $\beta$ is related to the alkali-metal electron spin-polarization by $P_A=\tanh(\beta/2)$, and either of these numbers ($\beta$, $P_A$) is sufficient to completely describe the states of the alkali-metal spins.   It is often convenient to account for the alkali-metal nuclei by the "slowing-down factor" $q=\expect{F_z}/\expect{S_z}$ which is approximately \Thad{$2I_A+1$} (isotopic average of \Thad{5.4} for Rb) under the usual high polarization conditions of SEOP \citep{Appelt98}.   Sometime it is useful to note that,  at small magnetic fields where $F$ is a good quantum number, the electron spin-polarization arises solely from the populations of the two states \Thad{$F=I_A+1/2,m_F=\pm(I_A+1/2)$}, {\it i.e.} $P_A=\rho(F,F)-\rho(F,-F)$.  
 
 Since the rapid alkali-alkali spin-exchange collisions keep the electronic and nuclear spins in spin-temperature equilibrium, regardless of the collision mechanism we can write the optical pumping process using conservation of angular momentum:
\be
{d \expect{F_z}\over dt}={1\over 2}\left[R({\bf r})(P_\infty-P_A)-\Gamma_A P_A\right]
\ee
where $\Gamma_A$ is an effective electron spin-relaxation rate and $R({\bf r})$ is the local photon absorption rate \Thad{(Fig.~\ref{fig:optpump})} for unpolarized atoms at position $\bf r$ in the cell.  The local steady-state polarization in the bulk of the cell is therefore
\be
P_A({\bf r})=P_\infty {R({\bf r})\over R({\bf r})+\Gamma_A}
\label{PA}
\ee
 The photon scattering rate is
 \def\Ascatt{{\cal A}}
\be
\Ascatt({\bf r})&=&R({\bf r})(1-P_\infty P_A({\bf r}))\nonumber \\&=&\Thad{\Gamma_A} P_\infty P_A({\bf r})+R({\bf r})(1-P_\infty^2)
\ee
The first term is the scattering rate required to make up for spin-relaxation collisions, while the second term  represents the scattering rate from a maximally polarized atom and would be zero for idealized D1 pumping \cite{Happer1987,HJW,Lancor10prl}.

\subsubsection{Alkali spin relaxation}\label{sec:alkalispinrelaxation}

There are many collision processes that can relax the alkali-metal spin polarization in SEOP cells.  We have already mentioned \Thad{(Secs~\ref{sec:SpinExchange}, \ref{sec:eff})} the  relaxation due to the spin-rotation interaction $V_{\gamma N}=\gamma(\xi) {\bf S}\cdot {\bf N}$ that couples the electron spin to the rotational angular momentum $\bf N$ of a colliding alkali-He pair at distance $\xi$ \cite{WalkerThywissen97}.  {$V_{\gamma N}$ arises from spin-orbit interactions induced by s-p mixing during collisions, and is therefore proportional to the spin-orbit splitting in the alkali excited state.  This is a primary motivation for using K, with its smaller fine-structure interaction, over Rb as a preferred spin-exchange partner.  }
This collision process has a strong $T^4$ temperature dependence  \cite{Baranga98}.  Analogous interactions are presumably responsible for relaxation in alkali-N$_2$ collisions.

 Also of  practical importance is the small non-conservation of spin angular momentum in alkali-metal-alkali-metal interactions.  This relaxation arises from the spin-axis interaction $V_{SS}={2\lambda(\xi)} {\bf S}\cdot \left({\bf \hat \xi}{\bf \hat \xi}-{\bf 1}/3\right)\cdot {\bf S}$, with about 1/2 of the relaxation coming from binary collisions \cite{Kadlecek01} and 1/2 from formation of triplet molecules \cite{Kadlecek98,Erickson00}. The molecular contribution can be isolated by magnetic decoupling with 0.1 T-scale magnetic fields, and has a surprising and as yet unexplained persistence at high He pressures.  The low-field rate coefficients are $1.0\times 10^{-18}$ cm$^3$/s for K-K collisions and $9.3\times10^{-18}$ cm$^3$/s for Rb-Rb.  Again, the smaller rate for $K$ relaxation makes it attractive for \He3 SEOP.   At very low pressures, a few tens of mbar, relaxation from Rb$_2$ singlet molecules becomes important \cite{Kadlecek01b}; this is a minor contribution for most SEOP situations but has a similar pressure dependence to diffusion and care has to be taken to account for both in relaxation experiments.

Spin-relaxation rates and optical pumping rates are generally much larger than the characteristic diffusion \Thad{rates} of the alkali atoms to the cell walls.  Thus throughout most of the cell the alkali polarization varies slowly (cm length scales) as the pumping light is attenuated during propagation through the cell.  Near the cell walls, however, there is a thin "diffusion layer" of unpolarized atoms.  The alkali polarization is nearly zero at the walls, so to a good approximation {the polarization at a distance $z$ from the wall is modified from Eq.~\eqref{PA} to $P_A({\bf r}) (1-\exp(-z/\Lambda))$, where the diffusion layer length scale is approximately $\Lambda=\sqrt{q D/R }$ at the entrance to the cell \cite{Wagshul94,Walker97,Appelt98}.  At 3 amg, $D\approx 0.15$ cm$^2$/s \cite{Nelson01b}, and assuming $R=100\,\Thad{\Gamma_A}=40000$ s$^{-1}$ gives $\Lambda\sim 60$ $\mu$m.  }





\subsubsection{Light propagation and circular dichroism} \label{sec:lightprop}


As already noted, the slow spin-exchange rates for SEOP require optical pumping of high density alkali-metal vapors.  For unpolarized atoms, the optical depth $OD_0$ is typically of order 100, {\it i.e.} the transmission of light through the cell is $e^{-100}$.   For  uniformly polarized atoms, the optical depth of circularly polarized light is $OD=OD_0(1-P_AP_\infty)$.  Thus the light can only significantly penetrate the cell if $1-P_AP_\infty\ll 1$, so the \Thad{atoms must be polarized to nearly 100\% ($P_A\sim 1$) {\it and} they must become transparent when maximally polarized  ($P_\infty\sim 1$).  }These conditions are only met for pumping light tuned to the $P_{1/2}$ D1 resonance. 

\begin{figure}
\includegraphics[width=3.3 in]{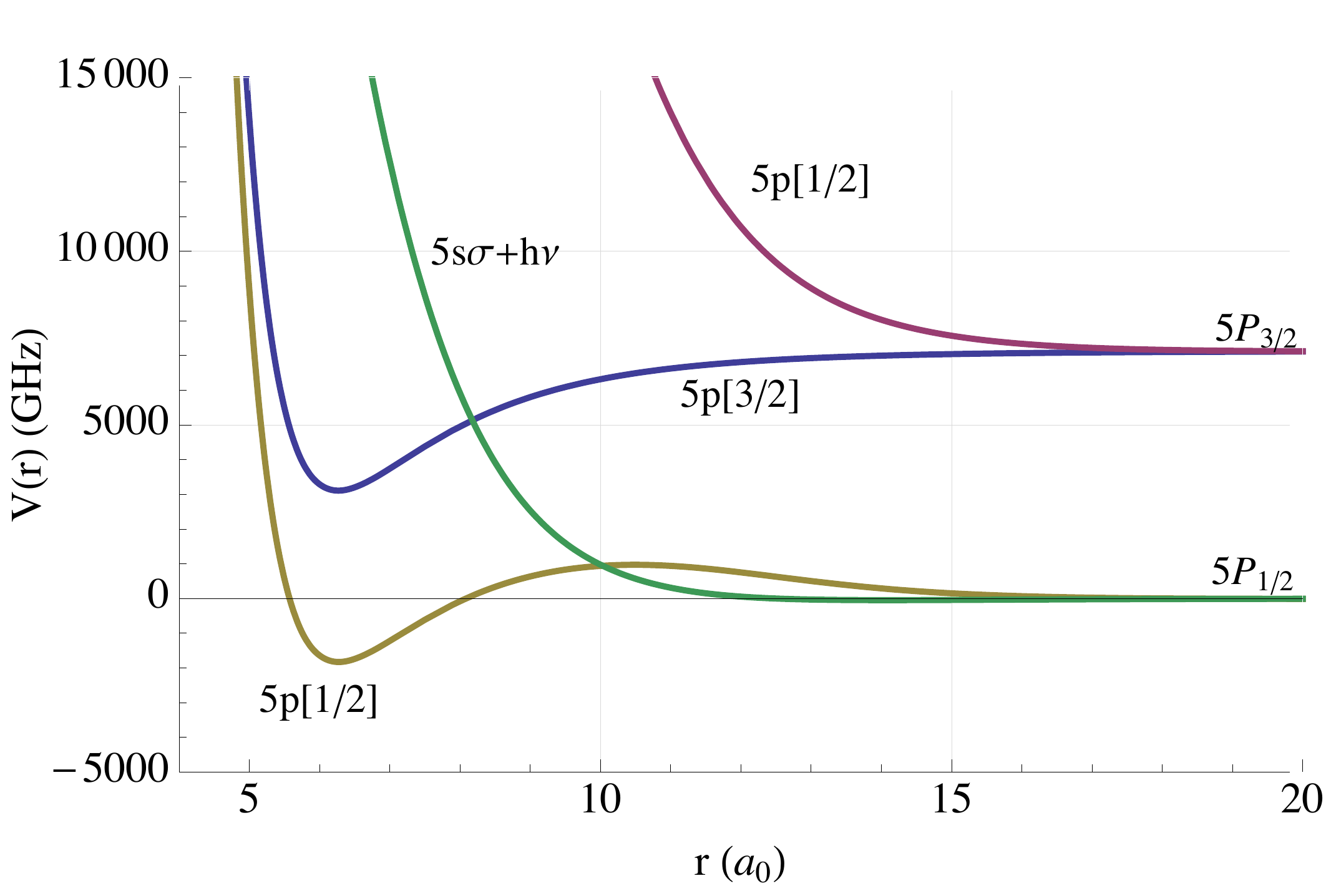}
\caption{Energy levels of RbHe molecules in the presence of optical pumping light \citep{Pascale83,Lancor10prl}.  The green curve, the ground state potential energy plus 1 photon, crosses 2 excited-state potentials at two different interatomic separations.  The 5p[$M=3/2$] curve is of purely P$_{3/2}$ nature while the 5p[1/2] is of mixed P$_{1/2}$-P$_{3/2}$ character.  The projection of the electronic angular momentum along the interatomic axis is $M$.  For both crossings the colliding atom pair can absorb the circularly polarized optical pumping light \Thad{even when each is  fully spin-polarized}.  From \citet{Lancor10prl}.
}
\label{fig:RbHeExcited}
\end{figure}

Under the high density conditions of SEOP, absorption of light during alkali-metal-He (or N$_2$) collisions does not obey the selection rules for isolated atoms \cite{Lancor10prl,Lancor10_circ}.  This is because the fine-structure states are mixed during collisions with He atoms, as illustrated in Fig.~\ref{fig:RbHeExcited}.  The normal D1 atomic selection rules are slightly violated, allowing absorption of circularly polarized light by fully polarized atoms.  Thus in the presence of He gas the alkali atoms do not become fully spin-polarized in the limit of high pumping rates; they instead acquire a maximum polarization $P_\infty<1$.

\begin{figure}
\includegraphics[width=3.3 in]{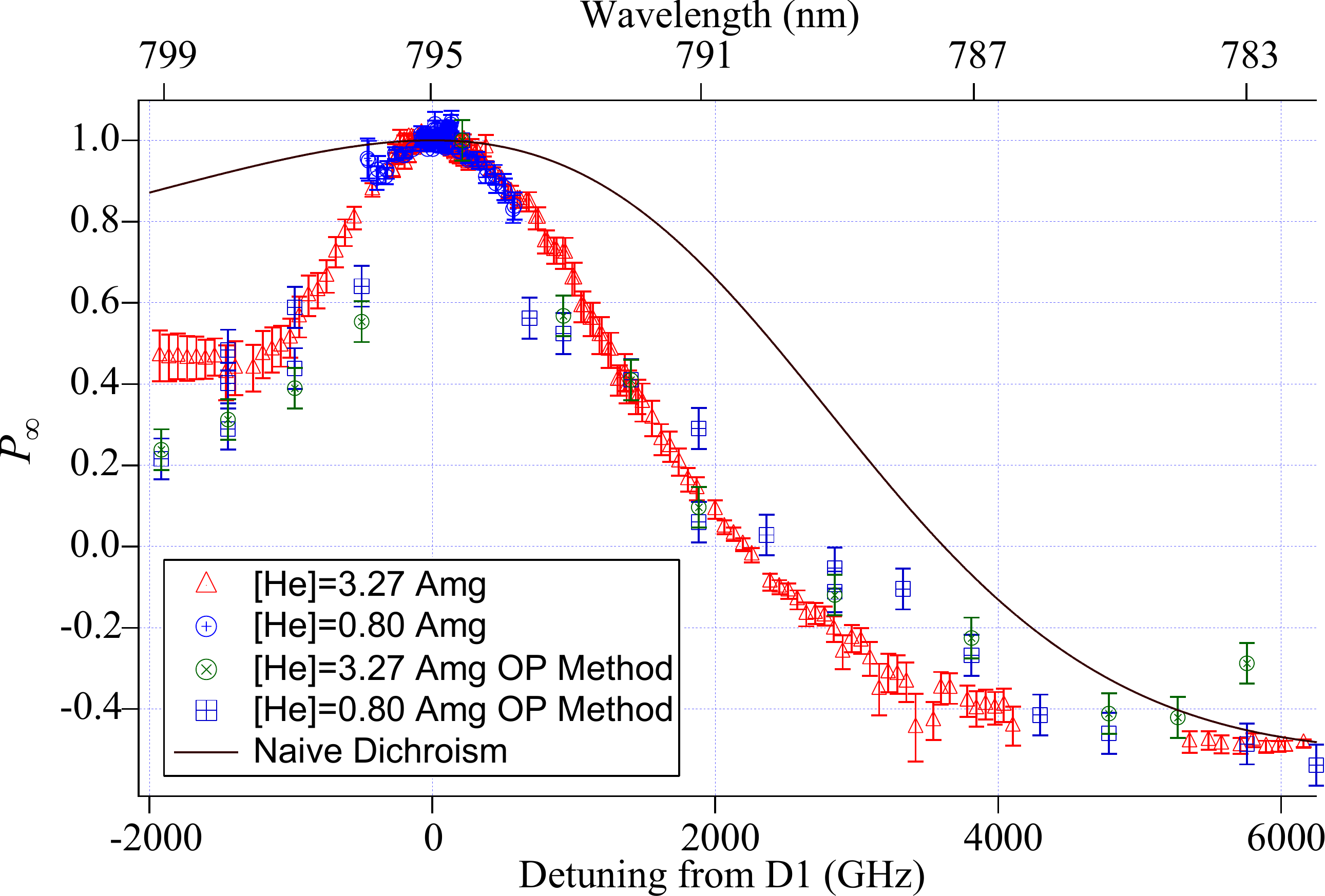}
\caption{Circular dichroism of Rb atoms in the presence of He  gas.  Near the D1 line, the dichroism approaches $1$, reaching $-1/2$ for the D2 line.  The solid line shows the dichroism neglecting He collisions.  The very significant reduction near the D1 line is responsible for excess photon absorption under SEOP conditions.  Adapted from \citet{Lancor10prl}.
}
\label{fig:pinf}
\end{figure}

Figure~\ref{fig:pinf} shows measurements of the circular dichroism $P_\infty$ of \Thad{maximally} polarized Rb atoms in the presence of \He3 gas \citep{Lancor10prl,Lancor10_circ}.  While the dichroism peaks at very close to 1 right on resonance,  at only 1 nm (\Thad{475} GHz) detuning from the line center, the dichroism has already dropped to 0.9. The strong reduction of the circular dichroism for 
off-resonant pumping implies that narrowband lasers will not only yield more efficient
optical pumping, but also higher maximum polarization.

For narrowband pumping in the bulk of the cell, the photon flux density $I(\bf r)$ (cm$^{-2}$s$^{-1}$) obeys (suppressing from now on the implied spatial dependence of $I$, $P_A$,  and $\cal A$)
\Thad{\be
{d I\over dz}=-\Ascatt\dens{A}=-\dens{A}\Gamma_A P_\infty P_A-\dens{A}\sigma I (1-P_\infty^2)
\label{lightprop}
\ee}
where the pumping rate and the photon flux density are related, $R=\sigma I$,  by the optical absorption cross section $\sigma$ for unpolarized atoms.  For pumping by resonant, narrowband light the first term, which leads to a linear decay $I(z)=I(0)-\dens{A}\Thad{\Gamma_A} P_\infty^2 z$ \cite{Bhaskar79,Walker97}, dominates.  The light is attenuated purely to compensate for loss of angular momentum in ground state collisions.  The transparency of nearly fully polarized atoms allows the light to be only weakly attenuated despite the great optical thickness of SEOP cells.

When there are significant off-resonant light components, however, the second term becomes important.  For high optical pumping rates, $P_A\approx P_\infty$, and Eq.~\eqref{lightprop} leads to exponential attenuation of the flux density in the usual fashion, $I=I(0)\exp[-\dens{A}\sigma(1-P_\infty^2) z]$.  The absorption length of the light is increased by $1/(1-P_\infty^2)$ over what it would be for unpolarized atoms, but can still be smaller than the cell length when $P_\infty$ deviates from 1.  

{The excess photon scattering due to the circular dichroism effect is exacerbated by pumping with un-narrowed diode array bars with linewidths on the order of 2 nm (1000 GHz).  One can define a photon efficiency analogous to the ideal spin-exchange efficiency of Eq.~\eqref{eff}, namely the ratio of the rate of production of polarized \He3 to the photon scattering rate:
\be
\eta_\gamma={V\dens{He}dP_{\rm He}/dt\over \dens{A}V {\cal A}}=\eta_{\rm SE}{\Gamma_A P_A\over {\cal A}}.
\ee
For ideal circular dichroism, $P_{\infty}=1$, $\eta_\gamma=\eta_{\rm SE}$.  At high optical pumping rates, $\eta_\gamma\approx \eta_{\rm SE} {\Gamma_A P_A\over R(1-P_\infty^2)}$ which can be much less than $\eta_{\rm SE}$.  Measured photon efficiencies for pumping with a particularly poor laser are shown in Fig.~\ref{fig:hybrideff}.
}

\subsubsection{Laser linewidth issues}

The early laser-based SEOP experiments (late 1980's, early 1990s) with \He3 used Ar$^+$-pumped tunable dye lasers~\cite{Chupp87,Coulter90} or Ti:Sapphire lasers~\cite{Larson91} to optically pump Rb vapor.  With the notable exception of the early SLAC\footnote{Stanford Linear Accelerator Center,  Menlo Park, CA} \He3 targets that used up to six Ti:Sapphire lasers~\cite{Johnson95}, these large, expensive and often unreliable sources were generally limited to $<10$ W of pumping light.  The first application to SEOP with diode lasers was performed
with relatively low power individual diodes~\cite{Wagshul89}.   

With the introduction of inexpensive, high power laser diode array bars in the early 1990s, SEOP experiments rapidly converted to these new sources that were readily available at a fraction of the cost~\cite{Cummings95}.  The drawback of laser diode array bars is their relatively broad spectral profile (typically 2 nm or 900 GHz) that is not well-matched to the $~20$ GHz linewidth of Rb atoms at 1 amg of He density.  Thus only a small fraction of the laser light is actually available for resonant pumping of the Rb atoms.  It thus became common practice to use higher He densities (between 3 and 10 amagat)
and rely on high optical thickness of the SEOP cells to absorb a significant fraction of the broad spectral profile of the pumping light \cite{Driehuys96}. 
Such pressures were well matched to high luminosity targets for electron scattering.
For MRI pressure of a few bar is convenient for gas \Thad{delivery;  for this reason and for optical pumping efficiency} high He densities were also employed~\cite{Middleton95,MacFall96}.  

In practice, though these high power lasers were able to polarize larger volumes of \He3,
the attainable \He3 polarizations using these lasers were generally found to be limited to 50~\% or less for these applications. 
Examination of Fig.~\ref{fig:pinf} shows that a polarization reduction \Thad{could} be at least qualitatively explained because
of optical pumping by off-resonant light.  In optically thick SEOP cells this effect is exacerbated because the central core of the spectral profile
of the light is depleted in the front portion of the cell, so that in the back of the cell the atoms are generally being pumped by off-resonant light
with a substantially reduced value of $P_\infty$.  The broad line width of diode lasers was a particularly bad match to neutron spin filters.
\Thad{Although pressures of 3 bar were employed in the early development of SEOP-based NSFs} ~\cite{Jones00,Gentile_sans00}, 
simpler construction and long lifetimes made pressures of closer to one bar preferable~\cite{Rich02}.

Attainable \He3 polarizations increased with the introduction of narrower diode laser sources, first using long external cavities \cite{Chann03,Babcock05laser} similar to those used in pulsed dye lasers and later using compact  volume holographic gratings  \Thad{(VHG)} \cite{Chenjap14,Vlodin04} 
Both narrowing methods generally reduce the laser linewidth to 0.2 nm (90 GHz), thus both giving a much better match to the atomic absorption spectrum and also increasing the value of $P_\infty$.  SEOP experiments utilizing such frequency narrowed lasers generally produce \He3 polarizations of 70\% or greater, and recently have demonstrated  85\% \cite{Chenjap14}.  \Thad{Scaling of narrowing techniques to stacks of diode laser bars \citep{Zhu05} have reached kilowatt levels \cite{Hersman16}.}

Full simulations of propagation of broad and narrowband laser light due to competing spin-relaxation and circular dichroism effects have been made for  SEOP in 1 to 3 bar spin-filter type cells  \citep{Lancor10_circ}, and for 8-10 bar cells \citep{Singh15}.  Such models generalize Eq.~\eqref{lightprop} to account for both spectral and spatial evolution of the optical pumping light, and do not yet include dual-sided pumping.  They also generally attempt to account for heating effects (Sec.~\ref{sec:heating}) in a simplified manner.  Such models generally predict higher polarizations than are observed experimentally, though they do semiquantitatively explain the much higher polarizations achieved with narrow-band pumping over broad-band pumping.  An important gap in the literature is a quantitative comparison of such models with three-dimensional alkali-metal polarization mapping techniques using  EPR spectroscopy \citep{Young97}.

\subsubsection{Heating} \label{sec:heating}

With the common use of 100 W scale diode lasers for SEOP, it is natural and important to ask about how this energy is dissipated in the vapor.  The first work addressing this topic was \citet{Walter01}.  They used Raman spectroscopy to measure the rotational and vibrational spectra of the N$_2$ molecules under SEOP conditions.  The picture painted by this study is as follows.  The N$_2$ quenching collisions \Thad{that are  so important for preventing radiation trapping} leave the N$_2$ molecules with 1.5 eV of excitation contained in 5 or 6 vibrational quanta.  The vibrational energy relaxes slowly, while the rotational degrees of freedom rapidly thermalize with the local translational temperature.  Therefore the rotational spectroscopy can be used to infer the internal temperature of the vapor.  The spatial dependence of the internal temperature was inconsistent with conductive heat transfer, indicating convective heat transport inside the cell.  While the convection tends to reduce temperature gradients, striking high temperature increases were \Thad{observed. For example, a 95$^\circ$C internal temperature increase with respect to the wall temperature occurred with only 22 W of deposited light power in an 8.4 bar cell. }

{\citet{Parnell10} measured the temperature rise in a 2.3 bar \He3 SEOP cell under illumination by a 100 W narrowed diode array bar.  Using a gradient spin-echo sequence, they were able to measure the spatial profile of the local diffusion coefficient and observed an increase in diffusion at the center of cell consistent with a 30 K temperature rise.  
Double cells provide a means for measuring the temperature rise during SEOP by the resulting drop in gas density in the SEOP cell, and a consequent increase in the non-SEOP cell.
\citet{Singh15} compared \He3 NMR signals with and without pumping light to deduce 20 to 50 K temperature increases for high pressure electron scattering cells.   \citet{Normand2016}  used neutron transmission to observe the reduction in the \He3 gas density in the SEOP cell under optical pumping conditions.  Their results indicated a remarkable 135 K temperature increase in a 1 bar  SEOP cell with 200 W of laser illumination.  Systematic studies using these kind of techniques \Thad{on } cells with varied nitrogen and \He3 pressures, hybrid mixtures, and laser illumination would fill in an important gap in our understanding of the the interior conditions of SEOP cells. }

\subsection{SEOP with pure K or Na}
As compared to traditional SEOP with rubidium, the efficiency of SEOP should be much greater for potassium 
due to its lower spin destruction rate~\cite{Baranga98,WalkerThywissen97}.  Increased spin-exchange efficiency for K, up to a factor of 10 above that of Rb for the same temperature, was observed \cite{Baranga98}. \Thad{  Enthusiasm for exploiting this substantial advantage has been tempered by potential pumping of the D2 line due to the small (3.4 nm) fine-structure splitting of the 4p state. The first reported SEOP application yielded 46~\% in a mid-sized double cell target,  using a Ti:Sapphire laser \cite{Wang03}
operating at the potassium 770 nm D1 transition wavelength. 770 nm diode lasers have had limited availability and lower power compared to their 795 nm counterparts for pumping Rb.  Thus hybrid SEOP with Rb/K mixtures has remained the preferred approach.  The single study of 770 nm pumping \citep{Chen07hyb}, which focused on NSF applications, did find increased efficiency as compared to pure Rb SEOP but also observed that the excess laser power demand and decline of alkali-metal polarization with increasing spin-exchange rate was similar to pure Rb cells, with poorer performance relative to Rb/K hybrid pumping.  These observations are not well-understood and merit future investigations.  Although 770 nm pumping was found to be preferable for the special case of high mixture ratio hybrid cells (see Sec. \ref{sec:hybrid}), it is generally desirable to use the same laser for all cells and so hybrid SEOP has remained favored.}
In principle, SEOP with Na  would yield even higher efficiency \cite{Borel03}
but practical application is hampered by cell browning at the required temperatures of $\ge$300~$^\circ$C
and the absence of convenient laser sources at 590 nm.

\subsection{Hybrid spin exchange}\label{sec:hybrid}

\begin{figure}
\includegraphics[width=3.3 in]{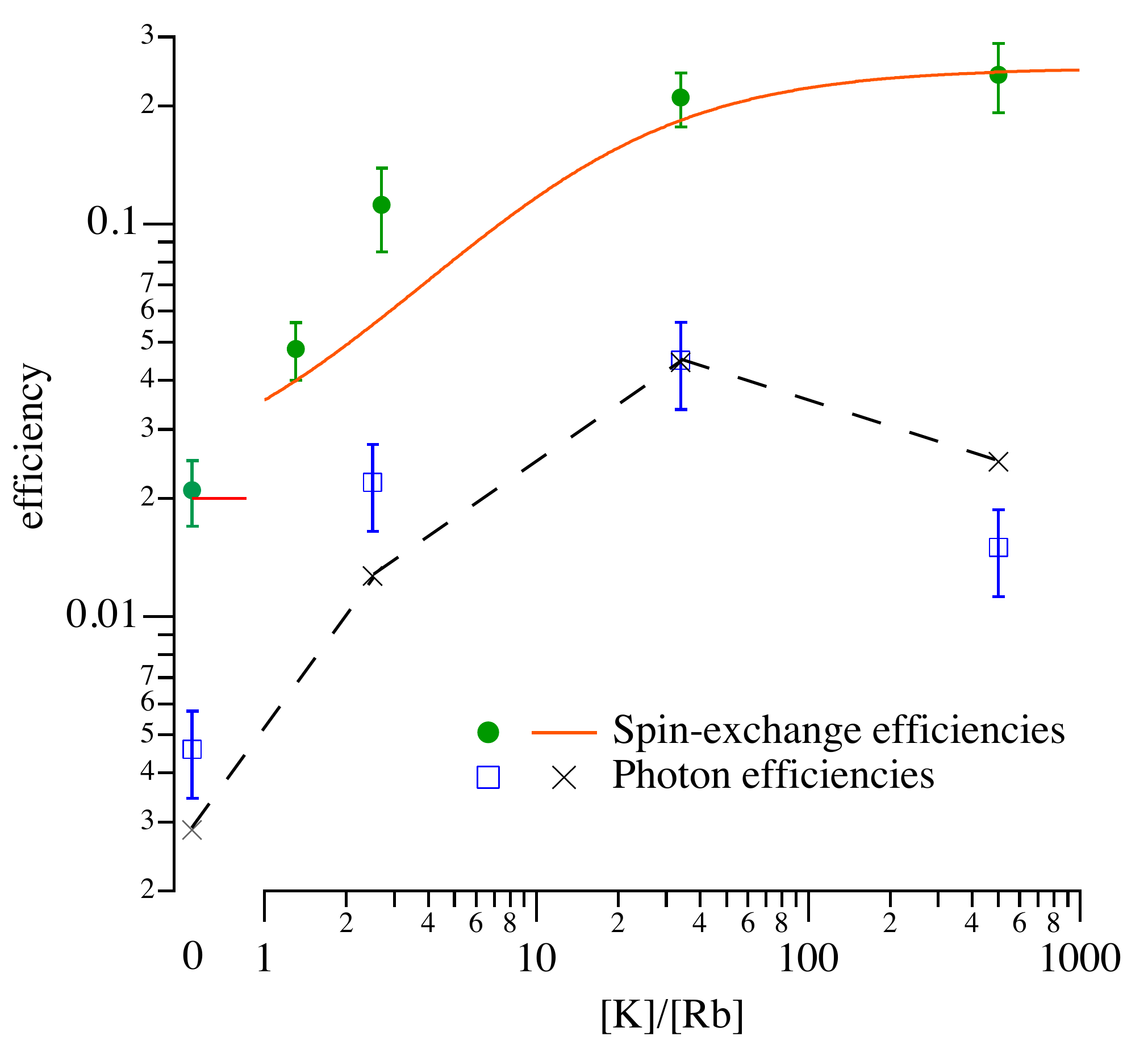}
\caption{Measured efficiencies at 190$^\circ$C as a function of \Thad{density ratio ${\cal D}=\dens{K}/\dens{Rb}$.  } The spin-exchange efficiency, $\eta_{SE}$, which is the maximum possible efficiency with which the angular momentum of the pumping light can be transferred to the nuclei,  shows the \Thad{clear increases predicted by Eq.~\eqref{KRbeff}, solid line, as the vapor approaches pure K.  Measured and modeled photon }efficiencies, $\eta_\gamma$, are much smaller, thanks to dichroism effects, see Sec~\ref{sec:lightprop}.  From \citet{Babcock03}.
}
\label{fig:hybrideff}
\end{figure}

The idea of hybrid SEOP is to optically pump Rb in the presence of a higher  density vapor of K.  Rapid spin-exchange collisions between the Rb and K atoms lead to spin-temperature equilibrium between them; thus the K atoms become collisionally polarized to a level equal to that of the Rb atoms.  The K-rich K-Rb vapor loses less angular momentum per atom due to spin-relaxation as compared to
a pure Rb vapor, so that the fraction of the angular momentum of the light that gets transferred to \He3 increases.   The spin-relaxation rate of the Rb atoms in the presence of K is increased due to spin-exchange collisions with the K atoms.
The  rapid K-Rb spin-exchange thereby causes the effective Rb spin-relaxation rate to increase from its K-free value
$\Gamma_{\rm Rb}$ to 
\begin{equation}
\Gamma'_{\rm Rb}=\Gamma_{\rm Rb}+\densrat\Gamma_{\rm K} \label{gkrb}
\end{equation}
where $\densrat={\dens{K}/\dens{Rb }}$ and the total K relaxation rate is $\Gamma_{\rm K}$.  The spin-exchange efficiency becomes
\def\se#1{{k_{SE}^{\rm #1}}}
\begin{equation}
\eta_{\rm SE}^{\rm KRb}={\left(\se{Rb}+\densrat  \se{K}\right)  \dens{^3He}\over \Gamma_{\rm Rb}+\densrat\Gamma_{\rm K}
}
\label{KRbeff}
\end{equation}
which approaches $\eta_{\rm SE}^{\rm K}$ at large $\densrat$.
Measurements of the spin-exchange efficiency as a function of $\densrat$ are shown in Fig.~\ref{fig:hybrideff}.

\begin{figure}
\includegraphics[width=3.0 in]{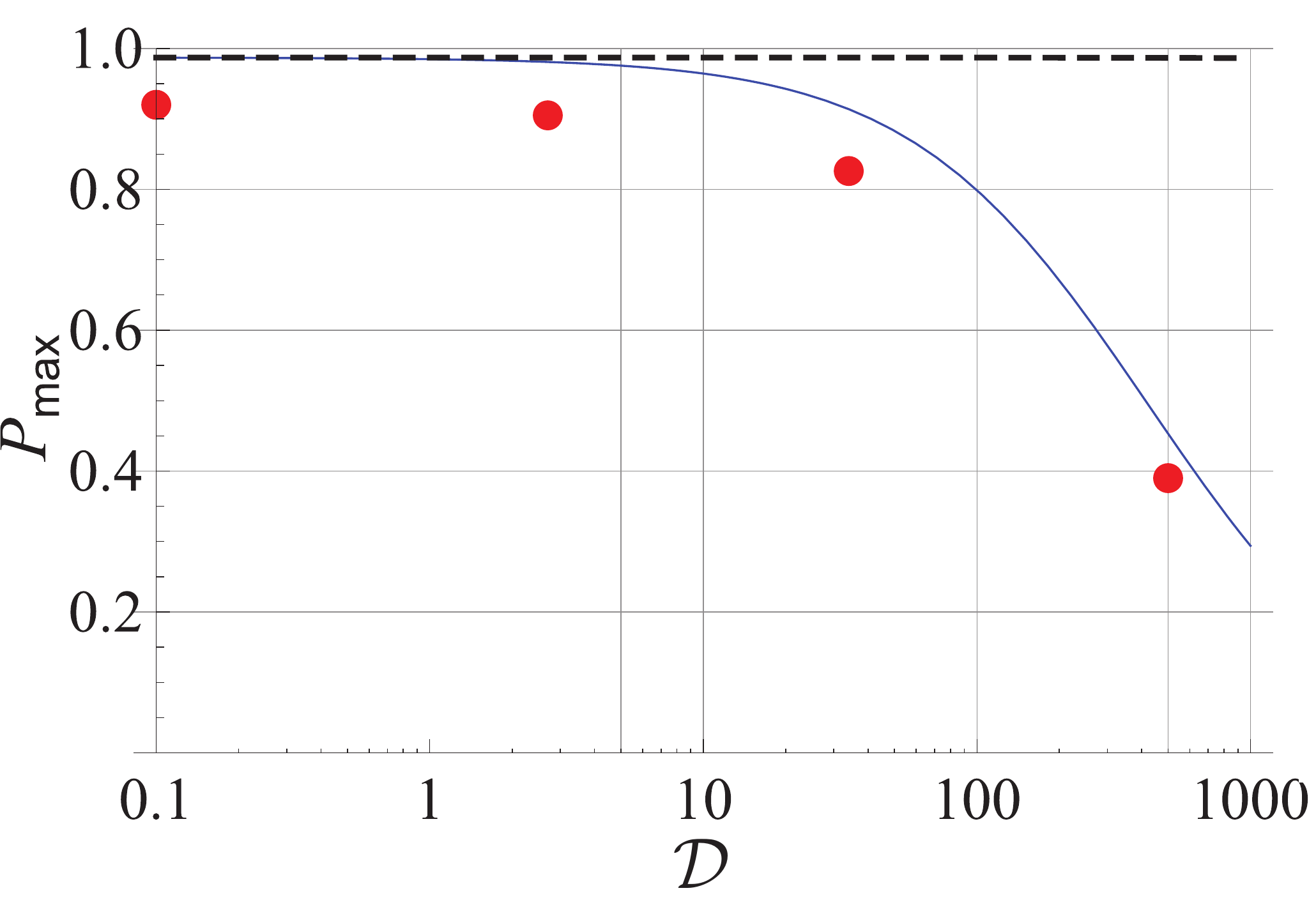}
\caption{Alkali polarization as a function of K-Rb density ratio.  The \Thad{ solid line} shows the density limit for a narrowband pumping laser, using the measured K absorption cross section at 795 nm.  The dots are experimental measurements using a broad pumping laser whose maximum polarization is limited by the dichroism effect to 0.92.  \Thad{The dashed line shows expected polarization limits predicted from a naive line-broadening model.}  From \citet{Lancor11}.
}
\label{hybridpol}
\end{figure}

The earliest experiments indicated that the maximum attainable alkali polarization drops at high $\cal D$ (Fig.~\ref{hybridpol}).  This comes from off-resonant pumping of the K atoms by the resonant Rb light \citep{Lancor11} and limits the useful density ratios to ${\cal D}<10$.  \citet{Chen07hyb} observed similar results and found
that the polarization decline was not observed if hybrid cells were optically pumped with 770 nm light.

The reduced collisional loss per atom for hybrid pumping means that for a given laser power the
volume of \He3\ can be increased or the alkali density and thus the polarizing rate can be increased. 
Practical application of hybrid SEOP has been investigated for neutron spin filters~\cite{Chen07hyb,Chen11}
and targets for electron scattering~\cite{Singh15,Ye13}.  Vapor mixture ratios, ${\cal D}$, between 2 and 7 were found
to yield the best results.  Because of the difference in vapor pressures for a given temperature,
$\approx25$ times more condensed phase K than Rb is required to yield ${\cal D}=4$.
In \citet{Chen07hyb,Chen11}, individual Rb and K distillation is described \Thad{whereas \citet{Singh15}
describe pre-mixing Rb and K in a glove box}.  Variations in ${\cal D}$
occur in both methods but can be minimized; in the individual approach ${\cal D}$ can be checked
before the cell is sealed off and in the pre-mix approach experimental feedback on the pre-mix ratio
to account for fractional distillation improves the reproduceability.   For NSFs ${\cal D}$ has been determined by white light
absorption and the pumping rate for a given temperature~\cite{Chen07hyb,Chen11} and for targets
by laser light absorption and Faraday rotation~\cite{Singh15}. \Thad{After a newly filled cell has been heated one or more times, the mixture ratio may increase, perhaps due to curing effects \cite{Ma09}.}

\indent  Although Fig.~\ref{fig:hybrideff} indicates six times higher efficiency for hybrid SEOP in the
optimum regime, hybrid cells are operated at typically 40$^\circ$C higher temperature than pure Rb cells in order to obtain the same spin-exchange rate.
For NSF cells at pressures near 1 bar, alkali-metal spin destruction from alkali-alkali collisions
dominates, which decreases the efficiency and/or rate gain.   In this regime, a resulting efficiency or rate gain
approaching 3 or $\sqrt{3}$, respectively, has been calculated and observed.
For high pressure targets, alkali-metal spin destruction from alkali- \He3\ collisions dominates,
and thus both the efficiency and rate gain is expected to be closer to that shown in Fig.~\ref{fig:hybrideff}.  
An additional benefit observed for hybrid cells is an observed slower decline of the
alkali-metal polarization with increasing spin-exchange rates.  
Whereas this result was found to be in agreement with modeling, a steeper decline
was observed for both pure Rb and pure K pumping in disagreement with modeling~\cite{Chen07hyb}.

Hybrid SEOP has substantially increased the quantity and production rate of polarized gas
for both NSFs and polarized targets.   For NSFs cells 85\% \He3\ polarization 
in cells approaching one liter in volume with pumping time constants of
between 4 h and 8 h~\cite{Chen14} has been achieved and for double cell polarized targets up to
70\% has been reached with similar time constants~\cite{Singh15}.

\subsection{Limits to \He3 polarization} \label{SEOPlimits}

Combining the capability of polarizing high density alkali vapors to near 100\% polarization, the developments of glass cells with hundreds of hour room temperature wall relaxation times, and rapid spin-exchange with hybrid pumping,  the attainable \He3 polarizations should be nearly 100\%.  The issues limiting the polarization are not yet fully understood.  We briefly review here what is known about this issue.  More details on most of these topics are discussed in other sections of this review.

\subsubsection{Anisotropic spin exchange}

\citet{Walter98} pointed out that the long-range portion of the \He3 nuclear magnetic field causes an additional term in the alkali-\He3 Hamiltonian:
\be
V_\beta=\beta(\xi){\bf S}\cdot(3\hat\xi\hat\xi-1)\cdot \Thad{{\bf I}_{\rm He}}
\label{aniso}
\ee
This anisotropic spin-exchange interaction tends to polarize the \He3 nuclei towards $P_{\rm He}=-P_A/2$, and so serves to limit the maximum \He3 polarization to
\be
P_{\rm He\infty}\le 1-{3k_\beta\over 2 k_\alpha}
\ee
where $k_\alpha$ and $k_\beta$ are the respective rate coefficients for the (isotropic) Fermi-contact interaction and the anisotropic interaction in Eq.~\eqref{aniso}.
A theoretical estimate of the effect of anisotropic spin-exchange yielded $P_{\rm He\infty}=0.95$
for SEOP with either Rb or K~\cite{Walter98}, and more recent calculations yielded a similar value of 0.96~\cite{Tscherbul11}.

There are no definitive experimental measurements of anisotropic spin-exchange.  Wall-independent techniques (\ref{semeas}) for measuring spin-exchange rates are sensitive to the combination $k_\alpha-k_\beta/2$.  As discussed below (Sec. \ref{sec:X}), spin-exchange transients are generally sensitive to $k_{\rm SE}\dens{A}+\Gamma_{\rm w}$, where $k_{\rm SE}=k_\alpha+k_\beta$, but the strong temperature dependence of $\Gamma_{\rm w}$
makes it difficult to isolate  $k_\beta$ from $\Gamma_{\rm w}$. However, \citet{Walker10} pointed out that comparison of spin-relaxation rates of K in \He3 and $^4$He, when combined with $k_{\rm SE}$ measurements, allow $k_\beta$ to be isolated without any assumptions about wall relaxation.  A first attempt gave $P_{\rm He\infty}=0.90\pm0.11$.

An upper limit on anisotropic spin-exchange can be deduced from absolute \He3 polarimetry.  The highest \He3 polarization obtained to date is $0.88\pm0.03$ \citep{Chen} in a ${\cal D}=4$ K-Rb hybrid cell,  giving $k_\beta\le(0.09\pm0.02) k_\alpha$.

\subsubsection{X-factor} \label{sec:X}

\begin{figure}
\includegraphics[width=3.3 in]{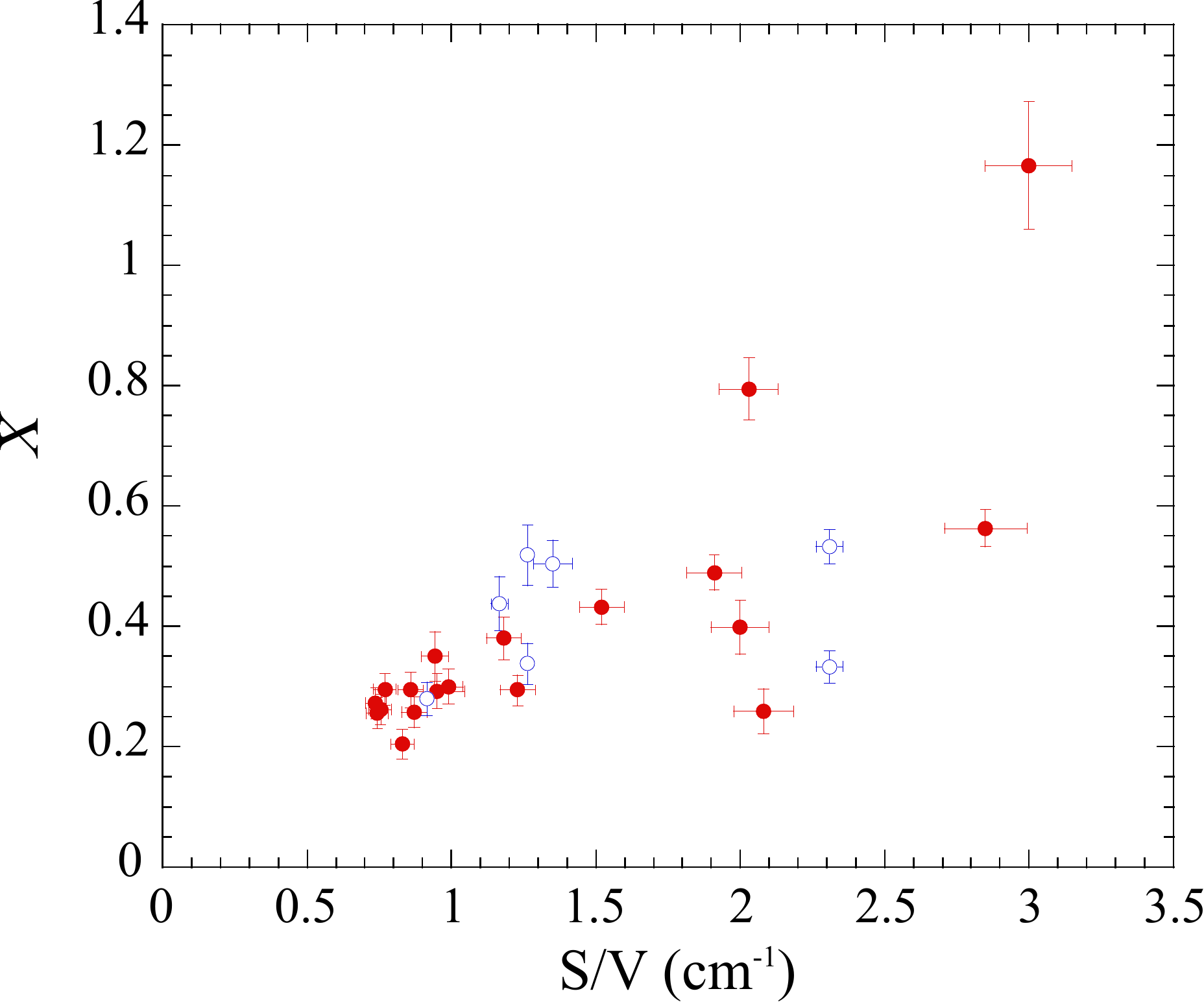}
\caption{NIST measurements of the X-factor, deduced from \He3 polarization limits at high temperature, for \Thad{ both blown (filled, red) and flat-windowed (open,blue) } neutron spin filter cells with a range of  surface to volume ratios $S/V$. Adapted from \citet{Babcock06}.
}
\label{XFactorData}
\end{figure}

Studies of the time constants for spin-exchange obey the phenomenological relation \citep{Chann02c,Chann03,Babcock06,Chen07hyb,Walker11,Chenjap14,Singh15}
\be
{1\over \tau}=k_{SE}(1+X)\dens{A}+\Gamma_{\rm r},
\ee
where $\Gamma_{\rm r}$ is the room temperature relaxation rate, taken to be independent of temperature.  This relation, experimentally verified for both Rb and K-Rb hybrid cells, indicates that the wall relaxation rate has an exponential increase with temperature that mimics the variation of alkali vapor pressure with temperature.  This limits the \He3 polarization to 
\be
P_{\rm He\infty}\le{1\over 1+X}
\ee
for 100\% Rb polarization and negligible $\Gamma_{\rm r}$.
That the "X-factor" originates from a temperature dependence of the wall relaxation rate is suggested by the tremendous variations in measured values of $X$ for many different cells.  Figure~\ref{XFactorData} shows a sample of such data.  If it is a wall relaxation effect, $X=\chi S/V$ should hold, where $S/V$ is the cell surface-to-volume ratio and the relaxivity $\chi$ is a poorly controlled parameter.  Indeed, there are greater fluctuations in $X$ for small cells (large $S/V$) than for large cells \cite{Walker11}. For large cells (eg. 
$S/V\le1$~\Thad{cm$^{-1}$})  $X$ was found to be typically $\approx$0.3,
thus limiting the maximum \He3\ polarization to between 75~\% and 80~\%.  Direct determinations of $X$
by measurements of the relaxation of heated cells were found to be consistent with 
measurements of the maximum \He3 \ polarization. 

Further determinations of $X$ in spin filter cells \cite{Parnell09} and double cells for \He3\ targets \cite{Ye10,Singh15} have been reported.
The temperature dependence of relaxation in quartz cells has also been studied
with deviation from the characteristic linear dependence on alkali-metal density \cite{Ino07}
and larger $X$ values observed \cite{Ye13}.  For this reason quartz presents issues for
achieving the best \He3\ polarization values for SEOP, even if a long relaxation time
at room temperature can be obtained.

Recently, the polarization of NSFs \cite{Chenjap14} pumped by 100 W \Thad{VHG-narrowed} lasers have found polarization limits over 85\%, compared to the 75 to 80\% found in earlier studies \cite{Babcock06,Chen07hyb}.  These polarization increases were also observed in cells that were studied over a decade time scale.   The polarization increases with time (or perhaps equivalently, narrowband laser power) suggest that at least some of the polarization limits attributed to the X-factor may be insufficiently polarized Rb vapor in those early experiments.  However, the Rb polarizations measured in those experiments using EPR spectroscopy were found to be conservatively in excess of 95\%, generally 98\% or above \cite{BabcockPhD}.

\subsubsection{Rb polarization limits}

When the spin-exchange rates dominate over any other relaxation mechanisms, the \He3 polarization can never exceed the spatial average of the alkali polarization.  

The first rather obvious implication of this statement is that due to the high pressures used in SEOP, so that there is little alkali polarization diffusion, care must be taken to provide sufficient pumping light to all regions of the cell, including ensuring that the light exiting the cell is not fully depleted of useful pumping light.  Because most SEOP experiments use lasers with spectral linewidths greater than the pressure-broadened line width of the atoms, it is essential to measure the spectral profile of the transmitted light, checking that the useful resonant portion of the light is not fully depleted.

Due to the imperfect dichroism of Rb pumped in the presence of high pressure \He3 (Sec.~\ref{sec:lightprop}),  spectrally broad lasers have been observed to have 10\% reduction in their maximum Rb polarization, even under low Rb density conditions \cite{Babcock03}.  This effect is exacerbated as the light propagates through the optically thick cell and the resonant portion is depleted, further lowering the dichroism.  These effects are now understood to be largely responsible for the 20\% increases in polarization observed when experimenters first used frequency narrowed diode array bars \cite{Chann03,Gentile05_jnist}.

Due to the locally enhanced spin-relaxation of alkali atoms within a diffusion length of the wall, correspondingly large light absorption occurs at the entrance to the cell, leading to reduced pumping rates and Rb polarizations within the bulk of the cell.  For single frequency pumping with dye or Ti:Sapphire lasers, where the spectrum of the light is much narrower than that of the atoms, this effect can be somewhat mitigated by purposely tuning the laser off resonance \citep{Wagshul94}.  For broad or frequency narrowed diode array bars, typically up to 10\% of the useful light can be absorbed by this layer, again reducing pumping rates and hence alkali polarizations within the bulk of the cell. 

Finally, even for narrow band pumping hybrid K:Rb mixtures exhibit reduced maximum polarizations at high $\cal D$ due to off-resonant absorption of the Rb pumping light by the K atoms \cite{Lancor11}, see Fig~\ref{hybridpol}.  This effect limits the useful values of $\cal D$ $<10$. 

\subsubsection{EPR spectroscopy}

It is often useful to complement \He3 polarimetry with diagnostics to measure and even image the Rb polarization.  In the very simplest version, one can simply monitor the transmission of the optical pumping light.  Assuming the pump light is approximately circularly polarized, the light transmission is minimum when the laser is tuned directly on resonance.  Then as the circular polarization of the light is adjusted, the transmission will be a maximum when the pumping light is maximally circularly polarized.  The sensitivity of this technique is enhanced by monitoring the transmitted pumping light with a diffraction grating spectrometer.  Then the greater polarization sensitivity of the on-resonant spectral components is easy to observe.  Such a diagnostic can however only identify the alkali polarization as "high".  For a more sensitive alkali polarization monitor, a linearly polarized Faraday probe, co-propagating with the pumping light but tuned near the D2 optical resonance (so it can be distinguished from the pumping light by a spectrometer or interference filter), can be used.  The high optical depth in SEOP make the polarization rotations large ($\sim$rad) even for light tuned a number of linewidths off resonance.  The Faraday rotation is proportional to $P_{A}\dens{A}$, so absolute calibration of $P_A$ is difficult without a precise alkali number density measurement.  Indeed, assuming $P_A\approx 1$, the Faraday rotation can be used to measure $\dens{A}$ \cite{Chann02c}.

For a more quantitative alkali polarization assessment, electron paramagnetic resonance (EPR) spectra can be obtained by monitoring the transmitted intensity of the pumping light, Faraday rotation of a co-propagating probe laser, or residual cell fluorescence as a transverse oscillating magnetic field is swept through the alkali Zeeman resonances, slightly changing the  alkali polarization \cite{Chann02c,Romalis98,Kramer07}.  These detection modalities are all proportional to the longitudinal alkali spin-polarization, so the "longitudinal EPR" signal is proportional to the square of the oscillating magnetic field strength.
At small magnetic fields, the splitting between adjacent EPR lines arises largely from the second-order Zeeman effect and is $q (\mu_B B)^2/(h^2\delta\nu)$, where $\mu_B$ is the Bohr magneton, $\delta\nu$ the hyperfine splitting, $h$ is Planck's constant, and $q=2/9$ for $^{85}$Rb  with nuclear spin \Thad{$I_A=5/2$} and $q=1/2$ for the \Thad{$I_A=3/2$} species $^{87}$Rb  and K.  This neglects the nuclear Zeeman interaction, which slightly splits the RF resonances for the two hyperfine levels.  Under the strong pumping conditions typical for SEOP, the two primary EPR lines are $m=F_{\rm max}\rightarrow F_{\rm max}-1$ and $F_{\rm max}-1\rightarrow F_{\rm max}-2$, where the maximum angular momentum of the atom is \Thad{$F_{max}=I_A+1/2$}.  The  $F_{\rm max}$ line is narrowed due to having little spin-exchange broadening, while the $F_{\rm max}-1$ line experiences substantial spin-exchange broadening \cite{Appelt99}.  The Rb polarization is simply related to the areas of the RF resonances.  In $^{85}$Rb, for example, assuming spin-temperature equilibrium one gets
\be
P_{\rm Rb}=\frac{7 A_3-3 A_2}{7 A_3+3 A_2}
\ee
where $A_{m}$ is the area of the EPR peak originating from $m_F=m$.  This is valid when the $F,m=3,2$ and $2,2$ lines are not resolved.

A more sensitive technique for EPR alkali polarimetry, and the first to be demonstrated, is to directly detect the transverse alkali polarization induced by the oscillating magnetic field\cite{Young97,Baranga98}.  This is done by demodulating the Faraday rotation of a transversely propagating probe laser.  The transverse polarization is linear in the oscillating magnetic field amplitude, so weaker oscillating fields are required to observe these "transverse EPR" signals.  Again, ratios of the areas of resonance peaks allow the alkali polarization to be measured.  
\citet{LancorPhD} compared simultaneous longitudinal and transverse EPR signals using Faraday rotation of a skew off-resonant laser.  He observed that the spatially averaged alkali polarizations deduced by the two methods differed, with the longitudinal EPR generally yielding 0-10\% higher alkali polarization estimates than transverse EPR.   The origin of this discrepancy is not known, though both transverse and longitudinal methods gave distributions consistent with a spin-temperature.  Thus earlier studies relying on longitudinal EPR for alkali polarimetry may have overestimated alkali polarizations.

\citet{Young97} and \citet{Baranga98} also demonstrated the use of EPR spectroscopy for alkali polarization imaging.  In this case the most convenient approach is to work at low fields where the individual EPR lines are not resolved, and apply a longitudinal magnetic field gradient $\partial B_z/\partial x$ so that the Faraday rotation of the probe beam, propagating along the x-direction, arises solely from a small voxel of atoms that satisfy the EPR resonance condition.  Sweeping the longitudinal field $B_z$ then produces a spatial map of the Rb polarization along the x-direction.  Moving the probe laser allows the polarization to be mapped in 3-dimensions.

\subsection{EPR frequency shift} \label{EPRpol}

\Thad{The spin-polarized alkali-metal atoms that are present in the SEOP cell can serve as a sensitive \Thad{\it in-situ} magnetometer, detecting the magnetic field produced by the polarized \He3.}
The magnetic field at position ${\bf x}$ with respect to a spin-polarized nucleus with magnetic moment $\bf m$ is the sum of the classical dipole field and a contact term:  ${\bf B}({\bf x})=[3{\bf x x}\cdot {\bf m}- x^2{\bf m} ]|x|^{-5}+8\pi{\bf m}\delta({\bf x})/3$ \citep{Jackson}.    The average field experienced by the alkali atoms due to the polarized \He3 is  \citep{Schaefer89,Barton94,Romalis98} 
\Thad{\be
{B}_{\rm He}&=&{B_{\rm cl}}+{8\pi\mu_{\rm He}\over 3} \kappa_0 \dens{He} P_{\rm He}\\
&=&{8\pi\mu_{\rm He}\over 3} \kappa \dens{He} P_{\rm He}
\label{EPRfield}
\ee}
where \Thad{$\mu_{\rm He}$} is the magnetic moment of \He3, and $\kappa_0$ is a frequency shift enhancement factor whose value is proportional to the average alkali electron spin-density at the \He3 nucleus.  For a gas that is uniformly polarized inside a spherical cell, the spatial average of ${ B}_{\rm cl}$ vanishes so that only the contact term, responsible for the hyperfine interaction (second term of Eq.~\eqref{VAHe}), contributes, \Thad{$\kappa=\kappa_0$}.  The  field $B_{\rm He}$ produces an EPR frequency shift \Thad{$\delta\nu=\gamma_A B_{\rm He}=\gamma_S B_{\rm He}/(2I_A+1)$}, where  $\gamma_S\approx 28$ MHz/mT is the \Thad{electron gyromagnetic} ratio.  For the most common $^{85}$Rb isotope, the numerical value of the shift is $1.13\kappa_0$ kHz for fully polarized \He3 at a density of 1 amg.

Using two orientations of a long cylindrical cell, for which \Thad{${ B_{\rm cl}}$} can be accurately calculated, \citet{Romalis98} isolated the classical and contact contributions, thus enabling a precision measurement of $\kappa_0$ for Rb-\He3, with an uncertainty of 1.5\%.  This result agreed with a prior calibrated absolute NMR polarimetry measurement \citep{Newbury93} and forms the basis of precision absolute polarimetry of electron scattering targets.  EPR frequency shift polarimetry  has also been indirectly tested  \Thad{using} neutron transmission \citep{Ye13}.

The resulting alkali EPR frequency shifts $\delta\nu$ \citep{Schaefer89} are typically tens of kHz for high density \He3 SEOP and are easily measured using any of the EPR methods.  AFP is \Thad{often} used to briefly reverse the \He3 polarization with respect to the magnetic field and hence isolate the EPR frequency shift.  The results are improved when the bias external magnetic field is stabilized using an auxiliary magnetometer (fluxgate or atomic). \Tom{  Since the temperature dependence of $\kappa_0$  yields a typical change in the EPR frequency shift of 0.14\%/$^\circ$C, 
the gas temperature must be carefully determined in order to retain the full precision in the absolute polarimetry.  The effect of internal heating (Sec.~\ref{sec:heating}) has not generally been investigated as a source of error and uncertainty.
If unaccounted for, a 10 $^\circ$C temperature rise of the gas sampled for the EPR measurement would yield a 1.4~\% error, which
would be comparable to the uncertainty in current values for $\kappa_0$.}


Using the carefully measured Rb\He3 enhancement factor as a reference, the enhancement factors for K and Na have also been measured, and the temperature range for Rb-\He3 extended as needed for hybrid pumping \citep{Babcock05}.  The values of $\kappa_0(T)$ for the various alkali-metal atoms are given in Table.~\ref{setable}.  \Thad{The temperature dependence $d\kappa_0/dT$, also given in Table~\ref{setable}, is sufficiently large that 10\% variations are seen over common SEOP temperature ranges.}

%
%
%


\section{Metastability-Exchange Optical Pumping}

\label{Sec_MEOP}
Metastability-exchange optical pumping mainly involves two processes: optical pumping on the \PJ{optically} closed \esS-\esP\ transition of He at 1083~nm, \PJ{described in Sec. \ref{Subsec_OPinMEOP},} and nuclear orientation transfer to the ground state of \Het\ through metastability-exchange (ME) collisions, \PJ{described in Sec. \ref{Subsec_MEcoll}}. MEOP is usually performed in pure \Het\ gas, and it is often sufficient to consider these two processes for \Het\ atoms only. However, the addition of \Hef\ to \Het\ gas can lead to higher nuclear polarizations and faster build-up rates, with potential applications whenever the admixture of  \Hef\ atoms has no adverse effect (e.g., for neutron spin filters). In isotopic mixtures, optical pumping advantageously operates on \Hef\ atoms \cite{Stoltz96}. From another point of view, traces of  \Hef\ are often found in  \Het\ cells, resulting \PJ{for} instance from cell preparation or from permeation through glass walls. The impact of even sub-\%  \Hef\ fractions on MEOP efficiency or optical measurement of polarization can be significant \cite{BatzPhD,Talbot11}. Therefore the knowledge of features and processes relevant for MEOP may be needed and they will be examined for both helium isotopes in this section, \PJ{starting with the relevant He level structures in Sec. \ref{Subsec_levels}}.

\subsection{Atomic levels involved in MEOP}

\label{Subsec_levels} 
\begin{figure}[tbh]
\includegraphics[width=3.0 in]{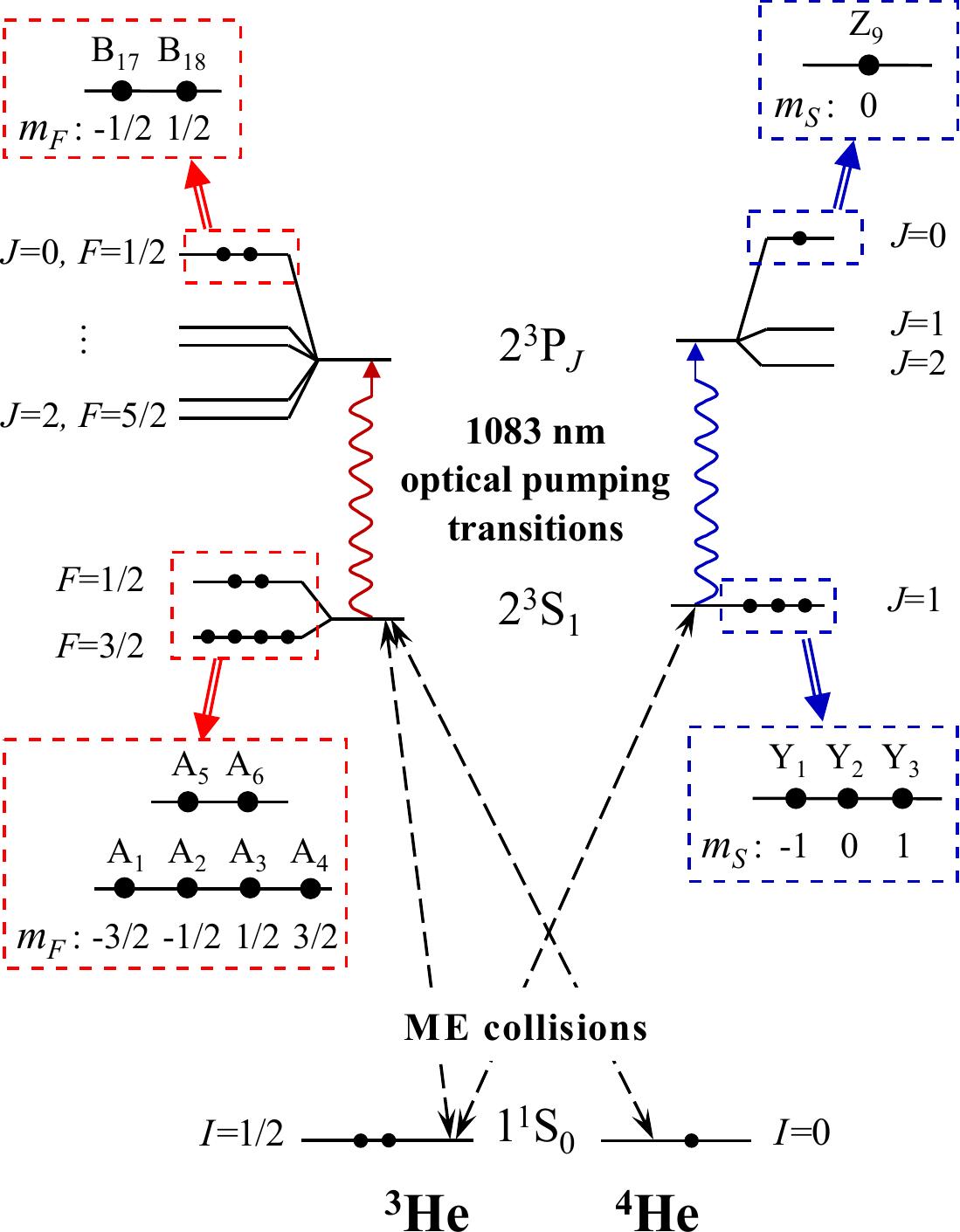}\caption{Fine- and
hyperfine-structures of the atomic states of He involved in the metastability-exchange optical pumping process,
for the \Het\ (left) and \Hef\ (right) isotopes, in low magnetic field (for
negligible magnetic Zeeman energies, i.e. below a few mT). The values of the
total angular momenta ($J$ for \Hef, $F$ for \Het) are indicated. Details and
names of the magnetic sublevels of the \esS\ and \esP$_{0}$ states are given
in blown-up boxes (with notations of \citealt{Courtade02}.) The shifts and splittings
are not displayed to scale.}%
\label{fig_levels}%
\end{figure}

Figure~\ref{fig_levels} schematically displays the optical pumping and ME collisions 
processes and the most relevant atomic levels for both He isotopes.
The \esS\ state of \Hef\ \PJ{($J=S=1$)} has three magnetic sublevels ($m_{S}=-1$,
0, and 1), linearly split at all values of the applied magnetic field $B$ by
the Zeeman energy. They are named Y$_{1}$ to Y$_{3}$ (for all sets of Zeeman sublevels indices increase with
increasing energies). The \esP\ state of \Hef\ has three fine-structure levels
with $J=0$, 1 and 2, hence nine Zeeman sublevels (Z$_{1}$ to Z$_{9}$). Due to its two nuclear spin states \Het\ has
twice as many Zeeman sublevels: six in the \esS\ state (A$_{1}$ to A$_{6}$) and eighteen in the \esP\ state (B$_{1}$ to
B$_{18}$). 
The magnetic sublevels A$_{i}$ can
be written using the decoupled basis states $\left\vert m_{S},m_{I}%
\right\rangle .$ A$_{1}=\left\vert -1,-\right\rangle $ and A$_{4}=\left\vert
1,+\right\rangle $ are pure states while the states for which $m_{F}=\pm1/2$
involve two mixing parameters $\theta_{\pm}$~\cite{Courtade02}:
\begin{align}
\mathrm{A}_{2} &  =\cos\theta_{-}\left\vert -1,+\right\rangle +\sin\theta
_{-}\left\vert 0,-\right\rangle \nonumber\\
\mathrm{A}_{3} &  =\cos\theta_{+}\left\vert 0,+\right\rangle +\sin\theta
_{+}\left\vert 1,-\right\rangle \nonumber\\
\mathrm{A}_{5} &  =\cos\theta_{-}\left\vert 0,-\right\rangle -\sin\theta
_{-}\left\vert -1,+\right\rangle \nonumber\\
\mathrm{A}_{6} &  =\cos\theta_{+}\left\vert 1,-\right\rangle -\sin\theta
_{+}\left\vert 0,+\right\rangle .\label{eq:Ai}%
\end{align}
For simplicity, only the
highest-lying Zeeman sublevels of the \esP$_{0}$ states are displayed in
Fig.~\ref{fig_levels}.

At low magnetic field, the $F=3/2$ and $F=1/2$ hyperfine levels of the
\esS\ state of \Het\ are well resolved and split by 6.74~GHz. Strong mixing of electronic and nuclear angular momenta yields large values of the mixing parameters $\theta_{\pm}$: $\sin^2\theta_{-}=2/3$ and $\sin^2\theta_{+}=1/3$. 
The \esP\ states
levels extend over 32.91~GHz for \Hef\ and 32.60~GHz for \Het. The isotope
shifts of the energy levels result in an increased optical pumping transition energy (by
33.67~GHz) and a higher \esS\ state energy (by 259.6~GHz, i.e. 12~K in temperature units) for
\Hef\ \cite{Drake2005}.

At high magnetic field, the Zeeman energy is no longer a small correction to the fine- and hyperfine-structure
energy contributions and the angular momentum structures of the \esS\ and \esP\ levels are deeply modified. For instance, the mixing parameters in Eq.~\eqref{eq:Ai} have small and almost equal values $\theta_{\pm}\approx 0.11/B,$ with \PJ{the field strength} $B$ in T. This has a strong impact on ME and optical pumping mechanisms alike. Therefore, in each of the following sections, the low-field and high-field situations will be separately discussed as limiting cases. Traditional low-field corresponds to field strengths $B\leq10 ~\mathrm{mT}$, with $\theta_{\pm}$ decreasing by less than 3\% and Zeeman shifts significantly lower than Doppler widths at room temperature. Conversely, $B\geq1~\mathrm{T}$ will be considered a high-field situation \PJ{with small mixing parameters $\theta_{\pm}$ and a strongly modified level structure (see Sec. \ref{secHiBOP}).}

\subsection{Metastability-exchange collisions}

\label{Subsec_MEcoll}

\subsubsection{Treatment of the \Thad{metastability-exchange} collision process}

Metastability exchange collisions are binary collisions in which the
electronic excitation is transferred from a \Hemet\ atom in the metastable
\esS\ state to one in the ground state. 
For a gas confined in a cell, the quantum states of the atoms can be statistically described by density operators depending on atomic positions and velocities. The position-dependent parameters include the local plasma characteristics and the optical pumping light intensity. This intensity has a spatial dependence because of the pumping beam transverse shape and divergence (if any) and of atomic absorption.
The velocity dependence mainly results from the velocity-selective character of optical pumping with a narrowband laser and will be addressed in Sec.~\ref{sec:LowBOP} below. Using a standard approach \cite{Partridge66}, the effect of a ME collision between
\Hemet\ and He atoms characterized by density operators $\rho_{\mathrm{m}}$
and $\rho_{\mathrm{g}}$ can be described as:
\begin{equation}%
\begin{array}
[c]{ccccccc}%
\mathrm{He}^{\mathrm{\ast}} & + & \mathrm{He} & \rightarrow & \mathrm{He} &
+ & \mathrm{He}^{\mathrm{\ast}}\\
\rho_{\mathrm{m}} &  & \rho_{\mathrm{g}} &  & \rho_{\mathrm{g}}^{\prime} &  & \rho_{\mathrm{m}%
}^{\prime}.
\end{array}
\label{MEcollprocess}%
\end{equation}
The \ ground state density operator of \Het\ is directly related to the 
populations of the two spin states $\left\vert +\right\rangle $ and
$\left\vert -\right\rangle ,$ which are functions of the nuclear polarization
$P_{\mathrm{He}}$:
\begin{equation}
\rho_{\mathrm{g}}=\frac{1+P_{\mathrm{He}}}{2}\left\vert +\right\rangle
\left\langle +\right\vert +\frac{1-P_{\mathrm{He}}}{2}\left\vert
-\right\rangle \left\langle -\right\vert ,\; \label{defrhog}%
\end{equation}
\newline while the \Hemet\ density operator can be written as:%
\begin{equation}
\rho_{\mathrm{m}}=\sum\nolimits_{1}^{3}y_{i}\left\vert \mathrm{Y}%
_{i}\right\rangle \left\langle \mathrm{Y}_{i}\right\vert \;%
\mbox{ or }%
\;\rho_{\mathrm{m}}=\sum\nolimits_{1}^{6}a_{i}\left\vert \mathrm{A}%
_{i}\right\rangle \left\langle \mathrm{A}_{i}\right\vert \label{defrhos}%
\end{equation}
for $^{4}\mathrm{He}^{\mathrm{\ast}}$ or $^{3}\mathrm{He}^{\mathrm{\ast}}$,
respectively. The relative populations of the \esS\ state sublevels $\mathrm{Y}_{i}$ and $\mathrm{A}_{i}$
are noted $y_{i}$ and $a_{i}$, respectively, with $\sum y_{i}=\sum a_{i}=1.$ 
The local number density of \esS\ state atoms in the Zeeman sublevels are thus $n_\mathrm{m}^{(4)}y_i$ and $n_\mathrm{m}^{(3)}a_i$, where $n_\mathrm{m}^{(4)}$ and $n_\mathrm{m}^{(3)}$ are the local number densities of \esS\ state atoms of each isotope. Their ratio depends on the gas composition and temperature, due to the lower energy of the $^{3}\mathrm{He}^{\mathrm{\ast}}$ state \cite{Courtade02,NacherPhD,Zhitnikov75}. At room temperature ($T\gg$12~K), $n_\mathrm{m}^{(4)}/n_\mathrm{m}^{(3)}=N_\mathrm{g}^{(4)}/N_\mathrm{g}^{(3)}$ where $N_\mathrm{g}^{(4)}$ and $N_\mathrm{g}^{(3)}$ are the number densities of ground state atoms of the isotopes. Number densities are simply noted $n_\mathrm{m}$ and $N_\mathrm{g}$ in a pure gas.

For an outgoing \Het\ atom:
\begin{equation}
\rho_{\mathrm{g}}^{\prime}=\operatorname{Tr}_{e}\rho_{\mathrm{m}} \label{defrhoprimg}%
\end{equation}
is simply obtained using the trace operator $\operatorname{Tr}_{e}$ over the
electronic variables of the incoming $^{3}$\Hemet\ atom. For the outgoing
\Hemet\ atom in \Het-$^{3}\mathrm{He}^{\mathrm{\ast}}$ collisions the density
operator is written as:
\begin{equation}
\rho_{\mathrm{m}}^{\prime}=\sum\nolimits_{F}P_{F}\rho_{\mathrm{m}}^{\prime\prime}P_{F}\;%
\mbox{ with }%
\;\rho_{\mathrm{m}}^{\prime\prime}=\rho_{\mathrm{g}}\otimes\operatorname{Tr}_{n}%
\rho_{\mathrm{m}}. \label{defrhoprims}%
\end{equation}
The tensor product $\rho_{\mathrm{m}}^{\prime\prime}$ is the density operator
immediately after the collision. Off-diagonal terms (coherences)
between the $F=1/2$ and $F=3/2$ hyperfine sublevels are created during metastability exchange collisions, but
they can be neglected for usual MEOP conditions, at low enough pressure or high enough field \cite{Courtade02}. 
Therefore the restricted expression of $\rho_{\mathrm{m}}^{\prime}$ can be used 
(Eq.~\eqref{defrhoprims}, where $P_{F}$ is the
projection operator onto the $F$ hyperfine substate).

Note that off-diagonal elements can also be created if
coherent light is used for pumping with a V-type or $\Lambda$-type scheme, with two optical transitions addressing at least
one common sublevel in the lower state or the upper state, respectively. However, such schemes are avoided in MEOP
since well-polarized light of given helicity is used.

For isotopic He mixtures, Eq.~\eqref{MEcollprocess} generically refers to three
different ME processes depending on the isotopic nature of the colliding
atoms \PJ{(see Fig.~\ref{fig_levels} , dashed lines)}. Two additional equations, similar to Eq.~\eqref{defrhoprims}, are used
for collisions between atoms of different isotopes \cite{Courtade02}. The
fourth type of collisions, between $^{4}\mathrm{He}^{\mathrm{\ast}}$ and
\Hef\ atoms, plays no role in the evolution of nuclear or electronic spin
variables, and therefore no direct link is sketched in Fig.~\ref{fig_levels}.

The initial approach of Partridge and Series was subsequently improved to
quantitatively link the time evolution of selected atomic observables with ME
collision \PJ{cross sections} in weak pumping and low polarization conditions
\cite{Dupont-Roc71,Dupont-Roc73ab,Pinard80}. It was then used in models
suitable for the description of MEOP with lasers, in which the evolution of
all Zeeman sublevel populations is evaluated for arbitrary pumping conditions
\cite{Nacher85,Batz11}. Rate equations are derived from Eq.~\eqref{defrhoprims}
for the populations in the \esS\ state. They explicitly depend linearly on
$P_{\mathrm{He}}$ due to the linear dependence of $\rho_{\mathrm{g}}$ on
$P_{\mathrm{He}}$ in Eq.~\eqref{defrhog}. For pure $^{3}$He they are written
as:
\begin{equation}
\left.  \frac{da_{i}}{dt}\right\vert _{\mathrm{ME}}=\gamma_{\mathrm{e}}\left[
-a_{i}+\sum_{k=1}^{6}(E_{ik}^{(3)}+P_{\mathrm{He}}F_{ik}^{(3)})a_{k}\right]  .
\label{eq:adotME}%
\end{equation}
The ME collision rate $\gamma_{\mathrm{e}}=N_{\mathrm{g}}k_{\mathrm{ME}}$ is
proportional to the number density of atoms in the ground state 
and depends on temperature through the ME rate coefficient
$k_{\mathrm{ME}}$ (Sec.~\ref{Subsec_MEcollvsT} below). The matrices $E^{(3)}$
and $F^{(3)}$ involve $B$-dependent parameters \cite[Tables 16 and
17]{Courtade02}. \PJ{For} the ground state, the contribution of ME
collisions to the rate equation describing the evolution of $P_{\mathrm{He}}$
is obtained by computing $\operatorname{Tr}_{n}\rho_{\mathrm{g}}I_{z}$ using
Eqs.~\eqref{defrhoprimg} and \eqref{defrhos}:%
\begin{align}
\left.  \frac{dP_{\mathrm{He}}}{dt}\right\vert _{\mathrm{ME}}  &
=\gamma_{\mathrm{e}}\int_{\mathrm{cell}}\frac{d^{3}\mathbf{r}}%
{V_{\mathrm{cell}}}\frac{n_{\mathrm{m}}}{N_{\mathrm{g}}}%
\left(  P_{\mathrm{He}^{\mathrm{\ast}}}-P_{\mathrm{He}}\right)  ,\;\label{eq:Mdot}\\%
\mbox{ where }%
\;P_{\mathrm{He}^{\mathrm{\ast}}}  &  =\sum_{k=1}^{6}L_{k}a_{k}.
\label{eq:Mdotline2}%
\end{align}
The nuclear polarization is usually uniform in the ground state since the diffusion
rate in a low pressure gas is much larger than typical rates of
evolution for $P_{\mathrm{He}}$. On the contrary, both the nuclear polarization
$P_{\mathrm{He}^{\mathrm{\ast}}}$ and the density $n_{\mathrm{m}}$
of atoms in the \esS\ state depend on local pumping light and discharge intensities and 
may strongly vary with position. Hence Eq.~\eqref{eq:Mdot} involves a spatial average over the cell volume
$V_{\mathrm{cell}}$. $P_{\mathrm{He}^{\mathrm{\ast}}}$ is directly derived from the set of
populations using the field-dependent parameters $L_{k}$ \cite[Table~13]%
{Courtade02}. 

This formalism has been extended to mixtures of \Het\ and \Hef, and the corresponding rate
equations have been established\ \cite{Courtade02}. If isotopic mixtures are optically pumped using a \Hef\ atomic transition, the electronic polarization created in the \esS\ state of \Hef\ atoms is first transferred to the \esS\ state of \Het\ atoms by ME collisions with ground state \Het\ atoms. %
\PJ{The nuclear polarization which subsequently develops in $^{3}\mathrm{He}^{\mathrm{\ast}}$ atoms due to hyperfine interactions} is then transferred to the ground state of  \Het\ atoms via further ME collisions. The larger light absorption probability of  \Hef\ atoms (see Sec.~\ref{sec:LowBOP}) contributes to make this indirect process more efficient \cite{Stoltz96}.


\subsubsection{Temperature dependence of \Thad{metastability-exchange}  collision rates}

\label{Subsec_MEcollvsT}

\begin{figure}[tbh]
\includegraphics[width=3.0 in]{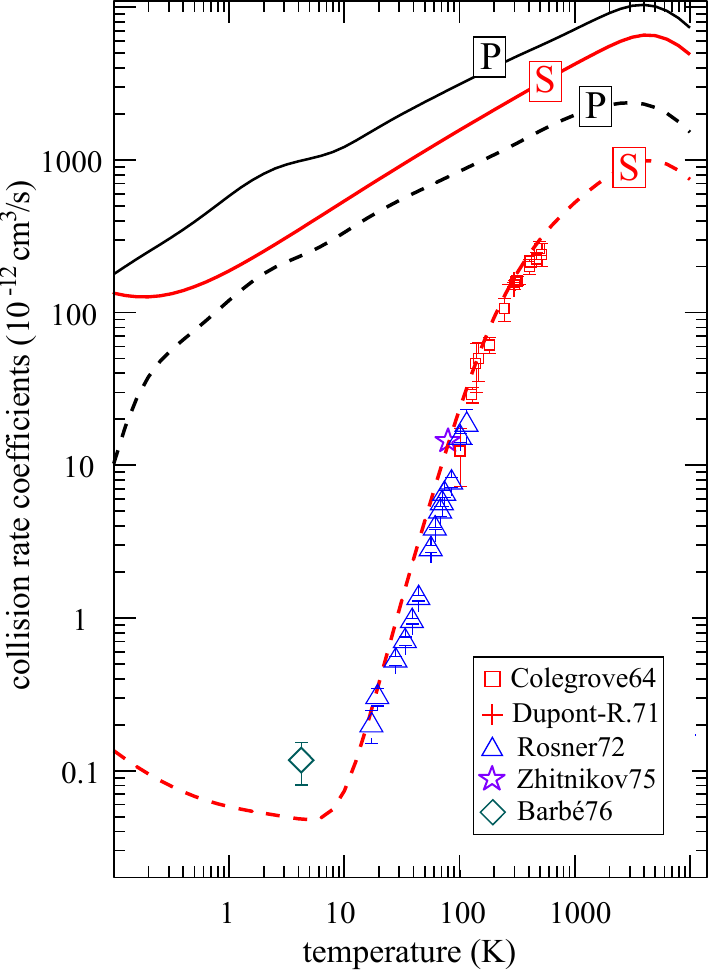}
\caption{\PJ{Temperature dependence of calculated and experimentally assessed collision rate coefficients in He. 
Ab-initio calculated rates are plotted for excitation transfer (dashed lines) and total (solid lines) rates of collisions for the \esS\ state (red curves, labeled `S') and the \esP\ state (black curves, `P'). Experimental $k_{\mathrm{ME}}$ data (symbols, see legend) are derived from published values of linewidths or ME cross sections in \Het\ (see text).}
\PJ{Three calculated rate coefficients slowly} decrease with
decreasing temperature and thermal velocities, but the rate coefficients
$k_{\mathrm{ME}}$ for ME collisions (the dashed red line and symbols) abruptly decrease below
room temperature. \PJ{This is attributed} to a weak repulsive barrier at large distance in the
\Hemet-He interaction potential. Adapted from \citet[Fig.~3]{Vrinceanu10} \PJ{for the calculated curves.}}%
\label{fig_MEvsT}%
\end{figure}

Early studies of \PJ{the} magnetic resonance linewidth in the \esS\ state \PJ{of \Het\ } have shown
that ME collision cross sections and collision rate coefficients strongly
decrease with decreasing temperatures between 4.2~K and 550~K \cite{Colegrove64}. 
These experimental data were found to be consistent with computations using
empirically determined potential parameters \cite{Fitzsimmons68}, but the
values \PJ{inferred for the cross sections need to be corrected to take into account the partial character of the loss of orientation in ME collisions} \cite{Dupont-Roc71}. These
data and the results of more accurate measurements performed at low
temperatures are compiled in Fig.~\ref{fig_MEvsT} together with the results of
ab-initio calculations of \PJ{different rates for collisions between \esS\ or \esP\ and ground state He atoms} \cite{Vrinceanu10}.
The first set of data \PJ{points} in Fig.~\ref{fig_MEvsT} is evaluated from
the plot of reduced linewidths $\Delta\nu/N$ in the temperature range
100~K to 550~K \cite[Fig.~7]{Colegrove64} using the correcting relation
$k_{\mathrm{ME}}=9/4\times\pi\Delta\nu/N.\footnote{The total number density
$N$ and the ground state number density $N_{\mathrm{g}}$ are almost equal in
He gas discharges under MEOP conditions, with at most a few ppm of the atoms in
an excited state. $N$ is used instead of $N_{\mathrm{g}}$ whenever necessary
or convenient.}$ Other experimental data for the exchange rate coefficients
$k_\text{ME}$ are derived from published values of linewidth $\Delta
\nu_\text{ex}$ \cite{Rosner72}, exchange rate $1/\tau$ \cite{Dupont-Roc71},
\PJ{cross section} $\sigma_\text{ex}$ \cite{Zhitnikov75,Barbe76}, and number densities $N$
using:
\begin{equation}
k_{\mathrm{ME}}=\pi\Delta\nu_{\mathrm{ex}}/N,\;k_{\mathrm{ME}}=1/N\tau,\;%
\mbox{ or }%
\;k_{\mathrm{ME}}=\sigma_{\mathrm{ex}}\bar{v}_{\mathrm{rel}}%
,\label{MEratesfromdata}%
\end{equation}
where $\bar{v}_\text{rel}=4\sqrt{k_{\mathrm{B}}T/\pi M}$ is the average relative velocity
of colliding He atoms at temperature $T$ ($k_{\mathrm{B}}$ is the Boltzmann constant, $M$ the \Het\ atomic mass). 
\PJ{The experimental and calculated values of ME rate coefficients are in fair agreement (the expected effect of isotope mass difference through $\bar{v}_{\mathrm{rel}}$ is of the order of symbol size). For convenience, values of $k_{\mathrm{ME}}$ }
and of their local temperature variations inferred from the compiled experimental data are listed in Table~\ref{tab-kME}. 
\begin{table}[htb]%
\begin{tabular}
[c]{|c|c|c|}\hline
$T$ [K] & $k_{\mathrm{ME}}$ [$10^{-12}$cm$^{3}$/s] & $1/\tau^{\mathrm{typ}}$
[$10^{6}$s$^{-1}$]\\\hline
around 300 & $154\times(T/300)^{1.09}$ & $4.09\times(T/300)^{1.09}$\\\hline
around 77 & $6.6\times(T/77)^{2.91}$ & $0.175\times(T/77)^{2.91}$\\\hline
4.2 & 0.12 & $0.32\times10^{-2}$\\\hline
\end{tabular}
\caption{Table of metastability-exchange rate coefficients $k_\textrm{ME}$ and of corresponding typical 
rates $1/\tau^\textrm{typ}=N^\textrm{typ}k_\textrm{ME}$ for a gas number
density $N^\textrm{typ}=2.653\times10^{16}$~cm$^{-3}$, i.e. $10^{-3}$~amg, which is
a typical value in metastability-exchange optical pumping experiments (1~mbar at 293~K)}%
\label{tab-kME}%
\end{table}

An interesting feature of the \PJ{ab-initio calculated} rates in Fig.~\ref{fig_MEvsT} is the comparatively high value of the exchange \PJ{rate} between ground state and \esP\ state atoms, approximately ten times larger than the ME rate at room temperature and more than 1000 times larger at a few Kelvin \PJ{(the dashed curves labeled P and S, respectively)}. In spite of this, this coupling channel is traditionally not considered in the transfer of angular momentum to the ground state, due to the comparatively low number density usually achieved in the \esP\ state. 

\subsubsection{\Thad{Metastability exchange}  and spin-temperature distribution}
\label{subsec:spinT}
An important property of ME collisions in helium is the negligible
depolarization that occurs during the very short collisions (10$^{-12}$~s),
due to the fact that all involved angular momenta are spins \cite{Pinard80}.
This accounts for the very efficient transfer of angular momentum in MEOP from
the absorbed polarized light to the nuclear orientation. This also enforces a
Boltzmann-like distribution in angular momentum for the relative populations
of all sublevels coupled by ME collisions: a so-called spin-temperature
distribution. Anderson et al. have proposed that such a distribution is
enforced, e.g., by fast spin-exchange \cite{Anderson59,Happer1972}. For He,
whenever the effects of pumping light and relaxation can be neglected, the
spin-temperature distribution can be derived from ME equations on density
operators \cite{Courtade02}. \PJ{Defining $e^{\beta}=(1+P_{\mathrm{He}}%
$)/(1-$P_{\mathrm{He}}$) as} the ratio of the populations in the ground state with nuclear
polarization $P_{\mathrm{He}}$ (1/$\beta$ plays the role of a spin
temperature), one finds that the ME-driven ratios of populations in the
\esS\ state of \Het\ are field-independent, and given by $e^{\beta\Delta
m_{F}}$ for any two sublevels with a magnetic quantum number difference
$\Delta m_{F}$. Similarly, in an isotopic mixture, the ME-driven ratios of
populations for the three magnetic sublevels in the \esS\ state of \Hef\ are
given by $e^{\beta\Delta m_{S}}$. Therefore, one \PJ{may write} for
populations $a_{i}^{\mathrm{st}}$ and $y_{i}^{\mathrm{st}},$ where the upper
index indicates spin-temperature distribution values:
\begin{align}
a_{1}^{\mathrm{st}}/a_{2}^{\mathrm{st}} &  =a_{2}^{\mathrm{st}}/a_{3}%
^{\mathrm{st}}=a_{3}^{\mathrm{st}}/a_{4}^{\mathrm{st}}=a_{5}^{\mathrm{st}%
}/a_{6}^{\mathrm{st}}=e^{\beta}\label{ratiosai}\\
y_{1}^{\mathrm{st}}/y_{2}^{\mathrm{st}} &  =y_{2}^{\mathrm{st}}/y_{3}%
^{\mathrm{st}}=e^{\beta}.\label{ratiosyi}%
\end{align}
This yields explicit expressions:
\begin{align}
a_{i}^{\mathrm{st}} &  =e^{\beta m_{\mathrm{F}}}/(e^{3\beta/2}+2e^{\beta
/2}+2e^{-\beta/2}+e^{-3\beta/2})\label{aivsbeta}\\
y_{i}^{\mathrm{st}} &  =e^{\beta m_{\mathrm{S}}}/(e^{\beta}+1+e^{-\beta
}).\label{yivsbeta}%
\end{align}
The spin-temperature values of populations of
Eq.~\eqref{aivsbeta} can be checked to yield, using Eq.~\eqref{eq:Mdotline2}, a nuclear polarization
in the \esS\ state 
of $P_{\mathrm{He}^{\mathrm{\ast}}}=P_{\mathrm{He}}$, as expected. 

These distributions of populations have a strong impact on the absorption of
light on the \esS--\esP\ optical pumping transition. The consequences on the efficiency of
MEOP at high nuclear polarization will be discussed in
Sec.~\ref{Subsec_OPinMEOP}, and the resulting features of polarimetry using
the optical pumping transition in Sec.~\ref{Subsec_Polarimetry}.

\subsection{Optical pumping of the  \esS\ - \esP\ transition}
\label{Subsec_OPinMEOP}
Besides the ME collisions described above, two kinds of processes jointly affect the populations in the \esS\ and \esP\ states: population transfers between Zeeman sublevels within the  \esS\ and \esP\ states, usually modeled as relaxation processes, and optical pumping that combines absorption and spontaneous or induced emission of light on the closed \esS\ - \esP\ transition. They are successively considered below.

\subsubsection{Relaxation in the \esS\ and \esP\ states}
\label{sec:relaxSP}
Population transfers between sublevels occur with very different rates for the two states \cite{Batz11}: 

In the \esS\ \PJ{state, the} $L=0$ orbital angular momentum is not affected by collisions: relaxation is slow, with rates $\gamma_\mathrm{r}^\mathrm{S}$ typically of order 10$^3$~s$^{-1}$. It is attributed to excitation quenching (e.g., at cell wall or in 3-body dimer-forming collisions) and re-excitation in the gas discharge. Its effect on populations is simply written as: 
\begin{equation}
\left.  \frac{da_{i}}{dt}\right\vert _{\mathrm{r}}=\gamma_{\mathrm{r}%
}^{\mathrm{S}}\left(  \frac{1}{6}-a_{i}\right)  .\label{eq:adotr}%
\end{equation}

In the \esP\ \PJ{state, $J$-changing} collisions occur at a much faster rate, proportional to gas pressure and of order a few $10^7$~s$^{-1}/\mathrm{mbar}$ \cite{Schearer67,Vrinceanu04}. \PJ{These collisions may induce significant population transfer during the radiative lifetime, which is }phenomenologically described in MEOP models using a single rate $\gamma_\mathrm{r}^\mathrm{P}$ in rate equations ruling the evolutions of the populations $b_j$ of the sublevels B$_j$: 
\PJ{\begin{equation}
\left.  \frac{db_{j}}{dt}\right\vert _{\mathrm{r}}=\gamma_{\mathrm{r}%
}^{\mathrm{P}}\left(  \frac{\sum\nolimits_{k=1}^{18}b_{k}}{18}-b_{j}\right)
.\label{eq:bdotr}%
\end{equation}}
For convenience, the $b_j$s are defined so that
the number density of atoms in each sublevel B$_j$ of the \esP\ state is $n_{\mathrm{m}}b_j.$ Consequently they are not true populations (i.e., diagonal elements of a trace-1 density matrix), and
$\sum b_k<1 $ depends on the pumping light intensity. When isotopic mixtures are considered, additional phenomenological equations similar to Eqs. \eqref{eq:adotr} and \eqref{eq:bdotr} are used, with possibly different rates for different isotopes due to the thermal velocity difference. 

A more realistic treatment of the effects of collisions in the \esP\ state, in particular at high magnetic field, remains to be implemented and tested in MEOP models \cite{Batz11}.

\subsubsection{Traditional low-field optical pumping}
\label{sec:LowBOP}

\begin{figure}[tbh]
\includegraphics[width=3.3 in]{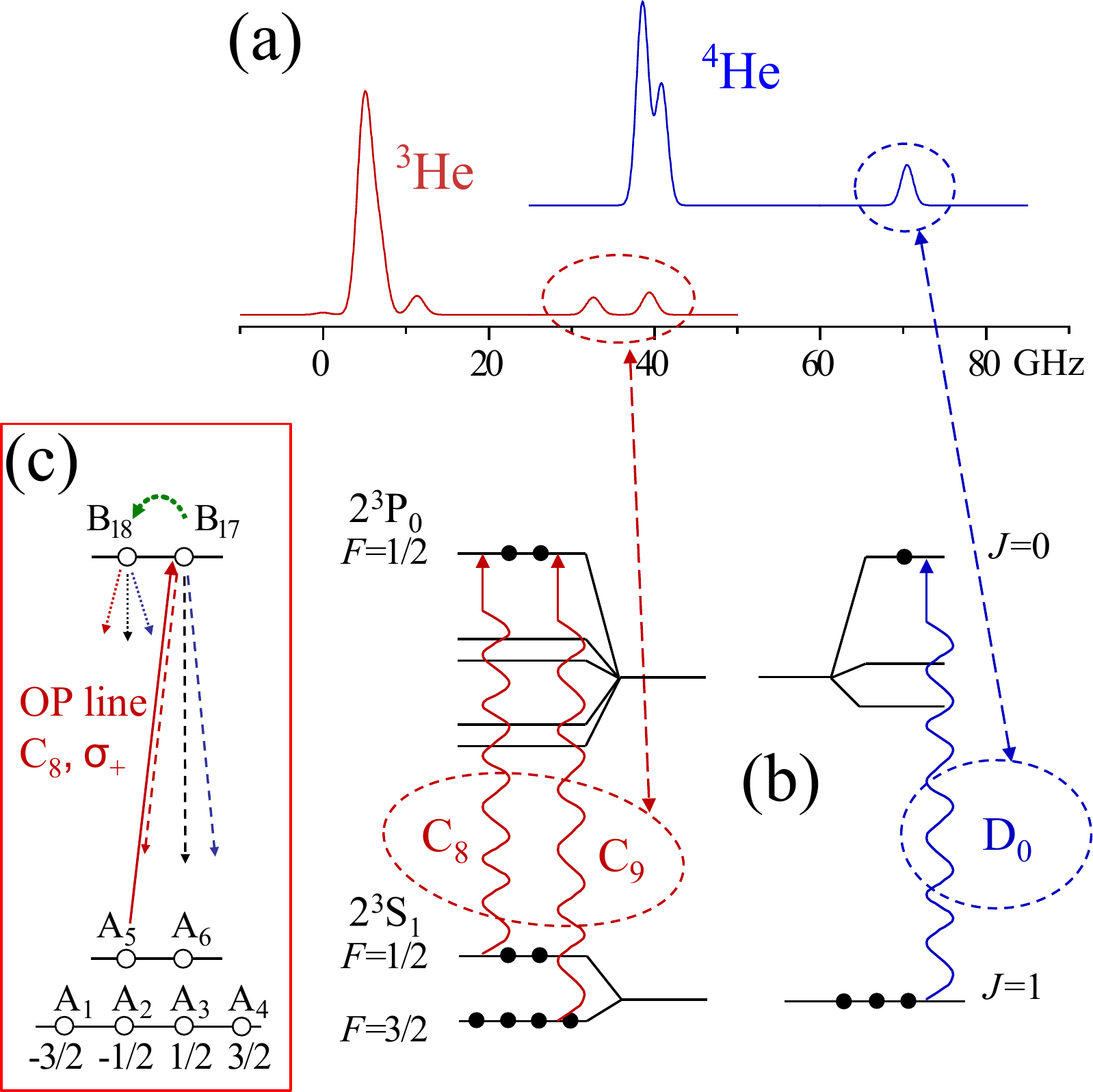}%
\caption{Low-field metastability-exchange optical pumping transitions: (a) computed absorption spectra for the \esS-\esP\ transition for a low-pressure, optically thin gas at room temperature. The linewidths essentially arise from Doppler broadening. (b) Fine- and hyperfine-sublevels involved in the transition (see Fig.~\ref{fig_levels}). The three transitions to the \esP$_0$ levels are schematically represented, and the names and positions of the corresponding line components in the spectra are highlighted. (c) The inset displays the optical transitions involved for C$_8$, $\sigma_+$ pumping (straight arrows, see text). The curved arrow represents population transfer between Zeeman sublevels of the \esP\ state. Note that no {\em direct} transfer, corresponding to a nuclear spin flip, is expected to occur between these two sublevels.}%
\label{fig_MEOP2}%
\end{figure}

Optical pumping selectively promotes atoms from a sub-set of Zeeman sublevels of the \esS\ state to corresponding sublevels of the \esP\ state according to selection rules depending on \PJ{the} frequency and polarization of the pumping light. Figure~\ref{fig_MEOP2}(a) 
displays typical absorption spectra for \Het\ and \Hef\ at 1083~nm. 

\PJ{The action of optical pumping in pure $^{4}\mathrm{He}$ is quite straightforward: each of the two strong low-energy components of the pumping transition (partly resolved in Fig.~\ref{fig_MEOP2}(a)) can be used to efficiently obtain a high \emph{electronic} polarization in the \esS\ state by depopulating, for instance, the Y$_1$ and Y$_2$ sublevels, whereas using the D$_0$ component depopulates only one sublevel and therefore yields a maximum polarization of 0.5 \cite{Giberson82,Wallace1995}. 
In a similar way, optical pumping can efficiently depopulate selected sublevels of $^{3}\mathrm{He}^{\mathrm{\ast}}$ atoms whenever ME collisions have a negligible rate, for instance at low temperature (see Fig.~\ref{fig_MEvsT}) or in atomic beams~\cite{Stas2004}. Achieving full nuclear polarization would require depopulating all sublevels except A$_1$ or A$_4$ by suitably addressing both hyperfine levels (see Eq.~\ref{eq:Ai} and Fig.~\ref{fig_levels}).}

\PJ{The effect of pumping light in a gas of \Het\ or in isotopic mixtures is quite different due to the strong coupling enforced by ME collisions between the Zeeman sublevels of the \esS\ state (Eq.~\ref{eq:adotME}). In a weak pumping limit, one may consider that $P_{\mathrm{He}^{\mathrm{\ast}}}\approx P_{\mathrm{He}}$ and derive a simple phenomenological model at low $P_{\mathrm{He}}$ \cite{Colegrove63}, or more generally consider the effect of optical pumping cycles on the $P_{\mathrm{He}}$-dependent spin-temperature populations of Eqs. \ref{aivsbeta} and \ref{yivsbeta} for arbitrary $P_{\mathrm{He}}$. In usual MEOP conditions, ME therefore not only transfers nuclear orientation to the ground-state atoms but also plays a key role in the way the pumping light is absorbed. In that case full polarization could in principle be achieved even if a single Zeeman sublevel is addressed by optical pumping. The actual limit for the polarization results from a balance of angular momentum gain from the pumping light and loss through relaxation. 
In practice, MEOP is usually performed by tuning the pumping light to one of the high-energy components of the spectra to selectively populate the \esP$_0$ level, using the D$_0$ component} for \Hef\ or the C$_8$ or C$_9$ component for \Het\ (see Fig.~\ref{fig_MEOP2}(b)). These are the most efficient and widely used transitions for pumping with tunable lasers. In the early days of MEOP, pumping light was usually obtained from \Hef\ lamps to polarize \Het\ gas with fair efficiency, thanks to the isotope shift \cite{Colegrove63}.

Figure~\ref{fig_MEOP2}(c) displays in more details the full optical pumping cycle process in the simple case of the C$_8$ component. Right-handed circular ($\sigma_+$) polarization excites atoms from the A$_5$ sublevel ($m_{F}=-1/2$) to the B$_{17}$ sublevel ($m_{F}=1/2$). Radiative decay brings atoms back to the \esS\ state by spontaneous and stimulated emissions with well-defined branching ratios from the directly populated B$_{17}$ sublevel and from any other sublevel that may have been indirectly populated by $J$-changing collisions, as described in the previous section \ref{sec:relaxSP}. 
There are two extreme situations for the impact of these collisions: low-pressure optical pumping \cite{Kastler57} for $\gamma_{\mathrm{r}}^P\ll \gamma,$ where $\gamma$=1.022$\times$10$^{7}$~\textrm{s}$^{-1}$ is the radiative decay rate of the \esP\ state, and depopulation optical pumping \cite{Dehmelt57}, a high pressure regime where all \esP\ populations are equalized, for $\gamma_{\mathrm{r}}^P\gg \gamma$.

All radiative processes can be modeled by sets of coupled rate equations for the populations. For \Het\ they are written as \cite{Nacher85}:
\begin{align}
\left.  \frac{da_{i}}{dt}\right\vert _{\mathrm{OP}}  &  =\gamma\sum
\nolimits_{j=1}^{18}T_{ij}b_{j}+\sum\nolimits_{j=1}^{18}\gamma_{ij}%
(b_{j}-a_{i})\label{eq:OPratea}\\
\left.  \frac{db_{j}}{dt}\right\vert _{\mathrm{OP}}  &  =-\gamma b_{j}%
+\sum\nolimits_{i=1}^{6}\gamma_{ij}(a_{i}-b_{j}), \label{eq:OPrateb}%
\end{align}
where $T_{ij}$ is the transition matrix element, and
$\gamma_{ij}$ is the optical pumping rate of the transition A$_{i}$
$\rightarrow$B$_{j}$.  $\gamma_{ij}$ is proportional to the pump light intensity and to $T_{ij}.$ On the right-hand sides of these \PJ{equations,} the first terms account for spontaneous emission and the second terms for the net difference between absorption and stimulated emission.

Adding up contributions of ME, relaxation, and radiative processes (Eqs. \eqref{eq:adotME} and  \eqref{eq:adotr}  to \eqref{eq:OPrateb}), one obtains a set of coupled rate equations for the populations of all sublevels of the \esS\ and \esP\ states. Given that the rate parameters are in the range 10$^3$ to 10$^7$~s$^{-1}$ in these equations, while the rate of change of the nuclear polarization is usually smaller than 1~s$^{-1}$, steady-state solutions for the populations are adiabatically computed taking $P_\mathrm{He}$ as a static input parameter.

\begin{figure}[tbh]
\includegraphics[width=3.3 in]{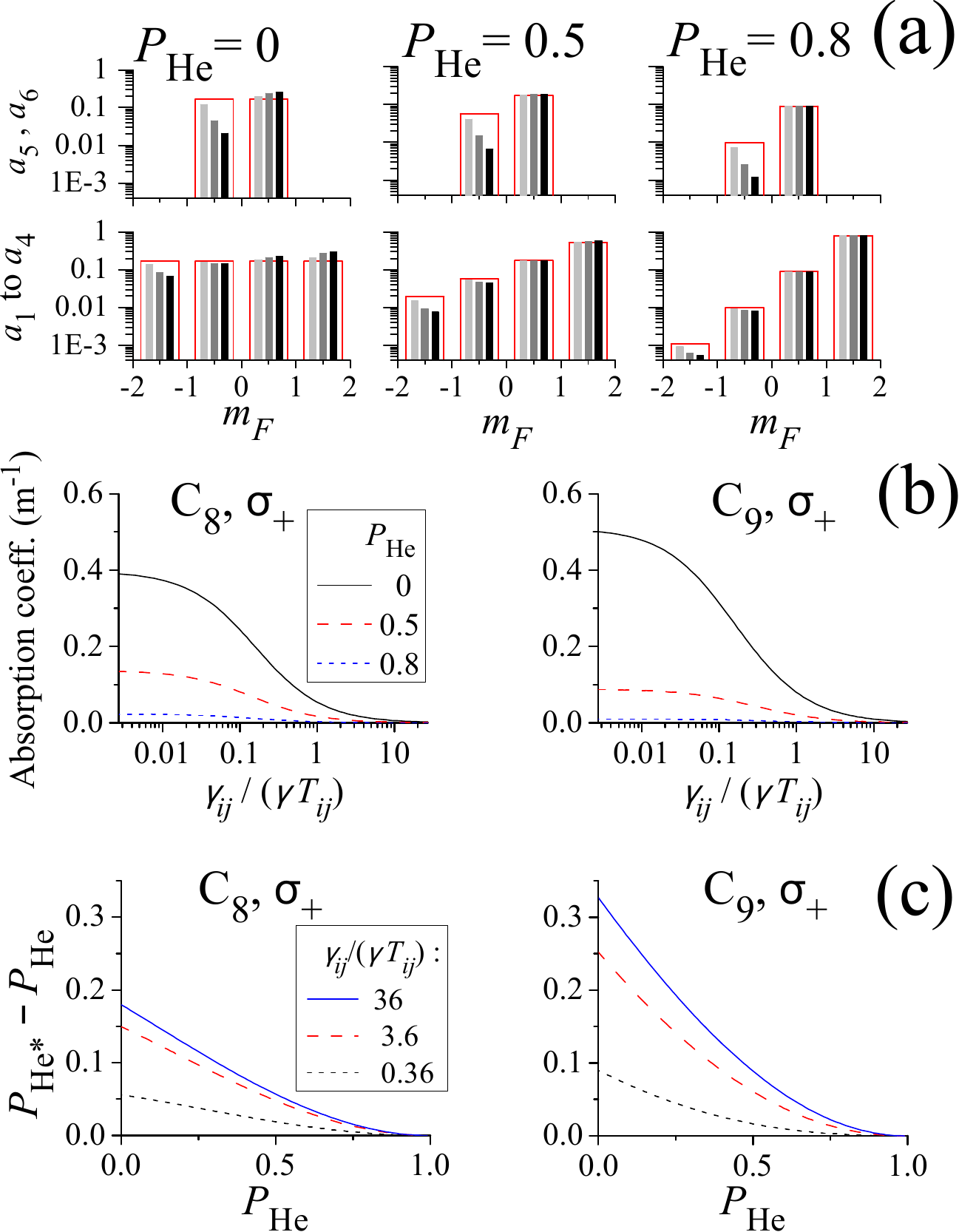}%
\caption{Results of the \PJ{MEOP} rate equations obtained using the same computer code as in \cite{Batz11} for typical operating conditions: 0.5~mbar \Het\ gas, $T=293~\mathrm{K}$, $B=1~\mathrm{mT}$, $n_{\mathrm{m}}=1.2\times10^{10}~\mathrm{cm^{-3}}$, $\gamma_{\mathrm{r}}^{\mathrm{S}}=10^3~\mathrm{s}^{-1}$, $\gamma_{\mathrm{r}}^{\mathrm{P}}=0.16\times10^7~\mathrm{s}^{-1}$. Atoms of all velocity classes are pumped with the same rates $\gamma_{ij}$ (see text). (a) Distributions of the \esS\ populations $a_i$ for C$_8$, $\sigma_+$ optical pumping and \PJ{$\gamma_{ij}=\gamma /10$, $\gamma$, and $10\gamma$ (bars from left to right and from light grey to black in each group) Surrounding red boxes are the spin-temperature populations. }(b) Computed absorption coefficients  are plotted as functions of \PJ{reduced} pumping rates for C$_8$ and C$_9$ optical pumping and different nuclear polarizations (see legend). For  C$_9$ the two pumping rates for the two pumped sublevels jointly scale with the pumping light intensity, with a fixed ratio $T_{1,18}/T_{2,17}=3$. Introducing $T_{ij}$ in the horizontal \PJ{scales makes the reduced pumping rates} identically proportional to the pump intensity. (c) Differences between nuclear polarizations in the \esS\ and ground states are plotted as a functions of $P_\mathrm{He}$ for C$_8$ and C$_9$ pumping and different \PJ{reduced} pumping rates (see legend)}%
\label{fig_populations}%
\end{figure}

Figure~\ref{fig_populations} displays examples of computed populations, pump absorption coefficients, and \esS\ state \PJ{laser-induced} over-polarization for parameters corresponding to typical MEOP experiments. 
Competition between metastability-exchange and optical pumping is illustrated in Fig.~\ref{fig_populations}(a), in which the steady-state populations of the \esS\ state are represented for different values of $P_\mathrm{He}$ and of the \PJ{pumping rate $\gamma_{ij}$} on the C$_8$ transition component. In that case, absorbed light directly depletes level A$_5$ while other populations are differently affected, which enforces systematic deviations from the spin-temperature distribution \PJ{(the red boxes)} that is obeyed in the absence of optical pumping  \PJ{and assumed in simple MEOP models}. The effect of relaxation in the \esS\ state on populations is too weak to be seen on this figure. 

Figure~\ref{fig_populations}(b) displays the computed variation of the gas absorption coefficient with  \PJ{the reduced pumping rate $\gamma_{ij}/\left(\gamma T_{ij}\right)$} for both C$_8$ and C$_9$ components. A strong decrease of the absorption results from the modifications of populations in the \esS\ state as well as from stimulated emission  of atoms promoted to the \esP\ state by optical pumping. Absorption also significantly decreases when nuclear polarization increases. It is therefore increasingly difficult to efficiently deposit angular momentum at increasing polarization and pumping intensity. 

Another meaningful quantity derived from the solutions of the rate equations is represented in Fig.~\ref{fig_populations}(c): the nuclear over-polarization $P_{\mathrm{He}^{\mathrm{\ast}}}-P_\mathrm{He}$, which acts as a driving term for the ground state polarization $P_\mathrm{He}$ in Eq.~\eqref{eq:Mdot}. 
The parameters in Eq.~\eqref{eq:Mdotline2} are $(L_{k})=(-1,L_-,L_+,1,-L_-,-L_+),$ with \PJ{$L_\pm \approx \pm 1/3$ in low field.}  Therefore $P_{\mathrm{He}^{\mathrm{\ast}}}$ is fairly sensitive to the modifications of all populations enforced by optical pumping in the \esS\ state. 
The decrease of this driving term with $P_\mathrm{He}$ and its saturating increase at high pumping rates mainly result from the decrease in pumping light absorption. Introducing the optical pumping photon efficiency $\eta$ as the net change of atomic angular momentum projection $m_F$ upon the absorption and emission of a photon, these correlated decreases can be quantitatively linked \cite{Batz11}. For C$_8$, $\eta$ has a fixed value ranging from 0.5 to 0.9 depending on $\gamma_\mathrm{r}^\mathrm{P}$, while for $C_9$ $\eta$ also depends on $P_\mathrm{He}$ and the optical pumping rate, ranging from 0.5 to 1.25.

So far, the pumping rates have not been explicitly related to the pumping light parameters. They depend on
the local characteristics of the pumping light (intensity, direction of propagation, spectral characteristics) and on the
atomic velocity projection $v$ on the direction of propagation of the pumping light. Velocity-selective optical pumping, i.e. a strong correlation between $v$ and populations or $P_{\mathrm{He}^{\mathrm{\ast}}}$, may occur if an intense narrowband laser is used, with spectral width $L$ smaller than the Doppler width $D$ \cite{Aminoff82,Pinard79}. Velocity-dependent pumping rates $\gamma_{ij}(v,\mathbf{r})$ can be computed for a single-frequency pump laser \cite{Courtade02}. The maximum rate is experienced by atoms in the velocity class $v^*$ that is resonant with the laser frequency, and is given by: 
\begin{equation}
\gamma_{ij}(v^*,\mathbf{r})=\frac{2\pi\alpha f}{m_{\mathrm{e}}\omega\gamma^2}~\frac{\gamma}{\Gamma^{\prime}/2}~I_\mathrm{las}(\mathbf{r})~T_{ij}\,\gamma,
\label{eq:gammaijSF}%
\end{equation}
where $I_\mathrm{las}(\mathbf{r})$ is the laser intensity, $\omega$ its angular frequency, $\alpha$ the fine structure constant, $f$ the oscillator
strength of the transition, and $m_{\mathrm{e}}$ the electron mass; the numerical value of the first fraction is $0.149~\mathrm{m^2/W}.$ The total damping rate of the optical coherence of the transition, $\Gamma^{\prime}/2=\gamma+\pi w, $ results from the combined effects of the radiative decay rate $\gamma$ and of the pressure-dependent collisional broadening $w$, with a value for $w/p$ of order 20~MHz/mbar \cite{Vrinceanu04} or 12~MHz/mbar \cite{Nikiel13} that remains to be confirmed. In the low-pressure limit ($w\ll \gamma$),  $\gamma_{ij}=T_{ij}\,\gamma$ for $I_\mathrm{las}=6.7~\mathrm{W/m^2}.$

Pumping rates can also be evaluated for broadband lasers of known spectral intensity distribution. For instance, for a laser of Gaussian width $L$ tuned to the center of the Doppler absorption profile, the rate is \cite{Batz11}:
\begin{equation}
\gamma_{ij}(v,\mathbf{r})=\frac{2\pi\alpha f}{m_{\mathrm{e}}\omega \gamma^2}%
\,\frac{\gamma}{2\sqrt{\pi}L}\,\exp\left[  -\left(  \frac{Dv}{L\bar{v}}\right)  ^{2}\right]\,I_\mathrm{las}(\mathbf{r})\,T_{ij}\,\gamma, 
\label{eq:gammaijL}%
\end{equation}
where $\bar{v}$=$\sqrt{2k_{\mathrm{B}}T/M}$ is the most probable
speed and $D$=$(\omega/2\pi)\bar{v}/c$ is the associated Doppler width. Due to the second factor on the right hand side, the rates are typically 100 to 1000 times smaller than in Eq.~\eqref{eq:gammaijSF} for a given pumping intensity. This accounts for the much higher total power absorption for a pump laser with a suitable bandwidth $L\approx D$ \cite{Tastevin04}.

One may be tempted to make an exact treatment of velocity-selective optical pumping effects by explicitly keeping a dependence of populations and collision rates on the atomic velocity projection $v$. However, solving sets of rate equations for various velocity classes would be a difficult task since their populations are coupled by velocity-changing and by
ME collisions, with ill-known collision rates. Instead, a coarse description
with only two broad velocity classes has been proposed to account for the Maxwell
distribution of atomic velocities in the pumped gas: strongly pumped atoms, in
the center of the velocity distribution, and weakly pumped atoms, in the wings of the velocity distribution \cite{Nacher85,Batz11}. Each class is pumped with an effective pumping rate depending on the pump spectral profile, and their populations are coupled by ME and velocity-changing collisions. This is a
crude model, however it is sufficient to capture key features of
velocity-selective optical pumping effects with few free parameters and it usually provides meaningful quantitative results.

Velocity-independent optical pumping models can be reliably used in specific cases: if (i)-
broadband (``white'') pumping light is used, with $L$ sufficiently larger than $D$, or if (ii)-the rate of change of populations from Eqs. \eqref{eq:OPratea} and \eqref{eq:OPrateb}, $\gamma\gamma_{ij}/(\gamma+\gamma_{ij}),$  
is much smaller than ME and velocity-changing collision rates. 
Condition (ii) is automatically fulfilled if $\gamma \ll \gamma_{\mathrm{e}}.$ Because $\gamma_{\mathrm{e}} $ scales with gas pressure $p$,
high pressure optical pumping is immune to velocity-selective effects and a single
set of populations can be used to locally characterize the
effect of MEOP on the gas. In pure \Het\ at room temperature, $\gamma_{\mathrm{e}}=\gamma~p/p^*,$ where $p^*=2.72~\mathrm{mbar}$ is the crossover pressure for which $\gamma_{\mathrm{e}}=\gamma$ (\PJ{from data in} Sec.~\ref{Subsec_MEcollvsT}).
At lower pressure, a velocity-independent optical pumping
regime is obtained only for laser intensities much smaller
than a crossover value depending on pressure and laser
linewidth, which can be evaluated using for instance a two-class model. 

Indeed, during experiments where
polarization decay occurs following the interruption
of optical pumping, velocity-independent values of the populations are expected to be enforced by ME collisions as
well.

\subsubsection{High-field optical pumping}
\label{secHiBOP}

\begin{figure}[tbh]
\includegraphics[width=3.0 in]{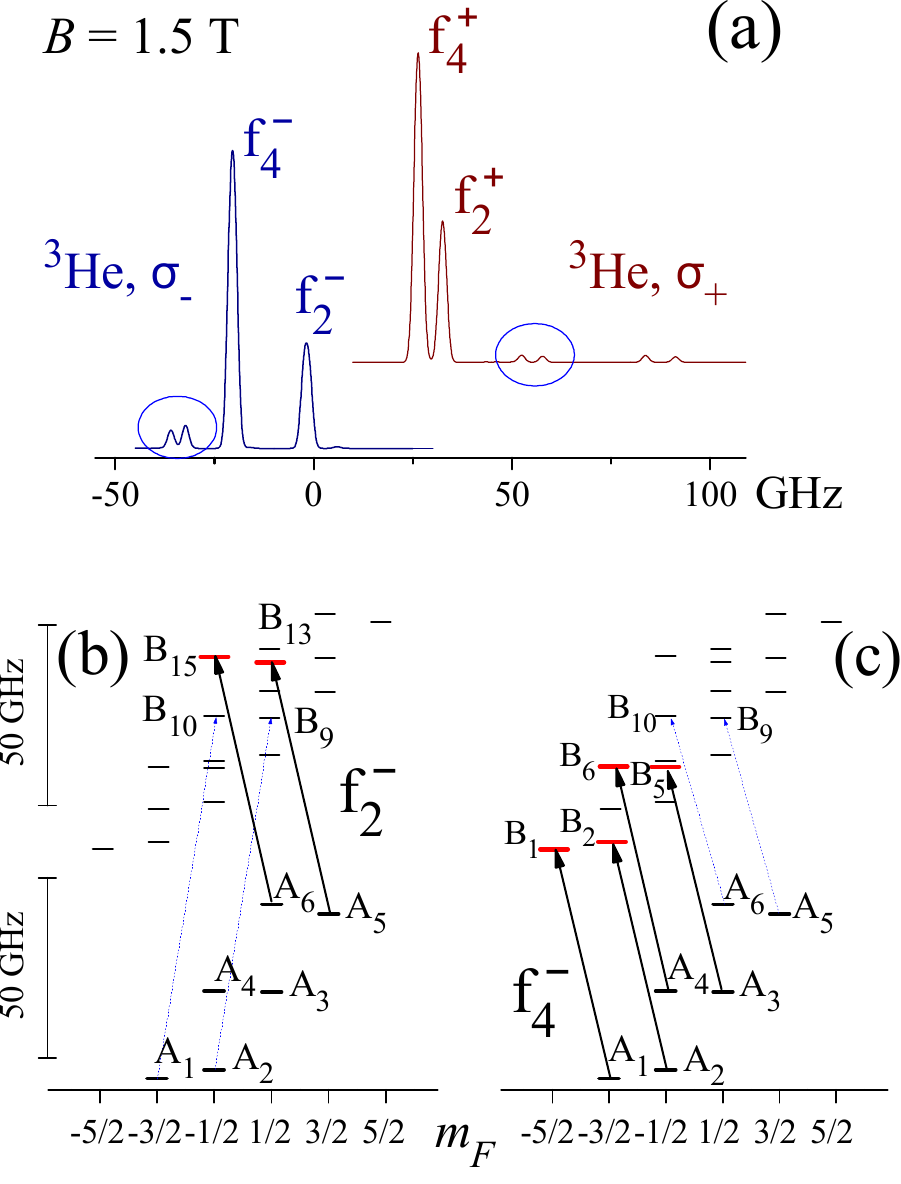}%
\caption{High-field metastability-exchange optical pumping transitions: (a) Computed absorption spectra at $B=1.5~\mathrm{T}$ for the \esS-\esP\ transition (low-pressure, optically thin gas at room temperature) for both circular polarizations. The strong unresolved components in the spectra are labeled
$\mathrm{f}_n^\pm$, where $n$=2 or 4 refers to the number of involved transitions and $\pm$ to the sign of the circular polarization. Doublets of resolved weaker transitions of interest for optical detection purpose are highlighted. 
(b) and (c) Energies of the \Het\ sublevels at 1.5~T for the \esS\ and \esP\ states. The transitions induced by the $\sigma_-$-polarized pumps (thick lines) and the suitable probes (thin lines) are displayed ($\mathrm{f}_2^-$ pump, $\sigma _+$ probe in (b); $\mathrm{f}_4^-$ pump, $\sigma _-$ probe in (c)). }%
\label{fig:MEOP_HiB}%
\end{figure}

As mentioned above (Sec.~\ref{Subsec_levels}), the angular momentum structures of the \esS\ and \esP\ levels and the 1083~nm transition are deeply modified in high magnetic field. Figure~\ref{fig:MEOP_HiB} displays the absorption spectra  and the energies of all Zeeman sublevels of \Het\ for $B=1.5~\mathrm{T},$ a field strength commonly met in MRI systems and thus of practical importance for applications. 
The six sublevels of the \esS\ state are organized in three pairs of states (Fig.~\ref{fig:MEOP_HiB}(b) and (c), bottom
graphs). In each pair the sublevel energy is mostly determined by the common dominant value of
$m_S$ and the nuclear spin projections are almost opposite. Similarly, hyperfine coupling only weakly mixes levels of different $m_I$ values in the \esP\ state (Fig.~\ref{fig:MEOP_HiB}(b) and (c), upper graphs). As a result, high-field spectra for a given light polarization comprise six main components clustered in two groups: a pair and a quartet, each group being unresolved at room temperature  (Fig.~\ref{fig:MEOP_HiB}(a)). 

In high-field, the parameters in Eq.~\eqref{eq:Mdotline2} are $(L_{k})=(-1,L_-,L_+,-L_-,1,-L_+),$ with $L_\pm \simeq 1$; the change in the order of levels results from crossings occurring around 0.16~T and 4.7~T. Simplified models as well as full resolution of the rate equations of MEOP show that a large number of ME collisions, proportional to $B^2$, are required to transfer a single unit of angular momentum between ground state and \esS\ state atoms \cite{Courtade02,Abboud04}. For instance, at null ground state polarization, strong pumping with the $\mathrm{f}_2^-$ pair of lines may depopulate the A$_5$ and A$_6$ sublevels while $a_1=...=a_4=1/4$ if spin-temperature is still obeyed for the unpumped sublevels. Hence $P_{\mathrm{He}^{\mathrm{\ast}}}=(L_+-1)/4=-\sin^2\theta_+/2$ with $\sin^2\theta_+=5.62\times10^{-4}$ at 1.5~T. Similarly, strong $\mathrm{f}_4^-$ pumping depopulates A$_1$ to A$_4$, so that $a_5=a_6=1/2$ and $P_{\mathrm{He}^{\mathrm{\ast}}}=(1-L_+)/2$. $\mathrm{f}_4^-$ pumping is thus expected to yield {\em positive} nuclear polarization in spite of its negative circular polarization, with a (twice) faster pump-up rate, in fair agreement with experimental observations \cite{Abboud05a,Nikiel07}. Over a wide range of field strengths and gas pressures, $\sigma_-$ pumping has been reported to be more efficient, possibly because the $\mathrm{f}_4^+$ and $\mathrm{f}_2^+$ groups are imperfectly resolved, especially at high pressure, and have opposite polarizing actions. The $\sigma_-$ pumping schemes are thus detailed in Fig.~\ref{fig:MEOP_HiB}(b) and (c), together with the corresponding probing transitions that will be discussed in Sec.~\ref{Subsubsec_1083Polarimetry}.

The photon efficiency of these high-field optical pumping schemes can be as high as in low field ($\eta\approx 1$ for $\mathrm{f}_2^-$ pumping at high pressure) but this remains to be systematically studied \cite{Batz11,Nikiel13}. With \PJ{modified He} level structures and only weak hyperfine couplings remaining in the excited states, high-field MEOP has two distinct features: highly absorbing lines can be used, with absorption coefficients $\simeq10$~times larger than for C$_8$ or C$_9$ pumping at low field, and the transfer of angular momentum to the ground state is orders of magnitudes slower than in low field at a given pressure. For this last point, ME collisions fail to enforce a strong coupling between the \esS\ state populations, which is reminiscent of the low-field, low-temperature situation with its ME \PJ{cross sections} orders of magnitude lower than at room temperature. 

It was initially argued that, due to the weak hyperfine coupling, MEOP would be inefficient in high field and early attempts were driven by applications and were limited to moderate field strengths: 0.1~T \cite{Flowers90,Darrasse97} and 0.6~T \cite{Flowers97}. It was only later that a key benefit of high-field MEOP was recognized: its efficient operation at high gas pressures \cite{Courtade00}. More recently, high-field MEOP \PJ{of \Het\ at low pressure has been shown to be efficient as well \cite{Abboud04,Nikiel14,Maxwell16}. To date, no study of high-field MEOP in isotopic mixtures has been reported.} Ultimately, the performance of MEOP for weakened ME coupling crucially depends on the balance between impeded polarization transfer and reduced polarization losses in the ground state, \PJ{which} will be discussed in Sec.~\ref{Subsec:Relax}.   

\subsection{\Thad{Optical methods of polarimetry}} 
\label{Subsec_Polarimetry}
The measurement of the nuclear polarization of \Het\ gas in optical pumping cells can be made using a direct magnetic measurement (see Sec.~\ref{metrology}). However, optical methods relying on polarimetry of the visible fluorescence light emitted by He gas or on 1083~nm light absorption measurements are more widely used since they are simple to perform, provide high sensitivity, and operate in the presence of the discharge sustained for MEOP.

\subsubsection{Fluorescence light polarimetry}

Nuclear spin polarization is conserved during collisions which excite atoms from the ground state and hyperfine interaction subsequently couples electronic and nuclear orientations in the excited states. The electronic 
orientation, and therefore the degree of circular polarization of the fluorescence light emitted by atoms losing their \PJ{excitation, both reflect the value} of the ground state polarization \cite{Pavlovic70}. The circular polarization $P_\mathrm{fluo}$ of the fluorescence light emitted along the direction of $B$ is experimentally monitored using a rotating \PJ{quarter-wave} plate and a fixed linear polarizer \cite{Pinard74,Bigelow92,Lorenzon93}, a 2-channel polarimeter consisting of a static \PJ{quarter-wave} plate and a beam-splitting polarizer \cite{Stoltz_pol96}, or an electro-modulated polarimeter comprising a liquid crystal wave retarder \cite{Maxwell14}. Light of two spectral lines, at 668~nm and 588~nm, are significantly polarized and yield high signal-to-noise-ratio (SNR) measurements, with typical polarization ratios $P_\mathrm{fluo}/P_\mathrm{He}\approx 0.1$ \cite{Stoltz_pol96}. 
The ratios significantly decrease with $B$ above about 10~mT due to the weakening of hyperfine coupling \PJ{efficiency and} decrease with gas pressure $p.$ 
In spite of these variations and of its sensitivity to the discharge strength for the 588~nm line, fluorescence polarimetry is a convenient and accurate polarization measurement technique for low-$p$ and low-$B$ MEOP experiments. The two existing calibrations of the technique, performed at 668~nm using different methods \cite{Bigelow92,Lorenzon93}, agree within typical uncertainties of 2\%.

\subsubsection{Light absorption polarimetry}
\label{Subsubsec_1083Polarimetry}
Since the early days of MEOP \cite{Colegrove63,Daniels71a,Greenhow64}, 1083~nm light absorption measurements have been used to evaluate nuclear polarization in cells. The method is based on the fact that the ME-enforced spin-temperature distribution establishes a strong link between $P_\mathrm{He}$ and \esS\ state populations (Sec.~\ref{subsec:spinT}, Eqs. \eqref{aivsbeta} and \eqref{yivsbeta}). Probing absorption by two suitable Zeeman sublevels, or combinations of sublevels, is sufficient to infer the spin temperature $\beta,$ and hence $P_\mathrm{He},$ from the ratio of measured absorption coefficients. Additionally, the \esS\ state density $n_{\mathrm{m}}$ integrated over the probe beam path can usually be inferred as well from such measurements.

At low $B$, two different probe beam polarizations are used to address different sublevels. For a longitudinal beam, propagating along \PJ{$\textbf{B},$} $\sigma_+$ and $\sigma_-$ polarizations selectively probe $a_5$ and $a_6$ when tuned to C$_8$ or $a_1+a_2/3$ and $a_4+a_3/3$ when tuned to C$_9$. For a transverse beam, the orthogonal linear polarizations ($\pi :\bot \PJ{\textbf{B}}$ and $\sigma : \parallel \PJ{\textbf{B}}$) provide independent combinations of populations only when tuned to C$_9$.  In isotopic mixtures the D$_0$ transition component can be used in both geometrical arrangements. For longitudinal probe polarimetry the ratios $r_i^\parallel =A_+/A_-$ of absorption coefficients for $\sigma_+$ and $\sigma_-$ light (the subscript $i$ is 8 or 9 for C$_8$ or C$_9$, 0 for D$_0$) are:
\begin{equation}
r_8^{\parallel }=\frac{1-P_\mathrm{He}}{1+P_\mathrm{He}},\: \: r_9^{\parallel }=\frac{2-P_\mathrm{He}}{2+P_\mathrm{He}}\:  r_8^{\parallel 2},\: \: %
r_0^{\parallel }= r_8^{\parallel 2}.
\label{eq:rabsparallel}
\end{equation}
This configuration offers the highest sensitivity at low polarizations, with  linear coefficients $-2,$ $-5,$ and $-4,$ respectively.  For transverse probe polarimetry the ratios $r_i^\bot =A_\pi /A_\sigma$ of absorption coefficients for $\pi $ and $\sigma $ light are:
\begin{equation}
r_9^{\bot }=\frac{1-P_\mathrm{He}^2}{1+2P_\mathrm{He}^2},\: \: %
r_0^{\bot }= \frac{1-P_\mathrm{He}^2}{1+P_\mathrm{He}^2}.
\label{eq:rabsperp}
\end{equation}
The transverse configuration has a weak sensitivity at low polarizations since it only depends on $P_\mathrm{He}^2$. This is actually a useful feature for measurements performed while an intense pumping light is applied: the method is almost insensitive to the potentially large pump-induced deviations of populations from their spin-temperature values, such as displayed in Fig.~\ref{fig_populations} \cite{Talbot11}.

As is done for fluorescence polarimetry, a mechanically or electrically modulated polarization or a 2-channel static scheme can be used for light absorption polarimetry. The probe beams must be weak enough so as not to locally affect the measured populations (low intensity) nor globally deposit angular momentum of the wrong sign (low absorbed power) \cite{Talbot11}. In static schemes, a small geometrical separation of the two probe beams should be used to avoid coherently addressing common levels with the two beams, which has been observed to induce sizable artifacts in the measurements \cite{Talbot11}.

At high $B$, all energy degeneracies of the Zeeman sublevels are lifted and two different probe beam frequencies are used to address different sublevels.  
Due to hyperfine decoupling, ME collisions imperfectly enforce the link between $P_\mathrm{He}$ and the \esS\ state populations, so that systematic measurement errors may appear even for moderate pump intensities. Fortunately, an adequate choice of optical transitions can be made to avoid such difficulties, by probing a pair of sublevels that are not addressed by the pump \cite{Abboud04,Nikiel13,Suchanek07}. This is achieved using the convenient line doublets highlighted in Fig.~\ref{fig:MEOP_HiB}a: a periodic frequency sweep of the probe (at fixed circular polarization) sequentially yields values of two absorption coefficients that are used to infer both $n_{\mathrm{m}}$ and $P_\mathrm{He}$.

In spite of the need for a probe laser at 1083~nm, absorption polarimetry is preferred for high field or high pressure situations, and for measurements in isotopic mixtures, since it yields accurate results if adequate care is taken.

\subsection{Relaxation in MEOP cells and steady-state polarizations }
\label{Subsec:Relax}
\subsubsection{Discharge-induced polarization decay}
MEOP apparatus and cells are designed so as to avoid significant polarization loss due to diffusion in field inhomogeneities or wall relaxation. When the discharge used to populate the \esS\ state is off, relaxation times of one to several hours are usually achieved. 
When the discharge is on and the pumping light is interrupted following polarization build-up, decay times $1/\Gamma_\mathrm{D}$ of order 1~min are typical. 
 As the discharge intensity
is increased, the metastable density $n_{\mathrm{m}}$ and thus the polarizing rate increases but the achievable
 \He3\ polarization usually decreases. These dependencies vary with gas pressure and discharge frequency;
\citealp{Gentile93}, reported on studies with pressures between 0.13~mbar and 6.5~mbar
and frequencies between 0.1~MHz and 10~MHz. Several other groups have studied the influence of cell size and shape, configuration of external discharge electrodes, and frequency: over a wide range of pressures (up to hundreds of mbars) and fields (up to 4.7~T), $n_{\mathrm{m}}$ and $\Gamma_\mathrm{D}$ are found to positively correlate, with $n_{\mathrm{m}}$ increasing less rapidly than $\Gamma_\mathrm{D}$ with the discharge excitation power \cite{BatzPhD,Nikiel13}. 

The effect of an applied field \PJ{exceeding usual low holding fields} depends on pressure. At low pressures ($p<1~\mathrm{mbar}$), the ratios $\Gamma_\mathrm{D}/n_{\mathrm{m}}$ increase \PJ{in the applied field} (at 30~mT and above). At higher pressures, on the contrary, $\Gamma_\mathrm{D}/n_{\mathrm{m}}$ decreases \PJ{in the applied field}, which is potentially beneficial for efficient MEOP. For instance, at $p=2.45~\mathrm{mbar},$ $\Gamma_\mathrm{D}$ is decreased by a factor of 2 to 4 at fixed $n_{\mathrm{m}}$ for $B=30~\mathrm{mT}$ \cite{BatzPhD}. At higher $p$ (tens of mbar) and $B$ (1.5~T) the decrease can exceed a factor of 10. However, at such high pressures, the radial distribution of $n_{\mathrm{m}}$ has an inverted distribution (with a minimum on cell axis) correlated to a plasma localisation near the cell wall \cite{Dohnalik11}. This distribution is sensitive to $B$, which makes it delicate to quantitatively compare $\Gamma_\mathrm{D}/n_{\mathrm{m}}$ ratios between low and high $B$.

Part of the polarization decay is due to relaxation in the \esS\ state, which drives a flow of angular momentum {\it from} the ground state reservoir with a rate $\Gamma_\mathrm{ME}.$ %
\PJ{The remainder of the decay, with a rate $\Gamma_\mathrm{g},$ is due to various relaxation processes directly affecting polarization in the ground state, for instance} collisions with excited ionic or molecular He, or electronic excitation and loss of angular momentum by emission of polarized light in the radiative cascade (see Sec.~\ref{Subsec_Polarimetry}). 
Altogether, the decay rate is:%
\begin{equation}
\Gamma_\mathrm{D}=\Gamma_\mathrm{ME}+\Gamma_\mathrm{g}.
\label{eq:gsdecay}
\end{equation}

$\Gamma_\mathrm{ME}$ scales as the volume average over the cell of $\gamma_\mathrm{r}^\mathrm{S}\,n_{\mathrm{m}}/N_{\mathrm{g}}$ with a $P_\mathrm{He}^2$-dependent computed factor equal to 11/3 in pure \Het\ at low polarization  \cite{Batz11}. Moreover, $n_{\mathrm{m}}$ is experimentally found to depend on $P_\mathrm{He}^2$ during polarization decays at fixed discharge excitation levels, changing by up to $\pm 20\% .$ $n_{\mathrm{m}}$ depends on the local balance of excitation to the \esS\ state in the discharge (involving ionization by electron impact, recombination, and radiative cascade), local decay through various processes (e.g., quenching by chemical impurities in the helium gas, 3-body conversion to metastable He$_2$ molecules, Penning ionizing collisions), and atomic diffusion in the helium gas combined with excitation loss at the cell wall. Optogalvanic effects may also play a role in the balance of processes occurring in the plasma. %
\PJ{Therefore $n_{\mathrm{m}}$ may depend on $P_\mathrm{He}^2$ and on MEOP conditions due, for example,  to the influence of electronic orientation on the cross section of Penning collisions \cite{Hill72,Fedichev96}.}
In spite of the expected $P_\mathrm{He}^2$-dependency of $\Gamma_\mathrm{ME}$, decays are systematically observed to be accurately exponential with $P_\mathrm{He}$-independent rates $\Gamma_\mathrm{D}$ over a wide range of gas pressures and field strengths, a so far unexplained observation \cite{Batz11,Nikiel13}.

\subsubsection{Optical-pumping-induced polarization loss}
When the pumping light is applied, the time evolution of $P_\mathrm{He}$ as well as its steady-state value, $P_{\mathrm{He}\infty},$ can be inferred from the balance between the inflow of angular momentum of Eq.~\eqref{eq:Mdot} and the loss term $-\Gamma_\mathrm{g} P_\mathrm{He}.$ This approach provides a detailed insight on the key relevant processes at play in MEOP, but it requires providing a trusted MEOP model with values for all physical and phenomenological parameters. A more pragmatic approach can be preferred for its robustness: a global angular momentum budget in which  the rate of change of the ground state nuclear polarization is written as the net balance between the angular momentum actually transferred through optical pumping cycles to the \esS\ atoms and an angular momentum loss directly associated to the ground state atoms \cite{Batz11}:  
\begin{equation}
\frac{dP_\mathrm{He}}{dt} = 2\eta \frac{W_\mathrm{abs}}{N_\mathrm{g}V_\mathrm{cell}\hbar \omega}-\Gamma_\mathrm{R} P_\mathrm{He},
\label{eq:balance}
\end{equation}
where $W_\mathrm{abs}$ is the pumping light power absorbed by the gas. The global polarization loss rate $\Gamma_\mathrm{R}$ introduced in this approach is not a constant but may vary with $P_\mathrm{He}$ and MEOP conditions as does the photon efficiency $\eta .$ Equation~\eqref{eq:balance} involves known parameters and experimentally measurable quantities ($P_\mathrm{He}$ and $W_\mathrm{abs}$), plus two unknown ones: $\eta $ and $\Gamma_\mathrm{R}.$ The photon efficiency $\eta $ can be evaluated using a MEOP model or directly inferred from transient buildup measurements at \PJ{$P_\mathrm{He}=0$ for which} the loss term in Eq.~\eqref{eq:balance} vanishes. For C$_8$ or D$_0$ optical pumping in low field, where $\eta $ does not vary with $P_\mathrm{He}$, $\Gamma_\mathrm{R}$ can thus be fully inferred from experimentally determined quantities.

\begin{figure}%
\includegraphics[width=3.3in]{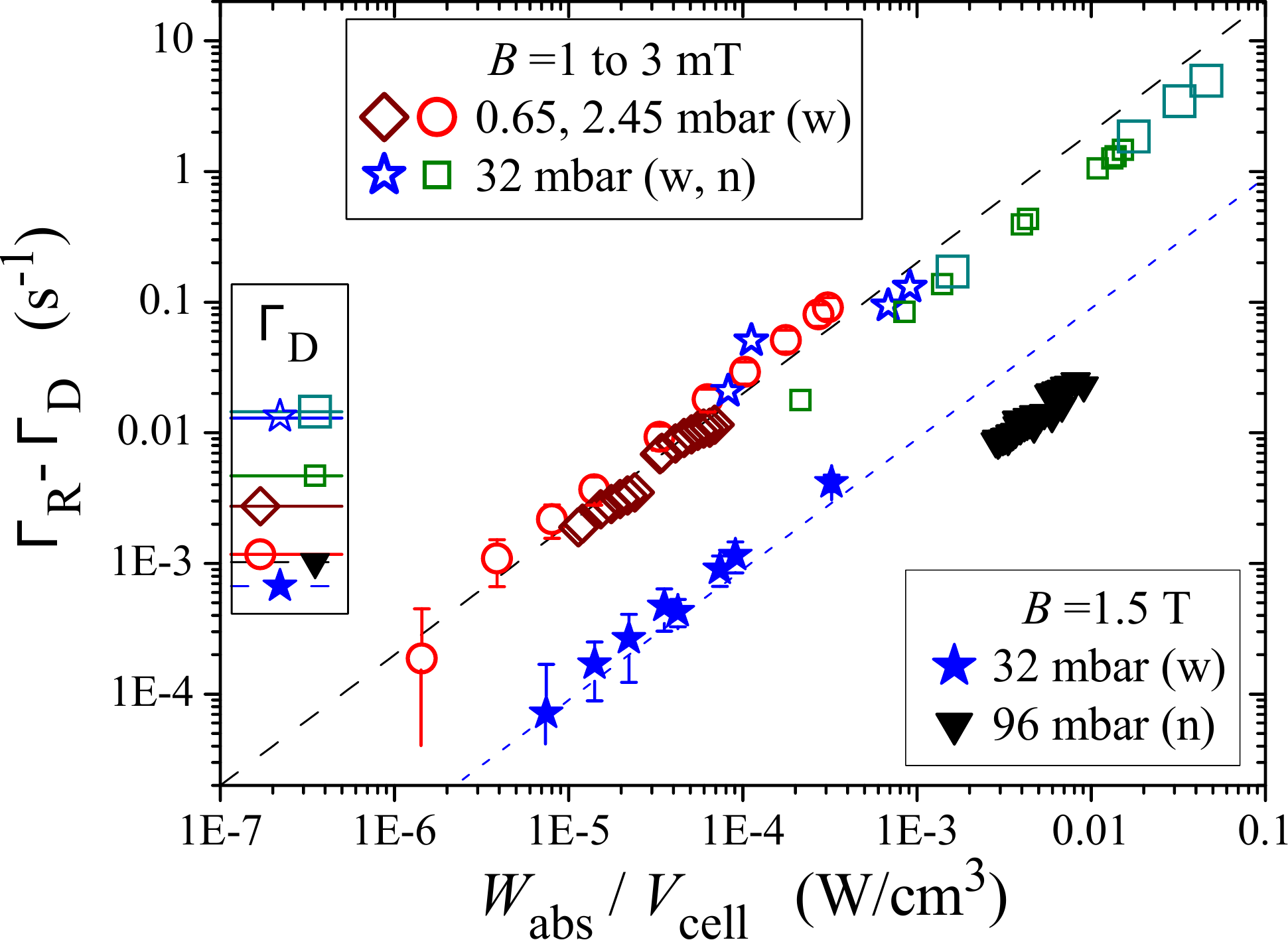}%
\caption{Optical-pumping-induced additional loss rates are plotted vs absorbed pumping power per unit volume for various gas pressures and fields (see legends) and cell diameters: wide (w: $\geq 5$~cm) and narrow (n: 1.5~cm). Each of the two logarithmic scales spans 6 decades. The lines are guides for the eye corresponding to linear variations \PJ{with linear coefficients 200~cm$^3$/J (dashed line, low B) and 9~cm$^3$/J (dotted line, 1.5~T). The ordinates of the horizontal lines in the box next to the left axis are the values of the pumping-free decay rates $\Gamma_\mathrm{D}$ for the different sets of data. They range from $0.67\times 10^{-3}~\mathrm{s}^{-1}$ (filled stars) to $14.5\times 10^{-3}~\mathrm{s}^{-1}$ (large open squares).} The figure is adapted from \cite[Fig.~6.62]{BatzPhD}; it compiles data from Mainz ({\color{Brown}$\meddiamond $}), Cracow ($\blacktriangledown $), and Paris (all other symbols; the small and large squares stand for weak and strong discharges in the same cell).}%
\label{fig:GammaR}%
\end{figure}

This pragmatic approach has been used to analyze experimental data for which the absorbed pump power was measured. Figure~\ref{fig:GammaR} displays selected results obtained at various pressures in low and high fields. Data obtained in steady-state (once \PJ{$P_{\mathrm{He}\infty}$} has been reached, with a null left-hand side term in Eq.~\eqref{eq:balance}) as well as data  collected during polarization build-ups towards \PJ{$P_{\mathrm{He}\infty}$} (three series of closely clustered symbols) are displayed. Despite the qualitatively different behavior of build-ups: exponential approaches towards \PJ{$P_{\mathrm{He}\infty}$} at high fields \cite{Nikiel13} contrasting with non-exponential build-ups having decreasing rates at high $P_\mathrm{He}$ \cite{Gentile93,Batz11}, all corresponding data nicely collapse with the steady-state data.
The decay rates $\Gamma_\mathrm{D}$ for the various experiments tend to increase with pressure, and are strongly reduced, at fixed $p$, in high field (inset in Fig.~\ref{fig:GammaR}). The polarization loss rates $\Gamma_\mathrm{R}$ are found to significantly differ from $\Gamma_\mathrm{D}$ for most of the data points. The differences $\Gamma_\mathrm{R}-\Gamma_\mathrm{D}$ plotted in Fig.~\ref{fig:GammaR} for low-field data (open symbols) span over 4 orders of magnitude when experimental parameters are varied. The absorbed power $W_\mathrm{abs}$ is varied by changing the incident power or the absorption coefficient, through $n_{\mathrm{m}}$ (in steady-state) or $P_\mathrm{He}$ (during build up); different cell diameters yield different \PJ{cell volumes} $V_\mathrm{cell}$. The pumping-induced additional relaxation rates reveal a consistent linear-like variation only when plotted versus the ratio $W_\mathrm{abs}/V_\mathrm{cell}.$   
The 1.5-T-data display a similar pumping-enhanced relaxation behavior, but with a significantly reduced rate coefficient (the two lines in Fig.~\ref{fig:GammaR} correspond to linear laws with a ratio 30 of their slopes). Data \PJ{obtained} at 30~mT (\citealp[Fig.~6.59]{BatzPhD}) for $p$ up to 2.45~mbar are identical to the corresponding low-$B$ results. This field \PJ{strength} is sufficient to suppress the angular momentum loss which occurs in the radiative cascade and plays a key role in fluorescence polarimetry. \PJ{The absence of effect of a 30-mT field on the data suggests that angular momentum loss in the radiative cascade plays no role in the pumping-induced polarization loss, }but that larger hyperfine structures are involved. 

The physical processes causing such strong pumping-enhanced polarization losses remain to be elucidated. Radiation trapping (i.e., re-absorption of 1083~nm fluorescence light) or plasma modification by light-enhanced creation of a relaxing long-lived species through the \esP\ state have been considered as possible origins of pumping-induced loss mechanisms, but they could not account for the observations \cite{BatzPhD}.

\label{Subsec:PHelimits}
\subsubsection{Steady-state polarization limits}
\begin{figure}%
\includegraphics[width=3.in]{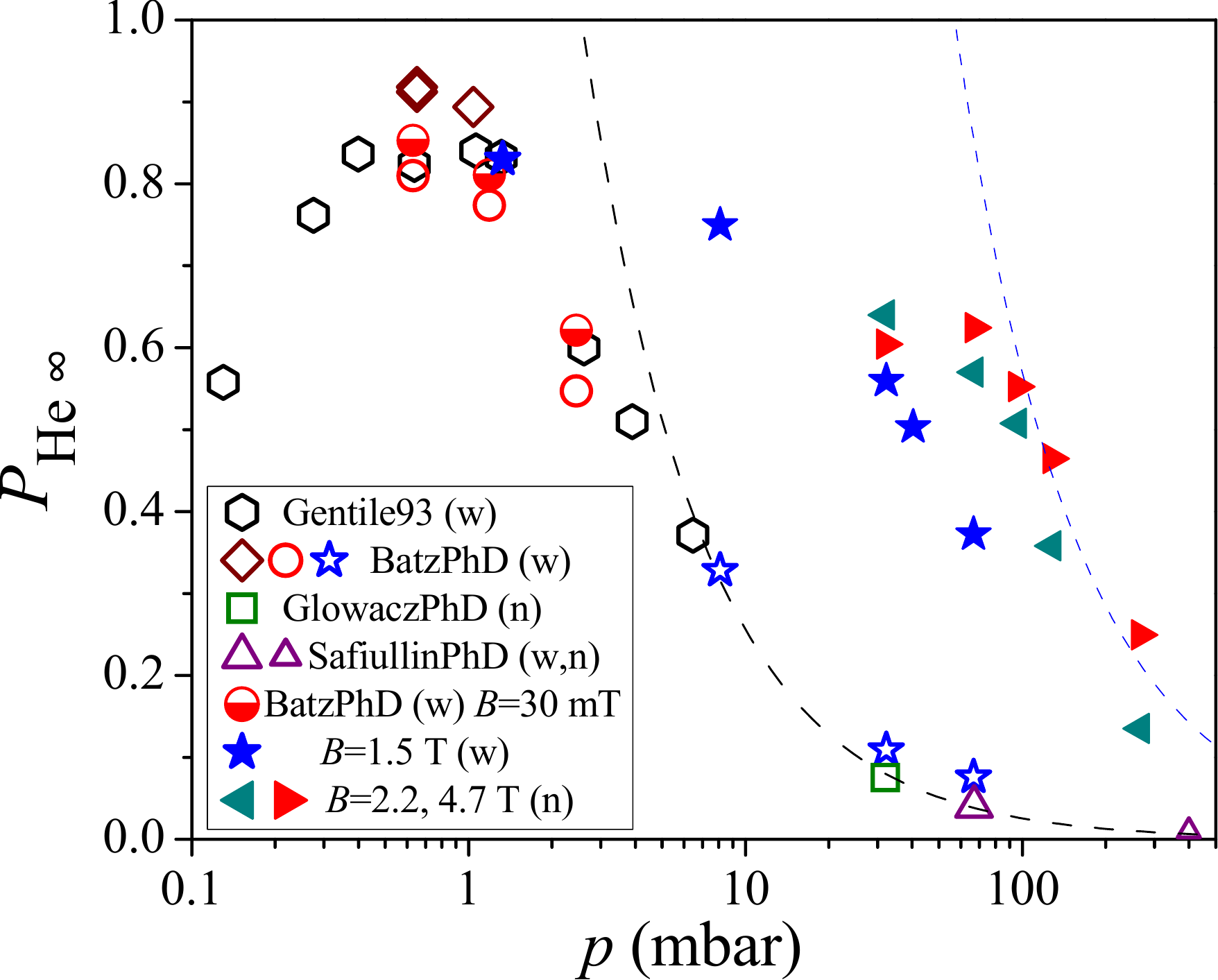}%
\caption{Variation with pressure of highest steady-state polarizations achieved by various groups at low fields (open symbols, 1 to 3~mT) and high fields (filled symbols, see legend). The two lines \PJ{($P_\mathrm{He}=2.56/p$ and $P_\mathrm{He}=57/p$) are upper bounds derived from the lines in Fig.~\ref{fig:GammaR} (see text).} The figure is adapted from \cite[Fig.~6.48]{BatzPhD}, with additional data from  \cite{GlowaczPhD} and  \cite{SafiullinPhD}; it compiles data from Caltech ($\hexagon $), Mainz ({\color{Brown}$\meddiamond $}), Cracow ({\color{PineGreen}$\filledmedtriangleleft $}, {\color{red} $\filledmedtriangleright $}), and Paris (all other symbols). Wide and narrow cells (w, n): see Fig.\ref{fig:GammaR}.}%
\label{fig:PHevsplin}%
\end{figure}

Since the development \PJ{of 
powerful} lasers for MEOP, it has been noted that the achievable steady-state polarizations $P_\mathrm{He\infty}$ are in practice obtained at moderate laser \PJ{intensities and} that higher intensities only yield faster build up rates. Moreover, very high $P_\mathrm{He\infty}>0.8$ can be obtained only in a narrow pressure range. Figure~\ref{fig:PHevsplin} displays a compilation of  steady-state polarizations \PJ{obtained at low fields (open symbols) and high fields (filled symbols, $B\geq 1$~T) over a wide range of pressures.} 
A moderate field increase, just sufficient to impede angular momentum loss in the radiative cascade (30~mT, half-filled circles), reduces $\Gamma_\mathrm{D}$ and increases $P_\mathrm{He\infty}$ for weak discharges, but the rapid decrease of $P_\mathrm{He\infty}$ with pressure is still observed. At higher fields, on the contrary, large polarizations can be achieved at 10- to 100-fold higher pressures.

Using \PJ{the steady-state solution of} Eq.~\eqref{eq:balance} to relate $P_\mathrm{He\infty}$ to the absorbed power $W_\mathrm{abs}$, one obtains strikingly different behaviors using for $\Gamma_\mathrm{R}$ an optical-pumping-independent loss rate, such as $\Gamma_\mathrm{D}$, or the pumping-enhanced loss rate which has been consistently inferred from observations at high laser powers. The steep decease of $P_\mathrm{He\infty}$ with $p$ above a few mbar is not reproduced by MEOP models assuming fixed losses. For instance, $P_\mathrm{He\infty}>0.8$ instead of 0.5 to 0.6 would be achieved for $p=2.45$~mbar \cite[Fig.~6.47]{BatzPhD}. On the contrary, if the approximate scaling \PJ{$\left(\Gamma_\mathrm{R}-\Gamma_\mathrm{D}\right)\propto W_\mathrm{abs}/V_\mathrm{cell}$ (the lines in Fig.~\ref{fig:GammaR}) is substituted in Eq.~\eqref{eq:balance} as a lower bound for $\Gamma_\mathrm{R},$ an upper bound for $P_\mathrm{He\infty}$ is found to be independent of  $W_\mathrm{abs}$, hence of the incident laser power, and to decrease as 1/$p$ with pressure. The lines in Fig.~\ref{fig:PHevsplin} are upper bounds corresponding to the lines in Fig.~\ref{fig:GammaR} using $\eta =1.25$ in Eq.~\eqref{eq:balance}. The angular momentum budget approach therefore provides a link} between the observed pumping-enhanced polarization losses of the previous Section and the limits in the experimentally achieved $P_\mathrm{He\infty}$, especially at high pressures. Additionally, the spectacular increase in efficiency of MEOP at high pressure in high field can be attributed to the increasingly strong reduction of the pumping-enhanced loss for $B>1~$T.

Achieving very high polarizations in low fields requires operating at a suitably low pressure with pumping light tuned to the most efficient transition component (usually C$_8$ or D$_0$) and carefully tailored to the needs: \PJ{
\begin{itemize}
	\item a bandwidth of order the Doppler width to avoid velocity-selective pumping and the associated optical saturation, 
	\item a spatial transverse profile matched to the pressure-dependent density profile of the \esS\ state atoms, and 
	\item a very high degree of circular polarization because sub-\% fractions of an opposite polarization are more efficiently absorbed and have thus a strong adverse effect at high $P_\mathrm{He}$ \cite{Leduc00}. 
\end{itemize}
} At higher fields and pressures, stronger transition components are the most efficient ($\mathrm{f}_2^-$ or $\mathrm{f}_4^-$ with a slight detuning from the composite line center, \citealp[Chap.~6]{AbboudPhD}) and the constraints 1 and 3 listed above can be somewhat relaxed thanks to collisional broadening and to large Zeeman splittings making the opposite polarization non-resonant. Recently, MEOP relying on the \esS\ -3$^3$P transition at 389~nm instead of the usual 1083-nm transition was demonstrated \cite{Maeda10}. It was found to yield much lower polarizations than standard MEOP for comparable laser powers; this may be due to the smaller line component splittings \cite{Sulai08} and larger Doppler width for this transition which prevents selecting an isolated efficient component.

\subsection{Lasers for MEOP}
\indent A succession of continually improving lasers has been used for MEOP.
For traditional, low field MEOP a match to the 2 GHz \PJ{Doppler-broadened} absorption width is ideal.
The first work in laser optical pumping was performed with color center lasers \cite{Nacher85},
and argon-ion pumped Nd:LMA lasers \cite{Eckert92}.  Velocity selective optical pumping was an issue
for narrowband lasers, but increased efficiency was obtained by frequency modulation~\cite{Elbel90}.
Frequency stabilization to the helium absorption lines was demonstrated~\cite{Heil90}.
These early approaches were superseded in both performance and simplicity by the arc-lamp pumped Nd:LMA laser,
which was constructed by replacing the Nd:YAG crystal rod in a commercial arc lamp-pumped laser and adding one or two etalons
to produce a 2 GHz linewidth \cite{Schearer_jap90,Aminoff91}.  Tuning the laser was accomplished
by varying the temperature and/or angle of the etalons.  These lasers produced an output of several watts,
but performance was highly dependent on crystal quality.  Although the spectrum was actually a series
of narrow peaks with substantial jitter the influence of velocity-changing collisions yielded efficient
optical pumping \cite{Elbel90,Gentile_josab03}, resulting in a substantial improvement in polarization
and pumping rate \cite{Gentile93}.  More recently Nd:LMA has been superseded by Yb fiber
lasers \cite{Chernikov97,Mueller01,Gentile_josab03,Tastevin04}. 
A Yb fiber oscillator and amplifier, properly constructed for MEOP, yields a fairly uniform spectrum with
a $\approx$2 GHz spectral width.  An alternative scheme is to use a diode laser oscillator and Yb fiber amplifier,
which requires modulation for efficient optical pumping~\cite{Mueller01}.  Comparable performance for a given power
is obtained for Nd:LMA and both types of Yb lasers~\cite{Gentile_josab03}, but Yb lasers have higher
available power as well as greater convenience and a smaller footprint. For small sealed cells, 1083 nm diode lasers ($\approx$50~mW) have been used to produce up to 40~\% \He3\ polarization
in pure \He3\ and 80~\% in \He3\ - $^4$He mixtures~\cite{Stoltz96}.

\subsection{Compression approaches}

Application to polarized targets for charged particle and photon scattering, MRI, and neutron spin filters requires
compressing \He3\ gas polarized by MEOP.  Following the path of an early attempt
to use a Toepler pump~\cite{Timsit71a}, a successful compressor was developed for electron scattering at MAMI\footnote{Mainz Microtron, Mainz, Germany}~\cite{Eckert92,Becker99}.
This compressor employed a mercury column to compress gas polarized by MEOP at a pressure of $\approx$~1~mbar 
into a 100 cm$^3$ target cell at $\approx$~1~bar. \Tom{For this device}  50~\% \He3 polarization was achieved.  At Mainz, this apparatus was replaced by
a titanium piston compression apparatus~\cite{Becker94,Surkau97} that achieved pressures of a few bar.
With subsequent improvements this compressor could produce between 60~\% and 80~\% \He3\ polarization
with corresponding polarizing rates of between 4 bar-L/h and 1 bar-L/h~\cite{Batz05}. 
\Tom{For this apparatus, the best achievable polarization in the open compression system under
static conditions was 84 \%, not far from the record high value of 91\% achieved in a sealed cell. }
At Indiana University an aluminum piston compressor developed for neutron spin filters reached 50~\% \He3 \ polarization \PJ{\cite{Hussey05}}.
\Tom{Both the Mainz and Indiana University apparatus} employed two stages of compression with a buffer cell between the stages.
In these and subsequent piston \PJ{compressors, the} polarization loss in the compressor
itself was a few percent or less, but additional polarization losses are incurred due to gas flow and relaxation in 
the buffer cell, storage cell, and interconnecting tubing. 
A similar compressor was installed at the ILL\footnote{Institut Laue-Langevin, Grenoble, France}, but eventually replaced by
a new apparatus, denoted \textquotedblleft Tyrex\textquotedblright, for filling neutron spin filter cells. 
With further improvements, a typical value of 75~\% \He3 \ polarization has been obtained in NSF cells~\cite{Lelievre07}.
A compression system denoted HELIOS was put into operation at the FRM II reactor in Munich~\cite{Hutanu_nn07}
and more recently a compression apparatus is also planned for use at ISIS \cite{Beecham11}.

In parallel with the development of these large scale piston compression apparatus,
smaller scale compressors based on peristaltic~\cite{Nacher99} and diaphragm pumps~\cite{Gentile01} have been employed for polarized gas MRI.
For the peristaltic approach with flow rates of 0.1 bar-L/h and polarization preservation near unity,
a polarization of between 30~\% and 40~\% could be achieved in the storage cell. 
Typically, 0.04 bar-L was used routinely for MRI applications~\cite{Bidinosti03}.
More recently a similar apparatus was reported~\cite{Collier12}.
The diaphragm pump~\cite{Gentile01} method produced between 20~\% and 35~\% \He3\ polarization
using pure \He3\ or between 35~\% and 50~\% polarization using \He3-$^4$He mixtures. 
The typical outlet pressure was one bar, polarization preservation \PJ{was 0.75, and} the polarizing rate was 0.4 bar-L/h.
However, further development of this approach has not been pursued.

\section{ \He3\ Relaxation}
\label{Relax}
\subsection{Dipole-dipole}\label{dipdip}
Newbury {\it et al}~ \cite{Newbury93} calculated that dipole-dipole interactions in bulk \He3 \ limit the longitudinal relaxation time of 
polarized \He3 \ to $T_1 = 807/p$~h (where $p$ is the \He3\ pressure in bar, for a cell temperature of 296~K),
where the polarization decays as $\exp(-t/T_1)$ for \He3\ polarized along the applied magnetic field.
This expression was verified for \He3\ pressure in the range between 4 bar and 12 bar
in sealed SEOP cells made from aluminosilicate glass (Corning 1720)\footnote{Corning Glass, Corning, NY.
}.
The observed limiting relaxation time $T_1$ was extended to lower pressures in borosilicate glass ($T_1=300$ h observed
for $p$=2.5 bar \citep{Smith98}) and sol-gel coated borosilicate glass (Corning Pyrex) ($T_1=344$ h observed
for $p$=2.1 bar \citep{Hsu00}).
The development of neutron spin filters led to practical interest in pressures near 1 bar,
in which relaxation times of several hundred hours have been observed \cite{Rich02,Parnell09,Chen11,Salhi14} in SEOP GE180 cells,
corresponding to relaxation times of order several thousand hours from wall
relaxation and other sources. All of these cells contained Rb or Rb/K mixtures for the SEOP process, films of which have been
shown to suppress wall relaxation as compared to bare glass cells~ \cite{Heil95}.
\subsection{Wall}
\subsubsection{Room temperature}\label{roomtemprelax}

In most practical situations, wall relaxation limits the achievable $T_1$ but is not well understood.
Nearly a half century ago\PJ{ it was found that,} for bare glass cells at room temperature, aluminosilicate glass yielded relaxation times
of tens of hours as compared to a few hours for the borosilicate glass Pyrex \cite{Fitzsimmons69}.   The temperature dependence of the relaxation times for both types of glass indicated that Pyrex relaxation at room temperature is dominated by permeation whereas adsorption dominates for aluminosilicate cells. In Pyrex, $T_1$ becomes shorter at high temperatures due to increased permeation of He into the glass.  A detailed model of \He3 relaxation on the surface of borosilicate glass which accurately predicted observed relaxation rates and their temperature dependence was reported \citep{Jacob03}.

 For bare aluminosilicate glasses, which have negligible permeation and adsorption energies of order 100K, higher temperatures result in longer $T_1$ due to reduced adsorption. This discovery \citep{Fitzsimmons69}, plus the much better alkali metal chemical resistance, led to the dominance of aluminosilicate glass for SEOP.  Nearly complete suppression of wall relaxation was achieved in the study of dipole-dipole relaxation
with SEOP cells made from aluminosilicate glass \cite{Newbury93}.  It was found that cells
made from fully blown glass yielded the best results and cleaning with nitric acid
was suggested if a fully blown cell was not possible. \PJ{The term ``fully blown'' indicates that all interior surfaces of the glass have been thoroughly melted; the resulting changes in these surfaces has generally been found to decrease \He3 relaxation.} Wall relaxation times of several thousand hours have also
been observed in SEOP neutron spin filter cells and the importance of fully blown glass verified~ \cite{Rich02,Parnell09,Chen11,Salhi14}.

In SEOP cells, alkali-metal coatings are always present.
In a series of measurements with various glasses and metal coatings, the importance of metal coatings for reducing wall relaxation was  demonstrated by \citet{Heil95}.   Relaxation times of 68 h and 120 h were reported for cesium-coated Pyrex and cesium-coated aluminosilicate glass (Schott supremax)\footnote{SCHOTT North America, Inc.,
Louisville, KY 
}, respectively.
(It is interesting to note that \citet{Fitzsimmons69} reported a relaxation time of 250 h  for a SEOP cell,    which was substantially longer than they reported for cells without alkali-metal.
    Whereas they recognized that the presence of Rb did not adversely affect the T1, the
    possible benefit of its presence did not appear to be recognized.)
In a later study that also included alkali-metal oxide coatings, it was found that despite the 
substantial role of the coating, the longest relaxation times were still obtained with aluminosilicate glass \cite{Deninger06}.
Hence the substrate still plays a role and it was speculated that this role could be associated with imperfections in the coating.  

\PJ{Whereas} for both methods the longest relaxation times are usually obtained with aluminosilicate glasses, 
Pyrex SEOP cells have been employed for MRI applications \cite{Jacob02} and sol-gel coatings developed for SEOP cells \cite{Hsu00}.
More recently this technique was employed for a \He3\ target to avoid the substantial barium content in GE180 \cite{Ye10}.
MEOP \PJ{storage} cells typically employ cesium coatings; long relaxation times have \PJ{also} been obtained in fused silica \cite{Deninger06}
and silicon-windowed Pyrex cells \cite{Lelievre07}.  

For compressor and target applications relaxation on various materials has been studied \cite{Gamblin65,Timsit71c,Jones93,Korsch97,Hussey05,Katabuchi05}. 

\subsubsection{Magnetic field and orientation effects} \label{relaxfield}
The advent of polarized gas MRI led to studies of relaxation in strong magnetic fields,
which has revealed yet further unexplained aspects of \He3\ relaxation. 
In the first study \cite{Jacob01}, it was found that the \He3\ relaxation rate could be increased by a factor of between 2 and 20 solely by exposure of
SEOP cells to a magnetic field of a few tenth of a Tesla (few-kG). The original $T_1$ could be restored by degaussing the cell, leading to the term
\textquotedblleft $T_1$ hysteresis\textquotedblright\ to describe the observed behavior.
Soon thereafter a significant dependence of $T_1$ in SEOP cells due only to the physical orientation of the cell in a 3 mT (30 G)
applied magnetic field was observed~ \cite{Jacob04}. The presence \PJ{of both alkali metal 
and heating of the cells associated with the SEOP process were} 
necessary to produce this low-field orientation dependence.  Later studies showed that the relaxation time
can depend on the direction and strength of the magnetic field~ \cite{Chen11}. More recently, the angular dependence of 
$T_1$ was measured and found to be characteristic of a dipolar effect \cite{Boag14}. Based on studies of boundary
collisions of random walks, \citet{Bicout13} concluded that the depolarizing effect of rare magnetic impurities may be much
larger than expected and speculated that this enhancement could explain $T_1$ hysteresis. \
In a test on a pure Rb, sealed SEOP \PJ{cell,} the $T_1$ was observed to decrease by an order of magnitude for a field strength of only 40 mT~ \cite{Chen11}.

$T_1$ hysteresis was also observed in both Cs-coated and \PJ{bare-glass}, valved, MEOP cells~ \cite{Hutanu07}.
The variation of $T_1$ with magnetic field was measured and found to decrease exponentially with a constant of (30 mT)$^{-1}$ [(300 G)$^{-1}$].
Superconducting quantum interference device (SQUID) measurements revealed that most of the magnetization was associated with the glass valves,
in particular the plastic parts and O-ring~ \cite{Hutanu_squid07}.  %


\subsection{Field gradients}
Relaxation due to static and oscillating magnetic field gradients in a variety of regimes has been addressed in several 
references~ \cite{Gamblin65,Schearer65,Cates88,Cates88a,McGregor90,Hasson90, Bohler94}.
At room temperature, a practical expression for the relaxation rate 1/$T_1$ 
\PJ{in a uniform field gradient} 
is given by~ \cite{McIver09}
\begin{equation}
{1 \over T_1^{\mathrm {gradient}}}={6700 \over p} \left({{\mid\vec\nabla B_x\mid^{2}} \over B_0^2} + {{\mid\vec\nabla B_y \mid^{2}} \over B_0^2}\right)\mathrm{h^{-1}}
\label{gradient}
\end{equation}
where $p$ is the gas pressure in bar and $\mid{\vec\nabla B_x \over B_{0}}\mid$ and
$\mid{\vec\nabla B_y \over B_{0}}\mid$ are the gradients in the transverse components of
the magnetic field (ie. the nuclear polarization is along the z-axis) in units of cm$^{-1}$~ \cite{Cates88}. 
For a cell at a pressure of one bar in a uniform gradient $\mid{\vec\nabla B_x \over B_{0}}\mid$=$\mid{\vec\nabla B_y \over 
B_{0}}\mid$=$3\times 10^{-4}$ cm$^{-1}$ the gradient-induced relaxation time is 830 h.  
{We note that Eq.~\eqref{gradient}  is valid as long as the product of Larmor precession frequency and the 
collision time is small compared to unity, valid for SEOP and also for MEOP at 20 G and typical
pressures of order 1 mbar.  For MEOP at very low pressures and high fields corrections to this are necessary \cite{Schearer65}}.

 In more recent work, \citet{Zheng_grad11} reported an approach based on calculating the autocorrelation function of spins to derive the magnetic field
gradient-induced transverse and longitudinal relaxation of spins undergoing restricted diffusion.
\citet{Guigue14}  performed a theoretical analysis of spin relaxation, for a polarized gas of spin 1/2 particles undergoing
restricted adiabatic diffusive motion within a container of arbitrary shape, due to magnetic field inhomogeneities
of arbitrary form.  This analysis provided a theoretical justification for the usual assumption that the
relaxation rate is determined by the volume average of the relevant gradients.  Studies of gradient-induced relaxation in the transfer of
gas into high magnetic fields has been also been reported \cite{Zheng_jmr11,Maxwell15}.

\subsection{Magnetostatic cavities}\label{sec:magcav}
Neutron spin filters have motivated a variety of magnetostatic cavities to provide a uniform
magnetic field for \He3\ gas on neutron beam lines.  For these applications, there may be space constraints and significant stray fields, as well as neutron spin transport fields to be matched (see Sec.  \ref{NSF_prince}).
In early applications of NSFs on the IN20 triple-axis spectrometer at the ILL, an end-compensated
solenoid surrounded by a $\mu$-metal box and a cubic iron box (40 cm$^3$ by 40 cm$^3$ by 40 cm$^3$)
was employed \cite{Heil99}.  Soon thereafter a superconducting magnetostatic cavity (Cryopol) was developed
for the D3 spectrometer that is capable of shielding an NSF from a nearby superconducting 
sample magnet operating at several Tesla \cite{Dreyer00}.  For moderate stray fields
a magnetic parallel plate capacitor was developed, which provided a uniform 1 mT field transverse
to the neutron beam  (known as a ``Magic Box'' because when first made, the performance was
better than predicted by finite element calculations) \cite{Petoukhov06}.  This cavity consisted
of \PJ{a} $\mu$-metal box with sides that are magnetized by either coils or permanent magnets.
The field lines enter the top plate and traverse the gap to the bottom plate.
The original box was 80 cm along the beam line, but in a \PJ{later, }permanent-magnet design this distance was reduced to 40 cm \cite{Hutanu08}
and the field increased to 1.7 mT.  An end-compensated design shortened the length along the beam line to 28.4 cm \cite{McIver09},
with field strengths up to 3.6 mT \cite{Chen14}. Magnetically shielded solenoids are also employed and  typically provide greater
protection from stray fields in space-constrained applications.  Typical achievable gradients
from magnetically shielded solenoids and magic boxes are between $2\times10^{-4}$ cm$^{-1}$ and $6\times10^{-4}$ cm$^{-1}$,
corresponding to gradient-induced relaxation times of between 3700 h and 600 h, respectively \cite{Chen14}. Polarized gas MRI also provided a motivation for compact, \PJ{light-weight} and inexpensive magnetized boxes for transporting polarized gas, in some cases via air freight~ \cite{Hiebel10}.

\subsection{Ionizing radiation}\label{IonRad}
Relaxation from 3 MeV protons for \He3\ polarized at mbar pressures by MEOP was reported in  \cite{Milner87}.
Soon thereafter relaxation from 18 MeV $^4$He alpha particles for \He3\ polarized at 0.8 bar by SEOP
was reported in  \cite{Coulter88}.  Theoretical studies~ \cite{BoninWalker88,BoninSaltzberg88}
showed that relaxation is caused by hyperfine coupling in $^3$He$^+$ and spin-rotation coupling
in $^3$He$_2^+$. 
\PJ{This relaxation source is greatly reduced in SEOP cells because nitrogen efficiently destroys these species.  For application of MEOP cells to electron scattering, addition of a small ([N$_2$]/[$^3$He] $\approx$ 10$^{-4}$) quantity of nitrogen after compression has been employed to quench these species~\cite{Meyerhoff94}.} 
Nevertheless, there is substantial relaxation induced by high current charged particle beams,
thus motivating the use of hybrid SEOP and rapid exchange of gas in double cell targets~ \cite{Dolph11,Singh15}.
Due to the much greater sensitivity of pure \He3\ cells to ionization, large effects
on \He3\ relaxation have been observed in MEOP cells due to neutron beams \citep{Petukhov16}.  For in-situ SEOP on neutron beam lines, beam-induced alkali-metal relaxation and 
other effects have been observed (see Sec. \ref{neutronbeameffects}).

\subsection{Low Temperatures}

Motivated by the need \PJ{for}dense polarized \Het\ targets, as well as by studies of thermodynamic and transport properties of polarized \Het\ liquid, attempts to dynamically polarize liquid \Het\ from optically pumped room temperature gas were made soon after the first MEOP experiments were reported. In this seminal work \cite{McAdams68}, polarized \Het\ gas was transported from a room temperature MEOP region through a connecting tube to a cold sample volume either by atomic diffusion or by transient gas flow during sample cool down. Most of the polarization was lost in the process, presumably due to fast wall relaxation in the cold part of the connecting tube or in the sample, with a maximum polarization of 0.15\% achieved in a small liquid \Het\ volume. This work triggered studies of wall relaxation of \Het\ gas and liquid at low temperatures by several groups, using either Boltzmann-polarized or MEOP- polarised \Het . It was confirmed that clean, bare Pyrex glass walls induce increasing relaxation rates $1/T_1$ at decreasing temperatures below $\approx 100~\rm{K}$ \cite{Fitzsimmons69,Lefevre88}, with $T_1<1~\rm{s}$ below 20~K for a low-density gas \cite{Barbe75,Chapman74}.  

In order to reduce the dwell time and polarization loss of \Het\ atoms colliding with cold glass walls, weak-binding diamagnetic cryogenic coatings have been added onto cell walls: Ne \cite{Chapman75,Chapman76}, H$_2$ \cite{Barbe75}, or liquid \Hef\ \cite{Himbert83}. The adsorption energies of \Het\ atoms on these coated \PJ{walls and} the relaxation mechanisms of atoms or adsorbed layers on weakly relaxing substrates have been extensively studied and \PJ{are} well understood \cite{Lusher88a,Lusher88b,Himbert89,Lefevre85,Lefevre88}. Depending on experimental conditions, different relaxation regimes can be observed. For instance, for low-density adsorbed \Het\ layers, $T_1$ does not depend on the bulk gas density $N$ and scales as $\exp(-2\Delta W/k_BT)$,  where $\Delta W$ is the adsorption energy and $k_BT$ the thermal energy. On the contrary, for high enough gas density or low enough temperature, a complete \Het\ monolayer is condensed and $T_1\propto N$ is essentially temperature-independent. H$_2$ coatings, for which $\Delta W \approx 12~\rm{K}$, yield long $T_1$s (up to several days) at 4~K (for comparison, $\Delta W \approx 150\rm{K}$ on bare glass). H$_2$ coatings are efficient up to $\approx 6~\rm{K}$ where they desorb and down to $\approx 2~\rm{K}$ where a \Het\ monolayer is formed. \Hef\ films extend the temperature range over which polarized \Het\ gas can be prepared or stored down to below 0.5~K (where the gas eventually liquefies if its pressure exceeds a few mbar).

Alternatively, cesium can be used to coat cell walls. With a \Het\ adsorption energy as low as $\Delta W \approx 2.3~\rm{K}$ \cite{Tastevin92}
cesiated glass (i.e. glass that has been in contact with cesium) is a weakly relaxing material from room temperature down to hundreds of mK. This property is plausibly linked with the non-wetting of alkali metal surfaces by \Hef\ liquid and films \cite{Nacher91}, and was initially demonstrated at low temperature before being assessed at room temperature \cite{Cheron95}.

Motivated by an experiment to measure the neutron electric dipole moment, new relaxation studies in liquid mixtures of \Het\ and \Hef\ have been performed for the relevant polymer-coated materials at cryogenic temperatures\cite{Ye08,Ye09}. 

With suitably coated walls, the way was open for reliable measurements of the bulk (dipole-dipole) relaxation in \Het\ gas \cite{Lusher88b} and \Het -\Hef\ liquid mixtures, for which $T_1$ exceeding 10~hours have been recorded at low \Het\ concentration \cite{Piegay02}. More importantly, McAdams' strategy to prepare polarized cold \Het\ samples from optically polarized gas could be successfully applied. For instance, polarization was nearly fully preserved when transferred to a gas sample at 4.2~K, therefore exceeding 50\% \cite{Crampton84a,Leduc84b}. 
Transient polarizations exceeding 40\% were reported for liquid \Het\ just after liquefaction \cite{Tastevin88}, subsequently decaying due to a bulk $T_1$ of $\approx 300~\rm{s}$. Thermally-driven convective flow between the low- and high-temperature regions was used to sustain up to 56\% nuclear polarization in steady-state in \Het -\Hef\ liquid mixtures \cite{Candela94}. Such high polarizations could not be achieved by MEOP directly performed at low temperature due to  the highly reduced rate of ME collisions (see Sec.~\ref{Subsec_MEcollvsT}), but room temperature OP with polarized gas transfer enabled a series of low-temperature studies of \Het\ as a quantum fluid: characterization of spin waves \cite{Nacher84,Tastevin85} and heat conduction changes \cite{Leduc87,Larat90} in \Het\ gas \PJ{and} of phase coexistence for liquid \Het\ \cite{Villard00,Tastevin_nmr92,Candela94}.  A similar technique was used with refillable cells to repeatedly prepare larger samples of polarized liquid in which magnetic interactions play a key role in non-linear NMR dynamics \cite{Hayden07,Baudin08}.

Let us also mention two alternative hyperpolarization techniques, not using MEOP but instead relying on nuclear relaxation and well-controlled phase transitions: rapid melting which yields up to 70\% transient polarization in liquid \Het\ \cite{Bonfait84,Bonfait87,Buu00} and spin distillation which  provides up to a seven-fold enhancement of the Boltzmann equilibrium polarization in \PJ{the} steady-state \cite{Nacher_prl91,Rodrigues97,vanSteenbergen98}.

\section{\He3\ polarization metrology and control}
\label{metrology}
Absolute polarization measurements methods for dense samples of polarized \He3\ include water-calibrated NMR \cite{Lorenzon93,Romalis_nim98},
EPR \cite{Romalis98,Babcock05}, neutron \PJ{transmission \cite{Jones00,Chupp07},}
and \PJ{magnetometry \cite{Noel96,Wilms97}}.  
In all of these methods except neutron transmission, the absolute \He3\ polarization is determined by measuring both
the \He3\ magnetization and the \He3\ density.   The \He3\ density is determined at the time a cell is filled with gas.
For SEOP cells the density can also be determined after the cell is filled from the width and/or shift of the alkali-metal absorption
lines \cite{Romalis97,Kluttz13}.  For SEOP \PJ{electron-scattering targets, the} two methods of determining
the density have been reported to agree within \PJ{2\% \cite{Romalis98}
and 1\% \cite{Singh15}.}
Since SEOP cells contain both \He3\ and N$_2$ gas, the pressure width and/or shift coefficients for both gases are required, although
the shift and width is typically dominated by the \He3\ gas. These coefficients were measured for the Rb $D_1$ and
$D_2$ lines with a typical accuracy of 2\% \cite{Romalis97}, and more recently for both Rb and K with a typical
accuracy for the width coefficient of \PJ{1\% \cite{Kluttz13}.} For neutron transmission, the density is replaced \PJ{with} the opacity,
which is determined from the transmission through an unpolarized cell.

\subsection {NMR}
\subsubsection{Adiabatic fast fassage}
In the technique of adiabatic fast passage (AFP) \PJ{NMR \cite{Abragam61},} an RF magnetic field is applied transverse
to the static magnetic field.  If the magnetic field is swept such that the Larmor frequency
passes through the RF frequency (or the RF frequency is swept so that it passes
through the Larmor frequency), the \He3\ polarization will invert. %
From a quantum-mechanical point of view, this process is essentially an avoided crossing due to perturbation from the RF field\PJ{~\cite{Rubbmark1981}.
The sweep rate and RF amplitude are adjusted to minimize losses resulting from polarization inversion.} 
If a pickup coil transverse to both the static and RF magnetic fields is employed,
a signal proportional to the magnetization is obtained.  %
\PJ{In order to minimize the large background from inadvertent pickup of the applied RF field, }the pickup coil
must be carefully adjusted to be orthogonal to the driving RF magnetic field.
An absolute \He3\ polarization measurement can be obtained if the response of the pickup coil and associated electronics
is calibrated using a thermally polarized water cell with exactly the same size and geometry as the \He3\ cell.  
Despite the extremely small thermal polarization of order 10$^{-8}$ in typical holding fields of a few millitesla,
these calibrations can be performed with uncertainties of a few percent.
(In principle the magnitude of the AFP signal could be determined absolutely for a given apparatus, but in 
a study of this approach discrepancies of between \PJ{20\%-50\%} were observed with cell to cell variations~\cite{Chen11}.)
Further descriptions of this technique and its application to electron scattering \PJ{targets \cite{Chupp87,Romalis_nim98}, }
polarimetry of low pressure MEOP \PJ{cells \cite{Lorenzon93},} and polarized gas \PJ{MRI\footnote{Polarean Inc., Durham, NC} have} been reported.

Whereas losses of a few tenths of a percent are typically encountered for AFP,
techniques have been developed to reduce these losses substantially.  The primary 
motivation has been for neutron spin filter cells that are not actively optically pumped on the beam line,
but in which the \He3 polarization may be \PJ{frequently} inverted during use so as to  invert the neutron polarization.
Besides the usual optimization of RF magnetic field strength and sweep rate, the RF field is modulated
by a Gaussian envelope during the \PJ{sweep \cite{Petoukhov_ILL06,McKetterick11}.}
Using this approach, losses of 10$^{-5}$ per flip have been obtained.
In a \PJ{compact RF solenoid} with shielding to confine the RF field, loss as low as \PJ{0.03\%} was reported~\cite{Ye13}.

\subsubsection{Free induction decay}
Monitoring of \He3\ polarization can also be performed by free induction decay (FID)~\cite{Bloch46},
in which an RF pulse tips the \He3\ spins and a pickup coil detects the freely precessing 
transverse component of the magnetization following the tip.  
This approach avoids the need for orthogonal drive and pickup coils.
The measurement can be non-destructive by either using small tip angles for a coil
of comparable size to the cell and/or a small coil. The decay of the signal
is given by the transverse relaxation time $T_2^*$, which is typically dominated by dephasing due to magnetic field gradients
and thus given \PJ{in high-pressure cells} by $T_2^* = (\gamma \Delta B)^{-1}$, where $\gamma/2\pi$=32.4~kHz/mT (3.24~kHz/G) is the gyromagnetic ratio
and $\Delta B$ is the variation in the magnetic field $B$ over the gas volume sampled by the coil.
Values of $T_2^*$ in typical applications range from a few milliseconds to a few hundred milliseconds.
{\Tom For the typical low pressures employed for low field MEOP, averaging of the field gradient
(motional narrowing~\cite{Pines55,Cates88a}) can increase $T_2^*$ substantially above this value.}
The use of FID is discussed in several references~\cite{Lorenzon93,Parnell08,Krimmer09,Chen11}.
Whereas FID is typically used for relative measurements, \citet{Gentile01} employed an FID system with large coils
that was calibrated against fluorescence light polarimetry in an MEOP cell and then applied
to determine absolute \He3\ polarization in compressed gas.

\subsubsection{\PJ{Radiation damping issues and control}}
\PJ{For pickup coils with a high filling factor, or dense samples with a high polarization, the nuclear magnetization can be significantly affected by radiation damping, \PJ{i.e.} the action
of the resonant RF field generated by NMR-precession-driven current in the coils.  For \He3\ gas polarized in the low Zeeman energy state,
the  lifetime of the observed FID signals is decreased whereas for the high energy state it is increased and unstable precession or maser operation can occur~\cite{Gentile01}. During AFP magnetization reversal, radiation damping affects the applied RF field and therefore modifies lineshapes and may increase losses, even if the unstable regime is avoided. Radiation damping is traditionally reduced using weakly coupled coils (at the expense of detected signal amplitudes) or applying a suitable field gradient to the sample~\cite{Romalis99,ZhengPhD}. Active feedback schemes can also be used to control radiation damping without signal-to-noise penalty~\cite{Hoult1979,Baudin2011}.}

\subsection{Neutron transmission}
\He3\ polarization measurement by neutron transmission relies on the simple equations governing a neutron spin filter~\cite{Coulter90,Jones00}.
The neutron polarization $P_{\mathrm n}$ is given by
\begin{equation}
P_{\mathrm n}^2(\lambda) = 1 - \frac{T^2_0(\lambda)}{T^2_{\mathrm n}(\lambda)}.
\label{eqn:Pn}
\end{equation}	
where $\lambda$ is the neutron wavelength and $T_0(\lambda)$ and $T_{\mathrm n}(\lambda)$ are the transmissions of unpolarized neutrons
through the unpolarized and polarized \He3\ cell, respectively.
Hence $P_{\mathrm n}$, which is typically what is of the greatest interest for spin filter applications,
can be accurately determined with only a relative transmission measurement.
Using the relationship 
\begin{equation}
P_{\mathrm n}(\lambda) = \tanh\left[ O(\lambda) P_{\mathrm{He}}\right]
\label{eqn:Pntanh}
\end{equation}	
where $O(\lambda)$ is the opacity, the \He3\ polarization $P_{\mathrm{He}}$ can be determined.  The opacity is determined from $T_0=T_{\mathrm{e}} \exp(-O(\lambda))$,
where $T_{\mathrm{e}}$ is the transmission of the cell without \He3\ gas.   
Hence to determine $P_{\mathrm{He}}$ from $P_{\mathrm n}$ requires an absolute measurement of $T_0(\lambda)$ and determination of $T_{\mathrm {e}}$.
The opacity is given by the product of the neutron absorption cross section (which typically increases linearly with neutron wavelength),
the \He3\ \PJ{density, and} the \He3\ path length.  Because of the dependence on neutron wavelength, neutron
transmission measurements are typically performed with monochromatic beam lines at reactors
or with the use of time-of-flight (TOF) analysis at pulsed neutron sources.  Neutron transmission
has been used to calibrate \PJ{NMR}-based polarization measurements for routine \He3\ polarimetry on reactor beams
\PJ{for} a wide variety of neutron spin filter cells with typical accuracy of a few percent~\cite{Chen11}. 
$T_{\mathrm{e}}$ can be determined by neutron transmission measurements before the cell is filled,
but for routine applications is often determined from measured values for glass transmission (eg. see \cite{Chupp07})
and estimated glass thickness; a typical value and uncertainty is $T_{\mathrm{e}}=0.88$ and 0.02, respectively.
With TOF \PJ{analysis, $T_0$} vs. wavelength can be fit, allowing extraction of the opacity and $T_{\mathrm{e}}$ for a filled cell.
Using this approach, determinations of $P_{\mathrm{He}}$ at pulsed sources with uncertainties of \PJ{2\%} have been reported~\cite{Chupp07,Tong12}.
\subsection{Detection by magnetometry}
Accurate magnetometry has also been applied to measure \He3\ polarization in a  5 bar MEOP-based target
for electron scattering with accuracy of \PJ{3\%}~\cite{Krimmer09}.  
This approach relies on the known dipolar field for a spherical sample of polarized gas.
Since the field due to the gas is only $\approx$1 mG or \PJ{0.02\%} of the holding field, the change
in field strength is determined by a fluxgate magnetometer upon inverting the \He3\ polarization with low-loss AFP~\cite{Wilms97}.
In addition, a spherical phantom with equidistant current loops was employed to calibrate the
exact location of the magnetic field sensor. 

\section{Charged particle and photon scattering targets}
\label{targets}

\begin{figure}[htbp]
\begin{center}
\includegraphics[width=9cm]{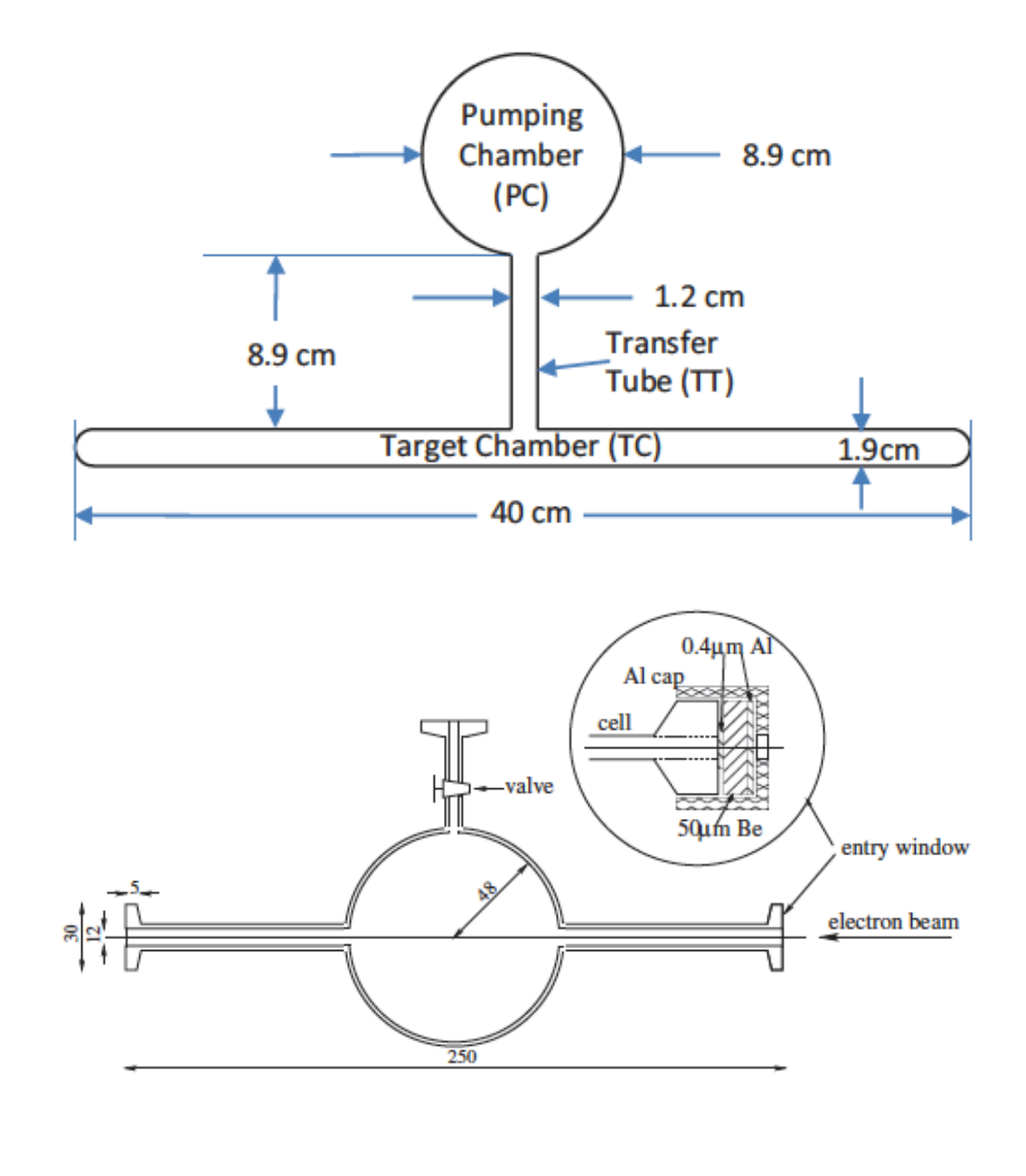}
\caption{Diagrams of current electron beam target designs. Top: JLAB, from  \citet{Dolph11}. Bottom:  Mainz, from \citet{Krimmer09}. \Tom{Dimensions are in mm.}}
\label{fig:escat}
\end{center}
\end{figure}

\Thad{Polarized \He3\ provides a reasonable
approximation to a polarized
neutron target because the proton spins are primarily paired off. Hence
about 87\% of the \He3 spin is carried by the lone neutron \cite{Friar90,Laskaris_prl13,Laskaris15}. } It is this property of the \He3\ nucleus
that yields its utility for studies of the spatial distribution of charge, magnetization, \cite{Bernauer12} and spin in the neutron.
\He3\ targets for nuclear and particle physics with charged particle and photon beams fall into four groups: 
1)  continuously polarized SEOP external targets for electron \cite{Singh15}, photon \cite{Ye10},
proton \cite{Katabuchi05,Hausser95}, and pion \cite{Larson_prl91} scattering, 
2)  MEOP external targets that are polarized remotely and transported to electron \cite{Krimmer09} and photon beam lines \cite{Krimmer11},
3)  MEOP internal targets for electron storage rings \cite{DeSchepper98_inttarget}, and
4)  continuously polarized  MEOP external targets for electron scattering \cite{Eckert92,Jones93}.
In addition, an MEOP-based polarized \He3\ ion source 
is under development for RHIC\footnote{Relativistic Heavy Ion Collider, Brookhaven, NY} \cite{Maxwell_ppn14}.
Fig. \ref{fig:escat} shows two current target designs.

\subsection{SEOP}
\subsubsection{TRIUMF, Bates LINAC, SLAC}

Early SEOP targets were constructed from Corning 1720 glass cells between 10 cm$^3$ and 35 cm$^3$ in volume, filled to pressures of several bar,
and optically pumped with several watts of laser light from Ti-Sapphire lasers.  Typical
\He3 \ polarization of 50~\% was obtained for pion~ \cite{Larson_prl91} and proton~ \cite{Hausser95} scattering at TRIUMF\footnote{Canada's national laboratory for particle and nuclear physics and accelerator-based science, Vancouver BC},
and production of polarized muonic helium at LAMPF\footnote{Los Alamos Clinton P. Anderson Meson Physics Facility,  Los Alamos, NM}~ \cite{Newbury_prl91}.  
Application to electron scattering followed soon thereafter but required the use of double cells~ \cite{Chupp87}
due to the strong ionization produced by electron beams with energies of several hundred MeV and several $\mu$A beam currents.  
In the double cell configuration, diffusion links the polarization produced in an optical pumping cell to the target cell
via a transfer tube.   The first double cell targets~ \cite{Chupp92} were employed for quasielastic scattering 
of polarized electrons at the Bates Linear Accelerator Center, which
provided some of the first double-polarization results for determination of the neutron's electric form factor~ \cite{Thompson92}.
Using a few watts of laser light from a Ti-sapphire laser, an $\approx$50 cm$^3$ cell was polarized up to 40~\% in tests and 
was maintained between 10~\% and 27~\% during the experiment.
Double cell targets were then applied for deep inelastic scattering experiments~ \cite{Johnson95}
with $\approx$25 GeV energy electrons
at SLAC for the first studies of the neutron's spin structure functions~ \cite{Anthony93,Anthony96}.
In this experiment, \He3\ polarization between 30~\% and 40~\% was maintained in a  9 bar, 200 cm$^3$ double cell 
with 20~W of laser light from five Ti-sapphire lasers, each pumped with a 20W argon-ion laser \cite{Johnson95}.
\Thad{With improved polarimetry, further studies of spin structure functions were performed with this target.} \cite{Abe_prl97,Abe_pl97}.

\subsubsection{JLAB and TUNL targets}
Double cell targets based on the design employed at SLAC were developed for electron scattering experiments at Jefferson Laboratory
with typical energies of $\approx$6~GeV, and have steadily improved in performance and capability.  
\Thad{As shown in Fig.~\ref{fig:escat}, these targets typically have an optical pumping cell between 6 cm and 9 cm in diameter, 
a $\sim$2 cm diameter target cell between 25 cm and 40 cm long, 
end windows $\sim$120$\mu$m thick, densities of 7-12 amg, 
and relaxation times of 20 h to 70 h}~ \cite{Alcorn04,Singh15}.  They have been constructed
from Corning 1720 \Thad{and, more recently, GE180 glass.}
Optical pumping is typically performed with $\approx$90~W of laser light from three { fiber-coupled diode laser systems}.
The typical \He3\ polarization achieved was 30\% to 40\% in the first experiments
on electromagnetic form factors~ \cite{Xu00,Xiong01} and neutron spin structure functions~ \cite{Amarian04,Amarian02,Zheng_prl04,Zheng04}.
With the use of hybrid SEOP, the \He3\ polarization was increased to $\sim$45\% for a more recent measurement
of the neutron's electric form factor~ \cite{Riordan10}.  For hybrid SEOP with spectrally narrowed lasers, the polarization has been increased
to between 50~\% and 55~\% for studies of \Thad{asymmetries in scattering from transversely  or vertically polarized
targets generally aimed 
at improved understanding of the origin} of the neutron's spin~ \cite{Qian11,Huang12,Allada14,Katich14,Zhang14,Zhao15,Parno15}.
In a detailed study of these targets, a maximum off-line value of 70~\% and a detailed analysis of achievable polarization was reported~ \cite{Singh15}.
Future experiments with even higher luminosity are expected to employ a new type of double cell in which 
exchange of gas between the optical pumping and target cells is increased by convection~ \cite{Dolph11}.
In this new design, there are two transfer tubes and gas circulation is produced by heating one tube.\\
\indent  Recently, two different SEOP targets have also been employed at TUNL\footnote{Triangle Universities Nuclear Laboratory, Durham, NC}
A target was developed for measurements of spin-correlation coefficients in $p$ + \He3\ elastic scattering~ \cite{Daniels10,Katabuchi05}, in which
the \He3\ was polarized by SEOP at 8 bar and transferred through a plastic tube to spherical Pyrex cells
with Kapton film covering the apertures for the beam and scattered particles.
For application to gamma-ray beams, a sol-gel coated Pyrex double cell was employed to avoid background {signals scattered off the barium present} in GE180 glass~ \cite{Ye10}.
The relaxation time in the 7 bar target was 35 h and a maximum off-line \He3\ polarization of 62~\% was obtained.
In experiments to study three-body physics, $\approx$~40~\% \He3  polarization was maintained~ \cite{Laskaris_prc14,Laskaris15}.

\subsection{MEOP}
\subsubsection{Early targets, Bates LINAC, internal targets}
The high efficiency of MEOP allowed for early targets based on optical pumping with lamps~ \cite{Phillips62,Baker69}.
The first laser-pumped targets were based on a double cell approach with a cryogenic target
cell to increase the density~ \cite{Milner89,Alley93}.  Such a target operated at a gas pressure of 2.6 mbar
and a target cell temperature of 17 K was employed for quasielastic scattering of polarized
electrons at the Bates Linear Accelerator Center, which provided some of the first double-polarization results
to determine the neutron's electric form factor~ \cite{Woodward90,Jones-Woodward91,Jones93}.
In this experiment, between 20~\% and 30~\% \He3 \ polarization was maintained using 0.3 W of laser light. 
The 10 cm long copper target cell was coated with frozen nitrogen to permit target cell relaxation times 
between 400 s and 1000 s, depending on beam current.  The same target design was employed
in  \Thad{subsequent experiments} to determine the neutron's magnetic form factor~ \cite{Gao94}
and \Thad{to  further  study quasielastic scattering} \cite{Hansen95}.  Employing optical pumping with a few watts
of laser light from a 
Nd:LMA laser~\cite{Gentile93}
allowed for over 38~\% \He3\ polarization at 2.5 times higher average electron beam current~ \cite{Gao94}.\\
\indent Concurrently with these external targets, MEOP was applied to internal targets for storage rings.
In these targets, gas flows from an optical pumping cell through a capillary to a open target cell.
The first polarized \He3\ internal  target~ \cite{Lee93} was employed at the Indiana University Cyclotron Facility
for measurements of quasielastic scattering of polarized protons from polarized \He3\ 
to study the ground state spin structure of the \He3\ nucleus~ \cite{Miller95,Bloch95,Lee93}.
A Nd:LMA laser was used for optical pumping and the average \He3\  polarization in the target
was 46~\% at a \He3\  {flow rate of $1.2\times10^{17}$ atoms/s.}
Soon thereafter, an internal target with a cryogenic target cell~ \cite{Kramer_int95,Kramer95,Korsch97,DeSchepper98_inttarget} was employed 
for spin-dependent deep inelastic scattering of 27.5 GeV polarized   positrons at DESY\footnote{Deutsches Elektronen Synchrotron, Hamburg, Germany}\cite{Ackerstaff97,DeSchepper98_expt}.  For an ultrapure aluminum target cell cooled to 25~K,
the \He3\ polarization was 46~\% during the experiments.  This target also used a variation of the fluorescence
polarimetry method:  rather than a discharge, high energy positrons provided the required atomic excitation
and the \He3\ polarization was determined from the circular polarization of 492 nm light from the 4 $^1$D - 2 $^1$P transition.
\Tom{An internal target at NIKHEF \footnote{ Dutch National Institute for Subatomic Physics, Amsterdam, Netherlands} operating at a nominal atomic flow rate of 1~$\times$10$^{17}$ s$^{-1}$ yielded a nuclear polarization of 0.50 for a target thickness of 0.7~$\times$10$^{15}$ cm$^{-2}$ at a target temperature of 17~K~\cite{Poolman00}.}

\subsubsection{{MAMI}}
Early \He3\ targets for electron scattering at MAMI 
employed a Toepler pump~ \cite{Eckert92},
in which a mercury column compressed gas polarized by MEOP at a pressure of $\approx$~1~mbar 
into a 100 cm$^3$ target cell at $\approx$~1~bar.  The polarized gas was continuously recirculated from the target 
cell to the optical pumping cell for repolarization {at a flow rate of 10$^{18}$ \Thad{atoms/s}. }
Optical pumping with a few watts of laser light from a Nd:LMA laser~ \cite{Eckert92}
yielded an average \He3\ polarization of 38~\% in the first MAMI experiment to determine the
neutron's electric form factor~ \cite{Meyerhoff94}.
In a later experiment on the same topic~ \cite{Becker99}, several improvements, including a cesium coated
cell with a relaxation time of 6 h, yielded 50~\% \He3\ polarization.   The continuous flow Toepler pump apparatus was replaced
by remotely polarized cells filled using a piston compression apparatus~ \cite{Becker94,Surkau97} to a pressure of 4 bar.
This approach was employed for further measurements of the neutron's electric form factor~ \cite{Bermuth03}.
More recently, an improved piston compression apparatus yielded 72~\% \He3\ polarization in a 25 cm long,
cesium-coated, quartz, 5 bar target cell with beryllium and aluminum foil windows~ \cite{Krimmer09}, see Fig. \ref{fig:escat}.  
Relaxation induced by the electron beam (see Sec. \ref{IonRad}) yielded typical beam-on relaxation times between 30 h and 40 h,
hence the cells were replaced twice a day. A time-averaged \He3\ polarization of 56~\%
 was maintained over the course of a three week measurement
of the neutron electric to magnetic form factor ratio~ \cite{Schlimme13}. 
Similar targets~ \cite{Krimmer11} have also been applied for the tagged photon beam facility at MAMI~ \cite{Costanza14}.


\subsection{{Brief summary of physics enabled}}
The complex and diverse subatomic physics studied with \He3\ targets is clearly beyond the scope of this review.
Nevertheless, we briefly summarize the overall topics of the majority of experiments. 
%
\indent The charge and magnetization distributions are typically characterized by the electric and magnetic form factors of the neutron,
$G_{\mathrm E}^{\mathrm n}(Q^2)$ and $G_{\mathrm M}^{\mathrm n}(Q^2)$, respectively, 
which nonrelativistically can be considered to be the Fourier transforms of the charge and magnetization distributions.
The first experiments with polarized \He3\ focused on measuring these form factors and on
testing the three-body nuclear physics calculations needed to quantitatively validate the approximation of
a polarized \He3\ target as a neutron target.  Since the neutron is neutral overall, measuring the 
electric form factor is particularly difficult.  In addition, measurements must be performed over a wide range
of momentum transfer to  test theoretical predictions.  In the last 25 years, these measurements have greatly
improved the knowledge of these form factors, yielding more precise tests of nucleon models~ \cite{Qattan12}.
Whereas the uncertainty for the earliest measurements
were larger than the values themselves and were performed at only a single value of relatively low momentum
transfer, recent measurements have fractional uncertainties of 15~\% or better
and cover a wide range of momentum transfer~ \cite{Gentile11}.  Although $G_{\mathrm E}^{\mathrm n}(Q^2)$ is generally
the more difficult and desired measurement, $G_{\mathrm M}^{\mathrm n}(Q^2)$ is also important
because typically $G_E^n$ is actually determined from a measurement of the ratio
$G_{\mathrm E}^{\mathrm n}(Q^2)/G_{\mathrm M}^{\mathrm n}(Q^2)$ combined with
separate measurements of $G_{\mathrm M}^{\mathrm n}(Q^2)$.
%


\Thad{Analogous to the electric and magnetic form factors are the more complex spin structure functions that are related to
the distribution of angular momentum} in the neutron~ \cite{Aidala13}.  These studies have been primarily motivated by
understanding the origin of the spin of the neutron.
Improvements in the polarized \He3\ targets have allowed for a substantial improvement in the precision
of tests of the fundamental sum rules for spin structure from quantum chromodynamics.
Decades ago it was determined that contrary to expectations, the intrinsic quark spins contribute
only a small fraction of the nucleon spin.  Despite many years of effort, studies of the origin of the spin of the nucleon are still incomplete.
Continuing studies with polarized \He3\ targets have focused on the possible contribution to the
nucleon spin from the orbital angular momentum of the quarks.\\

\section{Neutron Spin-Filters}
\label{NSFs}



{Neutron spin-filters (NSFs, \citet{JCNSworkshop}) produce highly spin-polarized beams of low energy neutrons, or analyze the neutron spin state,  by passing the neutrons through a glass cell of polarized \He3 in a uniform magnetic field.  The near 100\% contrast in the  
spin dependence of the \He3 neutron absorption cross section  \cite{Coulter90} results in highly polarized neutron beams or very efficient spin analyzers. } Although other devices such as supermirrors  \cite{Mezei76}
and Heusler alloy   \cite{Freund83} are employed for polarizing neutron beams  \cite{Williams88},
NSFs \Thad{ are advantageous for large area and large divergence beams, can be used for cold, thermal, and epithermal
neutrons,  and decouple polarization selection from wavelength selection. }
For both MEOP and SEOP, NSF cells are typically polarized off-line, transported to the neutron  beam line,
and stored in various magnetostatic cavities. In addition, there has been increasing use of SEOP-based in-situ NSF systems.

\subsection{Principles} \label{NSF_prince}
\begin{figure}[htbp]
\begin{center}
\includegraphics[width=8cm]{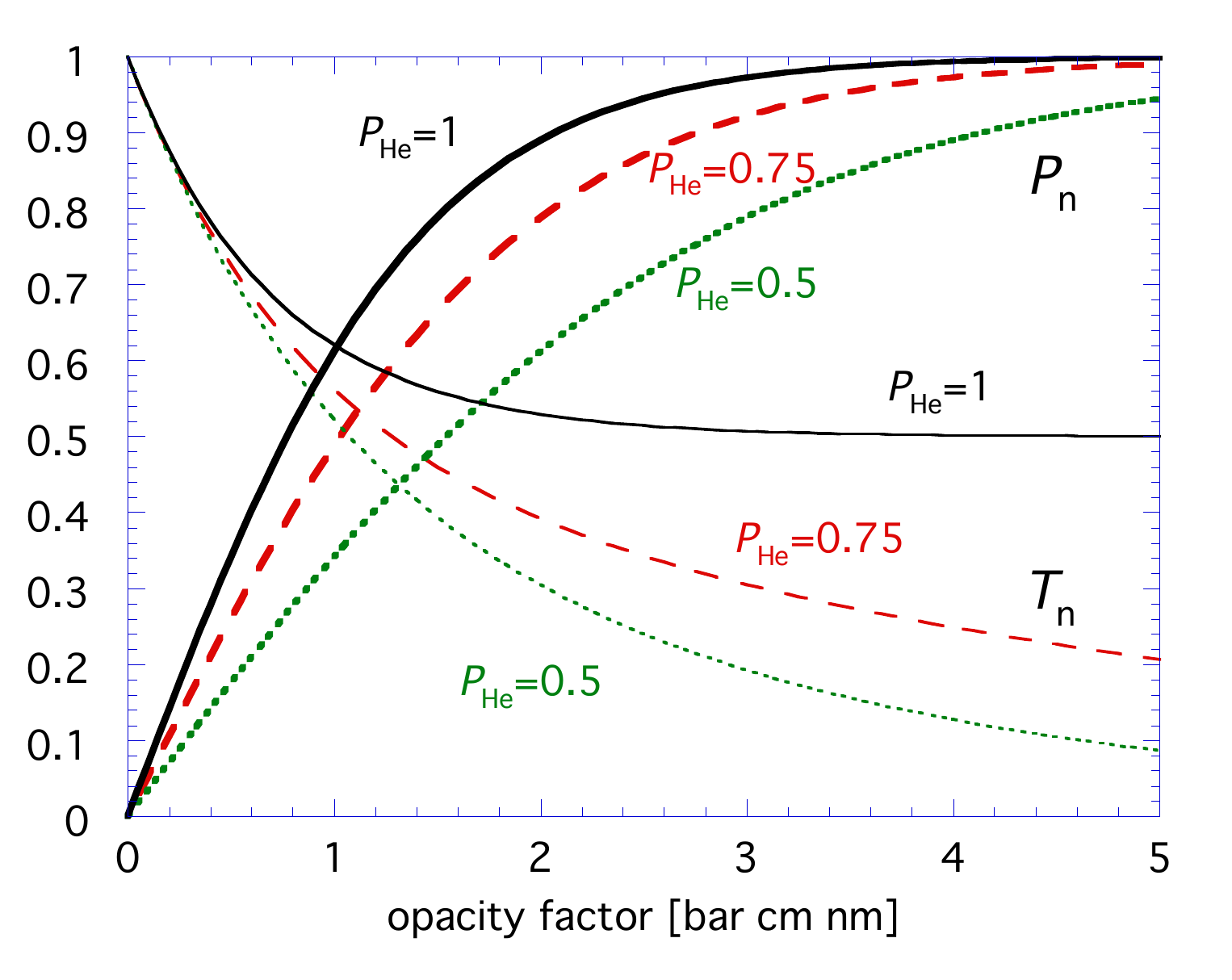}
\caption{The variation of the neutron polarization $P_{\mathrm n}$ and transmission $T_{\mathrm n}$ with 
the opacity factor, where the opacity factor is given by the pressure-length-wavelength product in bar cm nm,
for \He3 polarizations ($P_\mathrm{He}$) of 0.5, 0.75 and 1. The neutron transmissions
shown do not include the transmission of 0.88 due to neutron scattering from a typical glass NSF cell.}
\label{fig:PnTn}
\end{center}
\end{figure}

\begin{figure}[htbp]
\begin{center}
\includegraphics[width=8.5cm]{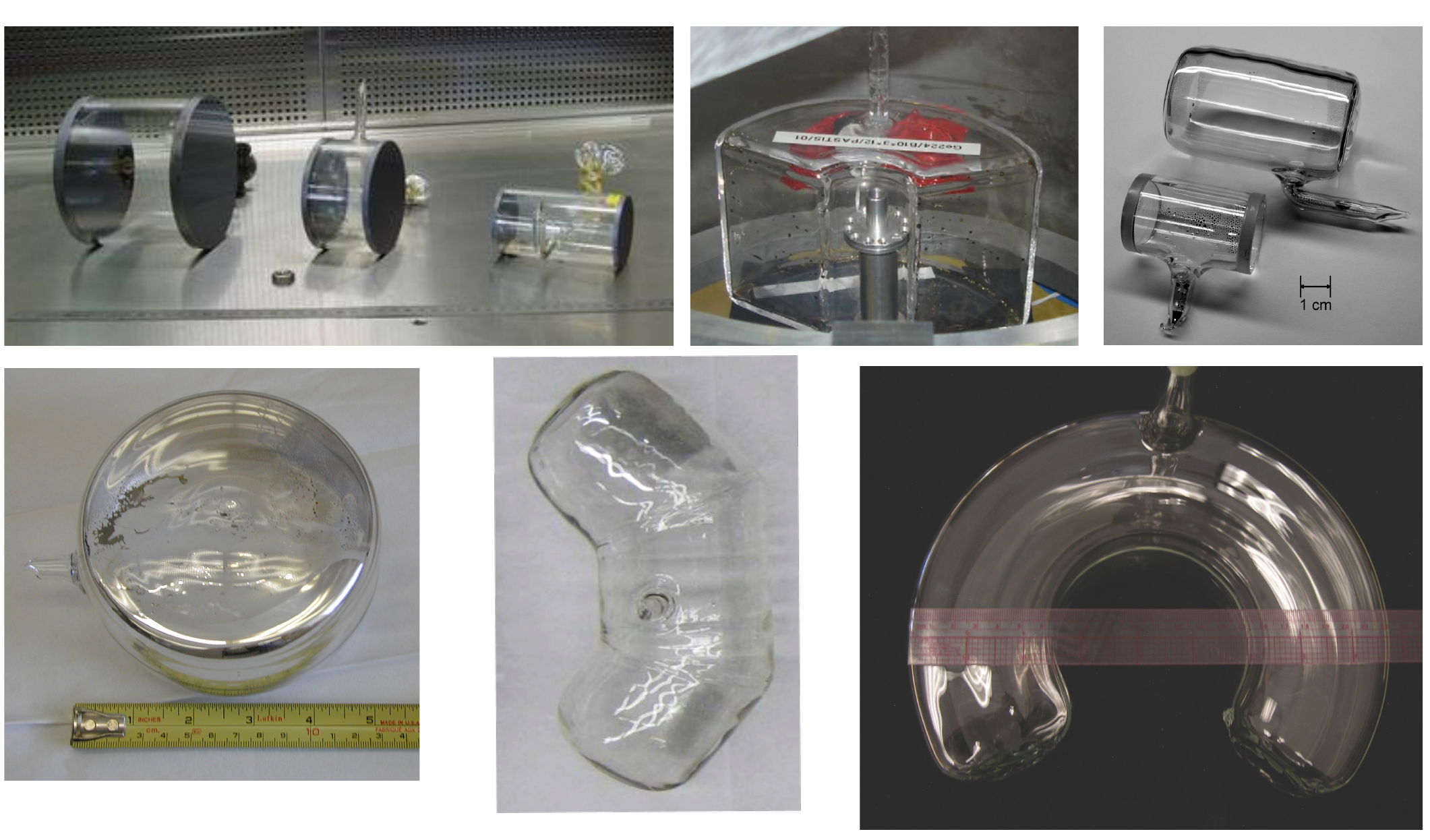}
\caption{\Tom{Neutron spin filter (NSF) cells. Clockwise from top left: MEOP silicon-windowed cells (largest cell is 14 cm diameter by 10 cm long, \citet{Lelievre07}),
 MEOP wide-angle cell (6 cm inner diameter, 20 cm outer diameter, 12 cm tall, \citet{Andersen09}), cells for neutron interferometry (larger cell is 4 cm diameter by 6 cm long, \citet{Huber14}),
horseshoe SEOP cell~ (9 cm inner diameter, 23 cm outer diameter, 7.5 cm tall, \citet{Chen}), SEOP wide-angle cell (14 cm inner diameter, 30 cm outer diameter, 8 cm tall, \citet{Ye13}), typical SEOP NSF cell (12 cm diameter by 7 cm long, \citet{Chen11}).}}
\label{fig:NSFcells}
\end{center}
\end{figure}

The large cross section for neutron absorption by \He3\ arises from a broad, unbound, resonance that yields a proton and a triton. 
The strong spin dependence arises because absorption only occurs if the neutron and \He3\ spins have their spins antiparallel.
Whereas the cross section for capture of 25 meV neutrons with spin antiparallel to the \He3\ nuclear spin is 10666 barns,
the ratio of the absorption for neutrons with parallel and antiparallel spin has been determined experimentally to be less than
a few percent and estimated theoretically to be less than 0.5~\%  \cite{Huber14}, and the scattering cross section is only a few barns  \cite{Mughabghab81}.
Hence for a sufficient opacity (product of gas density, cross section, and cell length) of 100~\% polarized \He3, all neutrons
with antiparallel spin would be absorbed,
while nearly all neutrons with parallel spin would be transmitted, resulting in 100~\% neutron polarization and 50~\% transmission. 
Although imperfect \He3\ polarization reduces the achievable neutron polarization and transmission for a given 
opacity, increasing the opacity allows any neutron polarization to be achieved at the expense of transmission.
The absorption cross section is directly proportional to wavelength, hence higher energy (shorter wavelength) neutrons
require greater opacities.  Typical room temperature pressure-length products vary between 4 bar-cm and 25 bar-cm for 
neutrons with energies between 2 meV and 80 meV (wavelengths between 0.6 nm and 0.1 nm).
For a \He3\ polarization of 75~\%, 90~\% neutron polarization with 28~\% neutron transmission can be obtained. \Thad{ These values include the transmission of 0.88 due to neutron scattering from a typical glass NSF cell.}
The basic equations governing the relationships between opacity, \He3\ polarization, and neutron polarization and transmission
appear in many publications  \cite{Coulter90,Jones00}.  These relationships allow both the neutron and \He3\
polarization to be determined by neutron transmission measurements.  NSFs are also used to analyze neutron polarization;
in this case an analyzing power of 90~\% with a transmission for the desired spin state of \Thad{54~\%} can be obtained.
Fig.~\ref{fig:PnTn} shows the variation of the neutron polarization and transmission with the pressure-length-wavelength
product for \He3 polarizations of 0.5, 0.75 and 1. The ideal opacity factor depends on the type of experiment and its
optimization has been addressed with different approaches  \cite{Tasset95,Williams99,Goossens_nim02,Gentile05};
a typical value is 3 bar-cm-nm. \\
\\
\indent  The strong neutron absorption cross section for boron restricts the choice of materials for NSFs.
For SEOP, GE180 is generally employed due to issues with \He3 permeation and temperature-dependent relaxation for quartz  \cite{Ino05,Ino07,Ye13}.  
Sapphire  \cite{Masuda05,Chen11} and silicon-windowed Pyrex cells  \cite{Chen11} have also been investigated for SEOP, but use has been limited due to
relaxation issues.  For MEOP cells, which do not need to be heated, both quartz and silicon-windowed Pyrex cells are routinely used \cite{Hutanu05,Lelievre07}.
Neutron transmission and scattering from glasses are important practical issues
for NSFs and have been studied by various groups  \cite{Sakaguchi_physicab11, Chupp07,Chen04,Babcock14}.
A variety of MEOP and SEOP cells are shown in Fig.~\ref{fig:NSFcells} and discussed below.\\
\indent  As discussed in Sec. \ref{sec:magcav}, various magnetostatic cavities are employed to provide the highly homogeneous magnetic field
required to maintain the \He3\ polarization.   Polarized neutron beam lines typically have magnetic fields to maintain
the neutron polarization from the polarizer to the analyzer (\textquotedblleft guide fields\textquotedblright) and fields for various types of neutron spin flippers. 
Often a sample under study may be immersed in a strong magnetic field provided by electromagnets
or superconducting magnets, hence protecting the NSF from field gradients is an important practical
consideration.  In addition, maintaining the polarization of the neutron beam during transitions between the 
homogeneous magnetic field for the \He3\ to the sample and/or neutron spin flipper fields can be difficult,
in particular on space-constrained apparatus not originally designed for NSFs.
In some cases AFP NMR is used to flip the \He3\ polarization (and thus the neutron polarization
or analyzer direction) so as to avoid the need for traditional neutron spin flippers.
For some applications, the polarizer NSF, sample, and analyzer NSF are all immersed in the same
field, thus bypassing neutron spin transport issues but also precluding the use of strong sample fields  \cite{Ye13}.
\\
\subsection{Neutron scattering implementation: SEOP}
\label{NSF_SEOP}
The first test of an NSF based on SEOP was performed at LANSCE\footnote{Los Alamos Neutron Scattering Center, Los Alamos, NM}  \cite{Coulter90}
and applied to the search for parity violation in neutron resonances.
A \He3 \ polarization of 70~\% was obtained in a  3 cm$^3$ cell optically pumped with 1~W of laser
light with a line width of 40 GHz that was provided by a standing-wave dye laser pumped by a krypton ion laser.
A neutron polarization of 20~\% was produced in a beam of epithermal neutrons (energy of 0.734 eV, 0.03 nm wavelength).
Soon thereafter an SEOP-based NSF was tested at the ILL  \cite{Tasset92}, but the 
ILL program shifted to MEOP (see Sec.~\ref{NSF_MEOP}). \\
\indent For neutron scattering, an ongoing program in SEOP-based NSFs
was begun at the NCNR\footnote{National Institute for Standards and Technology Center for Neutron Research, Gaithersburg, MD}   \cite{Jones00}, first demonstrated for small angle neutron scattering (SANS)  \cite{Gentile_sans00},
and subsequently applied to polarized neutron reflectometry (PNR)   \cite{Chen04}, triple-axis spectrometry (TAS)  \cite{Chen07}, and wide-angle
neutron polarization analysis  \cite{Ye13}. 
For these applications, SEOP cells are polarized off-line, transported to the neutron beam line, and stored in magnetostatic cavities.  
Typical NSF cells for SANS, PNR and TAS are blown from GE180 glass, between 8 cm and 12 cm diameter,
and filled to pressures between one and two bar.   Cells for wide-angle analysis have been formed by optically sealing three sections
of blown glass together so as to cover a 110 degree angular range  \cite{Ye13}.
More recently, fully blown cells for wide-angle analysis have been fabricated by 
first blowing a toroidal GE180 cell and then closing off a 270 degree section from the toroid  \cite{Chen16}.
The current state of the NCNR program has been summarized recently  \cite{Chen14}.\\
\indent The relatively small size of SEOP apparatus is well-matched to in-situ operation on neutron beam lines~ \cite{Jones06}.
An in-situ NSF has been operated routinely on the Magnetism Reflectometer at the SNS \footnote{Spallation Neutron Source, Oak Ridge, TN}  \cite{Tong12}
and smaller in-situ systems have been developed for the HB3 triple-axis spectrometer and the hybrid spectrometer (HYSPEC)
at the \Thad{SNS} \cite{Jiang14}.  At LENS\footnote{Low Energy Neutron Source, Bloomington, IN}, 
a continuously pumped analyzer has been demonstrated
for spin echo small angle neutron scattering (SESANS)~ \cite{Parnell15}.  Compact NSFs with low loss AFP NMR for on-beam operation
have been developed at  KEK\footnote{National Laboratory for High Energy Physics,  Tsukuba, Japan}  \cite{Ino12,Ino05}.
More recently, there has been spin filter development at the Materials and Life Science
Facility at J-PARC\footnote{Japan Proton Accelerator Research Complex, Tokai, Japan}
and at the High-flux Advanced Neutron Application Reactor at KAERI\footnote{Korean Atomic Energy Research Institute, Daejeon, Korea}~ \cite{Lee16}.
At J-PARC, in-situ SEOP neutron spin filters have been demonstrated
for SANS  \cite{Sakaguchi11}, PNR, and polarized neutron imaging~ \cite{Hayashida16}.

\indent The SEOP programs at the {Forschungzentrum J\"ulich  \cite{Babcock16} and the ISIS spallation source\footnote{Rutherford Appleton Laboratory, Oxfordshire, United Kingdom}
  \cite{Beecham11}} have also focused on in-situ systems for continuously polarized \He3.
An apparatus was developed in which a \He3\ polarization of 80~\% was maintained in a 100 cm$^3$
cell by SEOP with two 100 W diode lasers, each spectrally narrowed to 0.2 nm bandwidth with an external cavity  \cite{Babcock11hipol}.
{It has been applied to PNR and SANS}~ \cite{Babcock16}.
Cells for wide-angle polarization analysis  \cite{Salhi16} fabricated from sections
of fully blown doughnut-shaped GE180 cells  \cite{Salhi14} are under development.

\subsection{Neutron scattering implementation:  MEOP}
\label{NSF_MEOP}
Due to the large size of most compression apparatus, NSFs based on MEOP are typically polarized off-line
and transported to neutron beam lines.  The first demonstration of an MEOP-based NSF
was performed at the Univ. of Mainz with polarized \He3 \ from
a piston compression system  \cite{Surkau97}.  A similar compressor was installed at the
ILL, with the capability to produce between 50~\% and 55~\% \He3 \ polarization for a variety
of cell designs  \cite{Heil99}. The first application was on the IN20 triple-axis spectrometer  \cite{Kulda98}.
With the development of the superconducting magnetostatic cavity "Cryopol"  \cite{Dreyer00},
an NSF was applied for spherical neutron polarimetry (SNP) with hot (0.08 nm wavelength) neutrons on the D3 instrument  \cite{LelievreBerna05}.
Large solid-angle polarization analysis at thermal neutron wavelengths was demonstrated on the  two-axis
diffractometer D1B using a cesium-coated quartz cell that analyzed a 90 degree range
of scattering angles  \cite{Heil02}.  The 530 cm$^3$ banana-shaped cell exhibited a relaxation time of 92 h.
With improved optical pumping using a 30 W ytterbium fiber laser
instead of an 8 W Nd:LMA laser, a spin filter with a \He3 \ polarization of up to 70~\% was demonstrated
at the Mainz reactor  \cite{Batz05}.  Concurrently a new compressor, denoted "Tyrex", was 
constructed at the ILL and employed to fill spin filter cells with 65~\%~-~70~\% polarized \He3 \ gas
at rates of 1 bar-L/h to 2 bar-L/h~  \cite{Petoukhov_ILL06}.   An NSF employing a 14 cm diameter
silicon-windowed, valved, spin filter cell was demonstrated on the D17 reflectometer,
along with a "local filling" approach for refreshing the \He3 \ daily  \cite{Andersen06}.
With further improvements, a typical value of 75~\% \He3 \ polarization has been obtained in NSF cells,
enabling a large range of neutron scattering applications
  \cite{Stewart06,Lelievre07,Andersen09,Lelievre10}.
A similar compression system, denoted HELIOS,  was put into operation at the  FRM II reactor\footnote{Forschungs-Neutronquelle Heinz Maier-Leibnitz,  Munich, Germany}  \cite{Hutanu_nn07}
and has been applied to SNP at the hot neutrons single crystal diffractometer POLI-HEiDI  \cite{Hutanu11}.
More recently, two compression apparatus have been bult at the ILL for use
at the ISIS pulsed neutron and muon source  \citep{Beecham11}
and the Australian Nuclear Science and Technology Organization  \cite{Lee16b}.

\subsection{Brief summary of neutron scattering physics enabled}
The primary application of NSFs in neutron scattering is in the study of magnetic materials.
Neutron polarization analysis allows for separation of nuclear from magnetic scattering  \cite{Moon69,Williams88}.
SNP and TAS have been employed to study magnetic ordering in antiferromagnetic crystals~  \cite{Hiess01a,Hiess01,Zaharko06,Blanco06,Poole07},
magnetization distributions~  \cite{Brown02,Boehm03}, magnetoelectric and multiferroic crystals and films  \cite{Brown05,Tian08,Lee08,Cabrera09,Ratcliff11},
and the interplay between magnetism and superconductivity~  \cite{Schneider06,ChenY08,Wang11}.
The ability of NSFs to analyze the polarization of diffusely reflected neutrons has been applied
to magnetic multilayers  \cite{Nickel01}, domain walls in magnetic thin films~  \cite{Radu03,Radu05}, periodic magnetic rings~  \cite{Ogrin07},
and superlattices  \cite{Wildes08}.  For SANS, NSFs allow for polarization analysis 
which has been implemented via SEOP at the NCNR  \cite{Gentile_sans00,Krycka09}, and via MEOP at the Hahn-Meitner Institute in Berlin, Germany  \cite{Wiedenmann05,Keiderling08} and the ILL  \cite{Honecker10}.  At the NCNR, several recent studies have focused
on understanding the structure of magnetic nanoparticles  \cite{Krycka09,Krycka10prl,Krycka11,Krycka13,Krycka14,Hasz14}.
SANS with NSFs has also been employed to study multiferroics  \cite{Ueland10,Ramazanoglu11}, magnetostriction  \cite{Laver10},
exchange-bias  \cite{Dufour11}, and nanowires  \cite{Pimpinella13}.
\indent  Polarization analysis also allows for separation of coherent from spin-incoherent scattering  \cite{Moon69,Williams88},
which is relevant to biological samples due to the substantial spin-incoherent scattering from hydrogen.
Applications of NSFs to this technique are emerging~ \cite{Sakaguchi11,Babcock16}.\\
\indent  NSFs have advantages over neutron optical devices for polarized neutron imaging because they can 
provide a uniform analyzing power without optical distortions of the neutron beam.
At the ILL, polarized neutron imaging using NSFs from the Tyrex compressor was demonstrated
and applied to visualize small-scale magnetic features and trapped magnetic flux  \cite{Dawson11}.
In addition, depolarization imaging has been used to image ferromagnetic phase separation in real space  \cite{Pfleiderer10}.
\indent 
\label{NSFPhysics}
\subsection{Application to fundamental neutron physics}
\subsubsection{Accurate neutron polarimetry}
\label{neutronpolarimetry}
For a monochromatic neutron beam, the neutron polarization produced by an NSF can be
determined from the ratio of the transmission of unpolarized neutrons with the NSF polarized and unpolarized.
If the NSF is used as a polarization analyzer, the analyzing power can be determined 
by this method and then employed to accurately measure the polarization of a neutron beam.
This simple approach allows for highly accurate measurements of {neutron polarization or analyzing power  \cite{Coulter90,Jones00}}.
However, fundamental neutron physics experiments often require high neutron flux for statistics
and thus polychromatic beams.  With time of flight analysis, typically available at a pulsed neutron source, 
accurate polarimetry can be extended to polychromatic beams.  At LANSCE neutron beam polarization was
measured with an absolute accuracy of 0.3~\% in the neutron energy range from 40 meV to 10 eV~  \cite{Rich02nim}.
More recently,  \citet{Ino11} reported a measurement of the near-unity average neutron beam polarization produced by supermirror benders at  J-PARC
with an absolute accuracy of 0.03\% in the neutron wavelength range from 0.1 nm (81 meV) to 0.7 nm (1.7 meV).
At continuous sources, one can use "opaque" spin filters, i.e. NSFs with sufficient opacity
such that the analyzing power is near unity for the relevant wavelengths  \cite{Zimmer99a}.  This approach has
been applied to neutron beta-decay (see Sec.~\ref{fundapplications}).
\subsubsection{Applications}
\label{fundapplications}
The first application of NSFs to fundamental neutron physics was for measurements of the parity-violating neutron spin rotation
in the 0.734 eV resonance in lanthanum.  Measurements were performed at KEK in Japan  \cite{Sakai97,Sakai98},
LANSCE  \cite{Haseyama02}, and the ILL~  \cite{Heil_Pviolation99}. 
At LANSCE, SEOP with diode lasers was employed with 56\% (polarizer) and 29\% (analyzer) \He3 polarization
obtained in 6 bar cells.
At the ILL, the MEOP compression system was employed to obtain 50~\% \He3 \ polarization
in a 5 cm diameter, 20 cm long NSF analyzer filled to 3 bar.
More recently, a new experiment to search for time reversal violation in neutron transmission has been proposed  \cite{Bowman14}.
\\
\indent An SEOP-based, large area NSF for long term continuous operation  \cite{Chupp07} was employed for the study
of parity violation in the absorption of cold neutrons by compound nuclei  \cite{Gericke06} and hydrogen (the "NPDGamma"
experiment)   \cite{Gericke11}.  With two 30~W fiber-coupled diode lasers (between 1.5 nm  and 2 nm bandwidth), 57~\% \He3 \ polarization was obtained
in 11 cm diameter GE180 cells  \cite{Gentile05_jnist}.  The NSF was successfully operated for one year, but the \He3 \ polarization 
declined to $\approx$30~\% due to long term effects on the cell from the neutron beam (see Sec.~\ref{neutronbeameffects}.)
The neutron polarization was determined with an accuracy of 0.1~\% with time of flight analysis.\\
\indent Measurements of the electron ($A$), neutrino ($B$) and proton ($C$) asymmetries in polarized neutron
beta-decay provide accurate tests of the Standard Model \cite{Jackson57}.  These experiments require highly accurate
determinations of the neutron polarization.  For these experiments, the use of a series of opaque spin filters (see Sec.~\ref{neutronpolarimetry})
has been employed to determine neutron polarization to 0.1~\% accuracy, thus substantially reducing this contribution to
the overall uncertainty budget  \cite{Schumann07,Schumann08,Mund13}.  An NSF was also used to determine 
the neutron polarization in a measurement of the neutron electric dipole moment via spin rotation in a non-centrosymmetric crystal  \cite{Fedorov10}.

Polarized  \He3 \  has also been used to determine the spin dependence of the neutron scattering length for \He3,
which is important for nuclear few body models. Two different methods were used.   In the first method, the pseudomagnetic precession
of an unpolarized neutron beam passing through a sample of polarized \He3 was detected using a neutron
spin-echo spectrometer   \cite{Zimmer02}.   A 6 cm diameter, 10 cm long glass cell with flat, parallel, silicon
windows was filled with polarized gas using the ILL MEOP compression system. In the second method, 
the spin-dependence of the phase shift for polarized neutrons passing through a sample of polarized \He3 was determined
by neutron interferometry  \cite{Huber14}.  A 2.4 cm diameter, 4.2 cm long glass cell
with flat, parallel glass windows was polarized off-line by SEOP and installed inside a neutron interferometer. The
relaxation times observed during the experiments were between 75 h and 120 h  \cite{Zimmer02} and 
between 135 h and 150 h  \cite{Huber14}. For the interferometry experiment, the polarization
of the monochromatic neutron beam was determined with NSFs to better than 0.1\% absolute accuracy. The
two measurements differ by two standard deviations  \cite{Huber14}.

 
 \subsubsection{Neutron beam effects}
\label{neutronbeameffects}
Effects of the neutron beam on the operation of in-situ SEOP NSFs were first observed
in the NPDGamma experiment (see Sec.~\ref{fundapplications}) and soon thereafter
studied at LANSCE  \cite{Sharma08}.  Further studies  at the ILL
revealed that at a neutron particle flux density of $4.7\times10^9$ cm$^{-2}$s$^{-1}$,
the alkali-metal relaxation rate increased from 100 s$^{-1}$ to 1000 s$^{-1}$  \cite{Babcock09}. 
The relaxation rate was found to vary with the square root of the neutron flux, consistent
with the source being the ionization produced by the energetic triton and proton produced
in neutron absorption by \He3.  The alkali-metal relaxation rate was found to increase rapidly (time scale $<$1 s),
followed by a further slow rise on a time scale of hundreds of seconds.  Studies of
the cell whitening that had been previously reported after long term exposure to the neutron beam  \cite{Chupp07} indicated
that this problem may be reduced in K-Rb hybrid cells.  The origin of this whitening
is unknown, but speculated to be due to the production of RbH and/or alkali azides. 
It was demonstrated that a double cell configuration, an approach already long employed for electron scattering, 
is a practical method to bypass these neutron beam effects.  In a later study \citep{Babcock_jpcs11}, it was observed
that the slow component of alkali-metal relaxation increases with increasing nitrogen  density
in the SEOP cell. 

\section{Magnetic Resonance Imaging (MRI)}
\label{MRI}


\subsection{Introduction and history}
The application to MRI \Tom{\cite{Couch15,Leawoods01,Moller02} }arguably provides the most visible and salient connection of hyperpolarized noble gases to commercial (in this case, medical) technologies.  The two noble gases with nuclear spin 1/2, \He3\ and $^{129}$Xe, yield relatively long relaxation times due to absence of a nuclear quadrupole moment that can interact with electric fields at surfaces.  Conventional $^1$H MRI of lung tissues is very challenging, mainly due to the very short (microseconds) NMR-signal lifetimes \cite{Cutillo91} resulting from the large magnetic susceptibility broadening in the lung microstructures (alveolar spaces and connecting bronchioles). Rapidly moving gases average away this broadening, but their low density relative to protons in water introduces a severe sensitivity problem. A hyperpolarized inert gas is potentially the ideal signal source for investigating lung function: it serves as a tracer of gas-flow in an organ whose principal function is to move gas, it is not metabolized and interacts minimally with the body, and the enhanced magnetization overcomes the intrinsic sensitivity problem \cite{Kastler50,Cagnac58}.

In the early 1980's, even as MRI itself was still being developed for clinical use, \citet{Bhaskar82} noted that about 11 bar$\cdot$cm$^3$/h of highly polarized $^{129}$Xe might be produced with 1~W of laser power, although at the time, attention was focused on the ability this would afford to produce dense polarized targets. Some combination of the biological relevance of xenon (an anesthetic, which dissolves in lipid tissue), relatively fast spin-exchange rates \citep{Zeng85}, and the ability to freeze and transport xenon with minimal polarization loss \citep{Cates90} led to the first rudimentary MR images of excised mouse lungs \citep{Albert94}. However, the tipping point for the technology was not so much the marriage itself of hyperpolarized noble gases to MRI, as it was two major improvements in related technology. For SEOP, the advent in the early 1990's of inexpensive high-power, high-efficiency solid-state diode lasers increased the available photon flux from a few to many tens, even hundreds of watts (utilizing arrays of such lasers), at a fraction of the size and cost of Ti:Sapphire lasers \citep{Wagshul89}. For MEOP, where the intrinsic efficiency of the process does not put a premium on high-power lasers, the principal problem solved was compression of $^3$He gas polarized at very low pressure up to atmospheric pressure without substantial loss of polarization; this was accomplished at large scales through use of titanium piston pumps \citep{Becker94} and at smaller scales with peristaltic \citep{Nacher99} and diaphragm \citep{Gentile01} pumps.

The dramatic scaling up of production rates necessary for human-lung imaging occurred for both SEOP and MEOP at about the same time in the early 1990's, leading rapidly in both cases to the first human lung images using hyperpolarized $^3$He, reported in 1996 \citep{MacFall96,Ebert96}. For the next decade or so, the field was dominated by imaging with $^3$He, which was available for $\approx\$100 {\rm US}/({\rm bar L})$ at near 100\% isotopic abundance, has a large magnetic moment (75\% of the proton's moment) and for which the physics (and hence the scale-up) of SEOP and MEOP were generally better understood than for $^{129}$Xe SEOP. In comparison, naturally abundant xenon (26\% $^{129}$Xe) costs $\approx \$50$~/(bar L) (isotopically enriched samples cost about an order of magnitude more), and the $^{129}$Xe nuclear moment is roughly one-quarter that of the proton. Commercial development of SEOP began in 1996\footnote{Magnetic Imaging Technologies, Inc., later acquired by GE Healthcare}; currently two small companies \footnote{Polarean Inc. (Research Triangle Park, NC); Xemed LLC  (Durham, NH)} offer high volume \He3 and $^{129}$Xe SEOP systems.

\begin{figure*}
\includegraphics[width=7in]{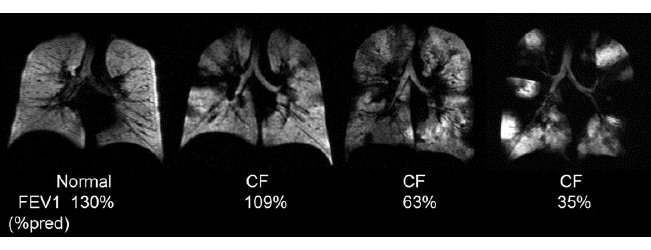}
\caption{Coronal (view from the front of the body) $^3$He magnetic resonance images from a healthy subject (left) and three patients with cystic fibrosis (CF). The number of ventilation defects increases with worsening results of a standard global ventilation test, FEV$_1$ (forced expiratory volume in one second). FEV$_1$ is shown as a percentage of the predicted value for a healthy subject. From \citet{Mentore05}.
}
\label{lungimage1}
\end{figure*}

In small-animal work, MR microscopy with $^3$He \citep{ChenXJ98a} was developed to resolve airways down to the fifth generation of branching in the guinea pig lung. In humans, stunning 3-D data sets of the lung with resolution of a few millimeters can now be acquired in a ten-second breath-hold \citep{Qing15}. A variety of diseases have been studied, among them cystic fibrosis  \citep{Paulin15,Mentore05,Flors16} (see Fig.~\ref{lungimage1}), asthma \citep{Kruger14,Tustison10}, and emphysema \citep{Kirby13,Quirk10,Spector05}.

\subsection{Gas handling and delivery}
Using SEOP, $^3$He can be polarized in reusable valved glass cells at pressures of up to 10~bar prior to being released for use in MRI. \citet{Jacob02} provided a recipe for making such cells from inexpensive borosilicate (Corning Pyrex) glass, with a length of capillary tube separating the main cell volume from the wetted valve materials. The high pressure limits $T_1$ due to the $^3$He-$^3$He dipolar mechanism \citep{Newbury93} { and cells are frequently dispensed and refilled}, so that using the low-permeability glasses that produce the longest wall-relaxation times (see Section \ref{roomtemprelax}) is not necessary. When polarized by MEOP for imaging, $^3$He is typically compressed to a few bar and stored in separate long-$T_1$ storage cells \cite{Heil95}. Such cells have not only been used to supply $^3$He for local experiments, but they have also been transported by air in compact magnetized boxes  \cite{Hiebel10}, for use in MRI at distant sites around the world \cite{Wild02,vanBeek13,Thien08}.

While more sophisticated ventilation systems have been employed, both for small-animal \cite{Nouls11} and human \cite{Guldner15} imaging, the most straightforward and widely used method to deliver hyperpolarized $^3$He to human subjects has been the use of plastic-valved flexible bags made of and or coated with one or more fluoropolymer materials; the most common among these is Tedlar\footnote{ Jensen Inert Products, Coral Springs, FL}. Both the cell containing polarized $^3$He and the bag are connected to a gas-handling manifold via a plastic valve built into the bag. Once the bag is filled with the requisite gas mixture, it is detached from the manifold and handed to the subject inside the magnet. We note that imaging protocols often call for $^3$He to be mixed with nitrogen to conserve it in cases where there is plenty of signal intensity available. The use of $^3$He for human lung MRI is subject to country-specific regulations, such as FDA regulation in the United States, where it requires an exemption as an investigational new drug (IND).

\subsection{Imaging modalities}
Several important characteristics of hyperpolarized $^3$He are of immediate consequence to MRI and necessitate rethinking of the pulse-sequence and signal-acquisition techniques commonly used for conventional $^1$H MRI. First, there is no relevant thermal recovery of the magnetization: the thermally polarized signal is negligible and typically characterized by a long recovery time $T_1$. { Hence,  while no (time-consuming) signal averaging is required, strategies are necessary to ration the magnetization inhaled in one breath to acquire the hundreds to thousands of data sets with different applied magnetic-field gradients needed to obtain an image.} The imaging sequences in early work employed successive small-angle excitations \citep{Kauczor97, deLange99} or successions of rapidly refocused gradient echoes \cite{Saam99}. Strategies for variable-strength excitations (variable flip angle) were also explored to make more efficient use of the magnetization \citep{Markstaller00, Santyr08, Deppe12}. Thus, while total imaging time is ultimately limited by $T_1$ relaxation of the inhaled gas, the fact that there is no waiting period due to thermal recovery means that image acquisition speed is not $T_1$-limited; this was exploited to make { frame-by-frame animations} of gas motion during breathing \citep{Saam99,Salerno01}, pointing toward the use of $^3$He to study lung {\em function} \citep{vanBeek04,Fain10} as well as structure. More recently, a wide range of more sophisticated pulse sequences and imaging protocols has been introduced \citep{Salerno03, Ajraoui10, Horn16}.

Second, the available $^3$He polarization depends only on the process (SEOP or MEOP) and the subsequent prevention of $T_1$ relaxation in the stored sample prior to the start of image acquisition. Contrary to the case in conventional $^1$H imaging, it does not depend on the applied magnetic field. Most human and animal MRI is ultimately dominated by conductive currents that produce Johnson noise in the sample and not in the probe. One can combine these facts to conclude that the SNR across a wide parameter space is approximately independent of applied field, although \citet{Parra-Robles05} identifies an optimal low field of $\approx 0.1$~T, with a factor of two loss in SNR occurring near the edges of the range 0.01 to 1~T. In the lung, there is the added prospect of improving image fidelity at low field by reducing artifacts due to susceptibility broadening created by multiple air-water interfaces \citep{Salerno05}. Low-field MRI using $^3$He has been developed and implemented both by \citet{Owers-Bradley13} at 0.15~T and by \citet{Durand02}, the latter using a commercial 0.1~T MR imager to obtain images with higher SNR and greater fidelity than achieved at the more standard 1.5~T field. In principle, one should be able to image with $^3$He at even lower fields ($<50$~mT). In practice, the results have been mixed, as one has to contend with increased pick-up and $1/f$-noise at lower Larmor frequencies, and concomitant gradients, i.e., the inability at very low fields to assume that gradients orthogonal to the applied-field direction are negligible \citep{Yablonskiy05}. There is often the need to develop dedicated pulse-sequences, probes, gradient coils, and other components, since commercial MRI has developed for several decades almost exclusively at ever-increasing applied fields. Despite these limitations, \Thad{very-low-field} MRI with $^3$He has been developed and implemented by \citet{Venkatesh03}, \citet{Bidinosti04}, and by \citet{Tsai08}, who studied the orientation-dependence of ventilation in the human lung with a custom-built 5-mT Helmholtz-coil pair in which a human subject could stand.

Third, paramagnetic molecular oxygen strongly relaxes $^3$He (at a rate of about 0.5~s$^{-1}$) for pure O$_2$ at 1~bar \cite{Saam95}, typically limiting imaging time {\it in vivo} to about 30~s for a single inhaled bolus of gas. Conventional MRI hardware is more than up to the task of using the magnetization during this time to acquire biologically relevant information, but one must be aware and account for the fact that a change in $^3$He density cannot easily be discerned from a change in magnetization, potentially confounding image interpretation. Schemes for continuous or quasi-continuous breathing of $^3$He have been developed for humans and for small animals. This apparent limitation can also be turned around to quantify regional pulmonary oxygen pressure in human lungs \citep{Deninger99, Miller10, Hamedani13}.

While state-of-the-art spin-density-weighted $^3$He lung images are quite spectacular in terms of brightness and resolution (see Fig.~\ref{lungimage1}), there are several other mechanisms that can be used to generate MRI contrast. $T_1$-weighting via interaction with O$_2$ has already been \Thad{mentioned};
highly mobile $^3$He is also particularly suited to diffusion imaging \citep{Schmidt97}, where the attenuation of signal due to diffusion through applied magnetic-field gradients \citep{Stejskal65} can be used to map the mean-squared displacement of spins during some characteristic time interval. This is particularly useful in the study of emphysema and related diseases, which are characterized by disintegration of the walls separating the many tiny ($\approx 300$~$\mu$m) alveolar (gas-exchange) spaces. The characteristic MRI diffusion time is easily chosen for $^3$He to be significantly greater than the diffusion time across alveolar spaces in a healthy lung, but on the order of or less than the diffusion time across the larger spaces that are created as the disease destroys the alveolar walls. Bright regions are those where $^3$He diffusion is restricted by many boundaries; diseased areas, where there is strong attenuation, show up darker. The quantitative measure of mean-squared displacement mapped in such images is known as the apparent diffusion coefficient (ADC). Several groups have developed and refined this technique \cite{WangC08,Conradi06,Fain05}. A quantitative model based on anisotropic $^3$He diffusion and ADC measurement was developed by \citet{Yablonskiy09} and supported through comparison to histologic sections of healthy and diseased tissue \citep{Woods06}. An alternate competing view of such modeling and its limitations was put forth by \citet{Parra-Robles13}.

\begin{figure}
\includegraphics[width=3.5 in]{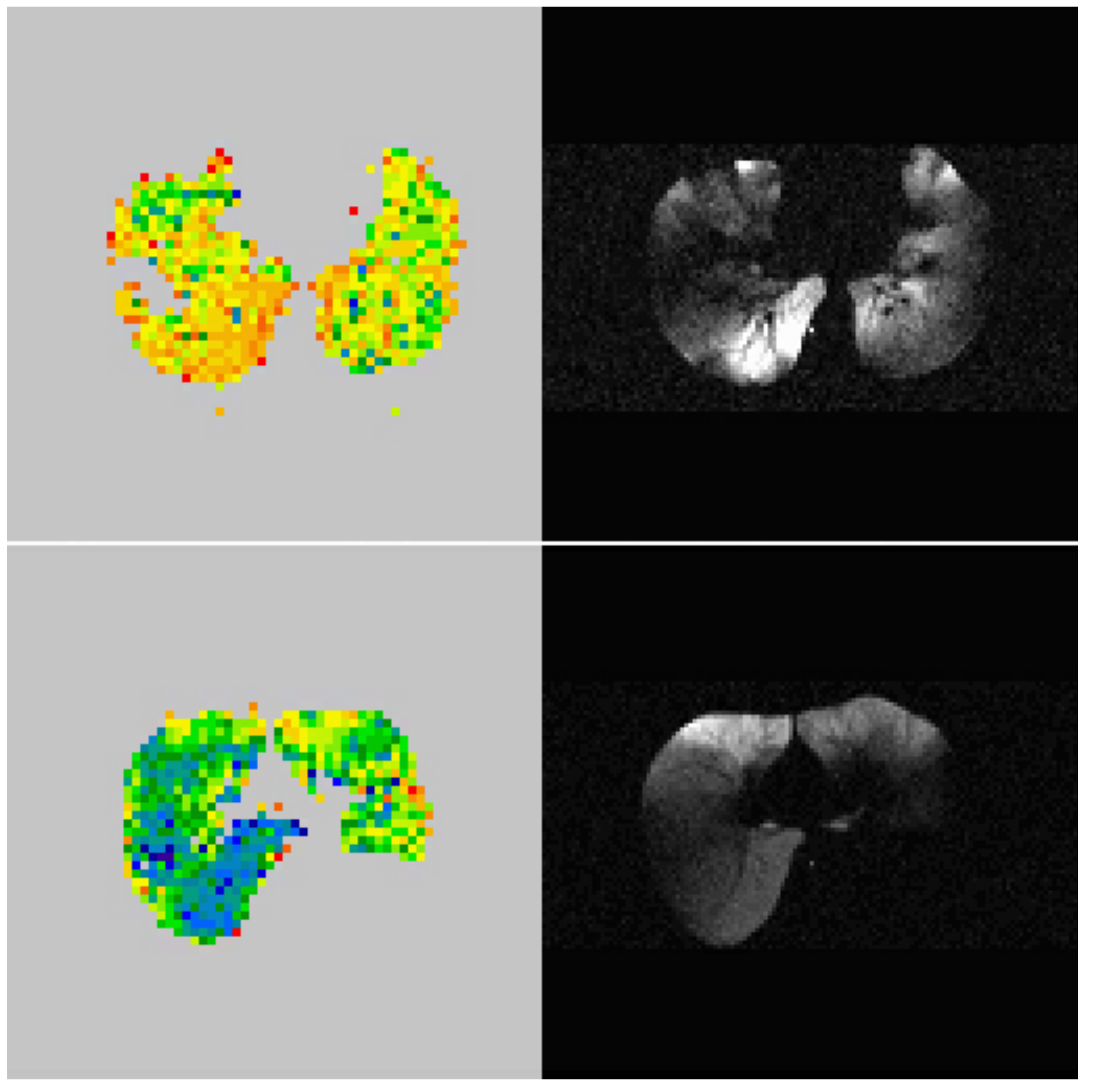}
\caption{Axial (perpendicular to the spine) $^3$He magnetic resonance images of a cross section of the ventilated lung in two emphysema patients, top and bottom. Images on the right are spin-density weighted while those on the left are maps of apparent diffusion coefficient (ADC), where deepest red (0~cm$^2$/s) represents the most restricted air spaces and deepest blue (0.88~cm$^2$/s) the least restricted. The blue regions correspond to the most diseased tissue, where the alveolar walls have been destroyed.  These regions do not necessarily correlate with the poorly ventilated regions seen in the spin-density weighted images, which demonstrates the potential for greater specificity with ADC mapping.
From \citet{Conradi06}.
}
\label{lungimage}
\end{figure}

\subsection{Current limitations and the future of $^3$He MRI}
While substantial progress has been made during the last two decades in improving image quality and applying various contrasts to study healthy and diseased lungs, these applications to date are generally termed ``pre-clinical,'' meaning that the technique has yet to find a specific application that warrants regular clinical use. One of the more promising avenues may be MR imaging of neonates \citep{Krjukov07,Tkach14,Walkup15}, where it is known that lung development is the limiting factor for survival of premature infants and where MRI is particularly attractive for longitudinal studies due to the absence of ionizing radiation.

Knowledge of the physics and correlated technological issues surrounding production of highly polarized $^{129}$Xe has improved somewhat in the last decade  \citep{Nelson01b,Nikolau13,Freeman14}. Coupled with native interest in the more biologically active xenon gas as a signal source and with the scarcer (and much more expensive) availability of $^3$He \citep{Shea10}, it would appear that, short of mining the moon for $^3$He \citep{Wittenberg92}, future further development of hyperpolarized-gas MRI as a clinical tool will primarily focus on $^{129}$Xe.

\section{Precision Measurements}
\label{spect}

Although most applications of hyperpolarized \He3 exploit the large achievable magnetizations, \He3 has tremendous potential itself for precision spectroscopy due to both large magnetization and long coherence times($T_2$). \Thad{ In a generic spectroscopy experiment, the uncertainty principle implies a measurement uncertainty $\sigma_\nu=1/T^*_2$ for a single measurement on a single particle.  For the $N=\dens{He}V$ polarized nuclei, and repeating the measurement $m=t/T^*_2$ times, the quantum-limited precision of frequency measurement becomes $\sigma_\nu=T^{*-1}_2/\sqrt{Nm}=1/\sqrt{T^*_2Nt}$.  Defining the equivalent frequency noise spectrum $\delta\nu$ through $\sigma_\nu=\delta\nu/\sqrt{t}$ implies a quantum projection frequency noise of }
\be
\delta \nu=\sqrt{1\over \dens{He}V T^*_2}=1.1\times 10^{-13} {{\rm Hz}\over\sqrt{\rm Hz}}
\ee
for a $V=1$ cm$^3$ volume if $T_2$ is limited by dipole-dipole relaxation (Sec.~\ref{dipdip}).  The corresponding magnetic \cite{Romalis2013} and rotation  \cite{Donley13,Walker2016}  sensitivities are $3\times 10^{-21}$ T/$\sqrt{\rm Hz}$ and $2.5\times 10^{-9}$ deg/$\sqrt{\rm h}$.  As such samples are now realizable, the primary challenges are detection and, for non-magnetic applications, compensation for environmental magnetic noise.  Detection strategies to date include  inductive pickup \citep{Chupp88}, SQUID detection \citep{Greenberg98,Savukov08,Gemmel10b}, external atomic magnetometers  \citep{Koch15b,CohenTannoudji69,Koch2015,Kraft2014}, fluxgate magnetometers \cite{Guigue2015,Wilms97} and, specific to SEOP implementations, using the embedded alkali atoms to detect the \He3 precession \citep{Kornack02,Zou16}.

The magnitude of the magnetic field produced by the precessing \He3 nuclei is of order
\be
B_{\rm He}=\kappa {8\pi\mu_{\rm He}\over 3}\dens{He}P_{\rm He}=\kappa\times 0.2 \mbox{ $\mu$T}
\ee
for 1 amg of hyperpolarized \He3.
The factor $\kappa\sim6$ is, for imbedded alkali detection, the frequency shift enhancement factor exploited for EPR polarimetry (Sec.~\ref{EPRpol}), while for the other methods that detect the classical magnetic field produced by the nuclei it is a geometrical factor, typically less than 1,  that depends on cell and detector geometry. Inside a multi-layer magnetic shield made of high permeability metal, the magnetic noise can be $10^{-14}$ T/$\sqrt{\rm Hz}$ or better, giving a potential SNR for the \He3 detection approaching 150 dB in a 1 Hz bandwidth.  No experiment has yet attained this value, but a recent experiment \cite{Allmendinger14,Gemmel10b} showed 5 nHz/$\sqrt{\rm Hz}$ for a 2 mbar cell with $T^*_2=60$ h in a He-Xe co-magnetometer experiment, with the He contribution likely substantially better than this.

Both SEOP and MEOP involve spin-polarized paramagnetic species (alkali atoms, metastable He atoms, free electrons) that produce \Thad{effective paramagnetic fields $B_A$ that in turn cause NMR frequency shifts of the \He3 analogous to the EPR frequency shift $\delta\nu_{A}=b_{\rm A}^{\rm He} P_{\rm He}$ of Eq.~\eqref{EPRfield}:
\be
\delta\nu_{\rm He}=-\gamma_{\rm He} { B}_{A}=-\gamma_{\rm He} b_{\rm He}^A P_A \label{NMRshift}
\ee} The field \Thad{$B_A=b_{\rm He}^A P_A$} is 2 mG for fully polarized Rb at a density of $10^{14}$ cm$^{-3}$.  This shift must be carefully managed in any precision measurement in which the \He3 is in the same region as the paramagnetic species.  Most precision NMR experiments to date manage the problem by transporting the polarized \He3 to a region free of paramagnetic species, with a key exception being the alkali co-magnetometer approach of Sec.~\ref{RomalisComag}.

\subsection{Co-magnetometry}

Any use of \He3 beyond magnetometry (symmetry violations, rotation) must account for the inevitable magnetic field fluctuations in any environment.  Thus a second species is required to separate out magnetic from non-magnetic interactions, generally by either locking one species to an atomic clock \cite{Bear98}, comparing the precession phase directly \cite{Allmendinger14,Chupp88}, or taking frequency ratios \cite{Chupp88}.  

A special case to note is the proposed use of \He3 as a comagnetometer for neutron \Thad{electric dipole moment (EDM)} experiments \cite{Tsentalovich14,Kim13,Chu15,Savukov08,Borisov2000}.  The cryogenic compatibility of \He3 is very attractive to co-locate with the neutrons, and the small 10\% difference in the magnetic moments of \He3 and neutrons allows dressing techniques to give them effectively the same moment, thus allowing common-mode rejection of magnetic field noise \cite{Golub1994}.  Recent experiments have shown the viability of this approach \citep{Eckel12,Chu11}.

Dual \He3-$^{129}$Xe magnetometry was demonstrated by \citet{Gemmel10b} using SQUID detection of MEOP-produced \He3 and SEOP-produced $^{129}$Xe.  They demonstrated magnetic sensitivity of 1 fT in about 200 s of integration.  Further developed versions of this approach, reaching coherence times of more than 100 hours \cite{Heil13} were used to set new limits on monopole-dipole interactions \cite{Tullney13} and charge-parity-time (CPT)/Lorentz invariance \citep{Allmendinger14}.  {In this latter experiment, an anomalous phase precession was attributed to interactions between the precessing nuclei due to the internal fields produced by the nuclei themselves, the magnitude of the effect being consistent with $\kappa_0=1$ (Eq.~\eqref{EPRfield}), implying a  significant contact interaction between the species \citep{Romalis14comment} that would call into question the geometrical calibration of the EPR frequency shift \cite{Romalis98}.}  This argument was disputed by \citet{Allmendinger14reply}. Similar Xe-He techniques are being developed for new Xe \Thad{EDM experiments \citep{Heil13,Kuchler2016}}.

He-Xe co-magnetometry using the embedded magnetometer possible in SEOP was studied by \citet{Sheng14}.  The alkali field shifts were nulled by strongly driving the Rb EPR resonance during free precession of the spins.  Of particular emphasis was systematic shifts caused by magnetic field and temperature gradients.

\subsection{Masers}

The first demonstration of the potential of hyperpolarized \He3 for precision measurements was in a maser configuration, by \citet{Robinson64}, who reported a statistical uncertainty of 500 nHz/$\sqrt{\rm Hz}$ using a lamp-pumped MEOP setup.  This experiment also pioneered the use of a dual cell system, with the MEOP process occurring in one bulb, connected by a few mm diameter tube to the spherical maser chamber.  The use of a relatively low frequency (100 kHz) means that the resonant cavity used for most electron spin masers is replaced by an open LRC circuit.  
 Nearly 25 years later, \citet{Richards88} and \citet{Flowers90} built similar systems for precision magnetometry at fields up to 0.1 T.  Later,  \citet{Gilles03} built a laser-pumped MEOP maser for earth field measurements, comparing the performance to a metastable $^4$He magnetometer and obtaining a field sensitivity of 20 pT/$\sqrt{\rm Hz}$.

\begin{figure}
\includegraphics[width=3.5 in]{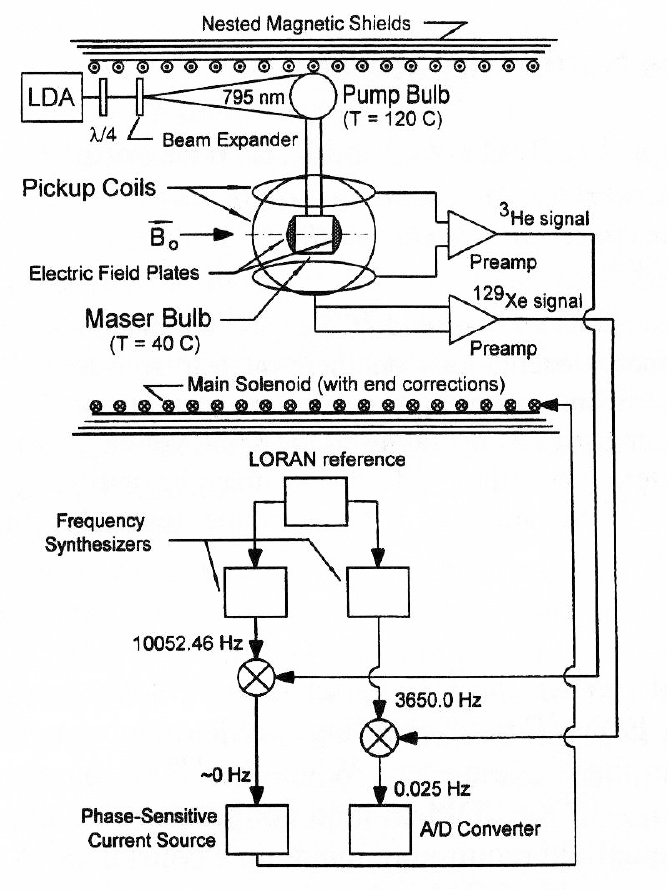}
\caption{Dual species maser, from \citet{Stoner96}}
\label{MaserFig}
\end{figure}

In 1994, dual species SEOP-pumped \He3 and $^{129}$Xe masers were introduced \cite{Chupp_maser94}.  The maser bulb was again isolated from the pumping bulb, in order to suppress frequency shifts from optically pumped Rb.  With dual species operation (shown in Fig.~\ref{MaserFig})  \cite{Stoner96,Bear98}, the common mode magnetic field fluctuations were cancelled to a high precision, enabling the dual species maser to be sensitive to non-magnetic interactions such as permanent electric dipole moments.  Using \He3 as comagnetometer, \citet{Rosenberry01} reported the most precise limits on a possible $^{129}$Xe \Thad{EDM} 
to date.  An earlier version  demonstrated the development of molybdenum electrodes with long relaxation times \citep{Oteiza92}.  A dual species maser under development for future \Thad{EDM} tests is described in \citet{Funayama2015}.  

The dual species maser was used in a series of symmetry violation tests by Walsworth and co-workers.  \citet{Bear00} set new limits on Lorentz and CPT violations, further improved by \citet{Cane04}.  These experiments search for sidereal variation in the relative precession of the two nuclei as the laboratory rotates with respect to the distant stars \citep{ISI:000165808700015,ISI:000240911700018} .  \citet{Glenday2008} used the dual species maser and a nearby SEOP \He3 cell inside a separate magnetic shield to search for anomalous interactions between neutrons.  Extensive modeling and details of the experimental implementation of the dual species maser, which reaches a precision of about 10 nHz, are given by \citet{Glendaythesis}.

\subsection{Alkali-He co-magnetometer}\label{RomalisComag}

{The maser and FID approaches to precision spectroscopy and comagnetometry with \He3 avoid systematic shifts due to alkali spin-exchange fields by making measurements in the absence of the alkali atoms.  Another approach, pioneered by Romalis and co-workers \citep{Kornack02,Kornack05} is to use the spin-exchange fields in an alkali-\He3 SEOP setup to null the magnetic sensitivity while maximizing sensitivity to non-magnetic interactions.  Taking advantage of both the spin-exchange field $B_{\rm He}$ from the \He3 nuclei (Eq.~\eqref{EPRfield}) and the analogous field $b_{\rm He}^AP_{A}$ from the \Thad{polarized} alkali atoms, they show that when $B_z+b_{\rm He}^AP_{A}+b^{\rm He}_AP_{\rm He}=0$ the alkali metal atoms experience no  response to transverse magnetic fields.  The application of a transverse magnetic field causes precession of the \He3 whose spin-exchange field cancels the applied field.  However, when a non-magnetic interaction is present, the \Thad{ nuclei and alkali metal atoms do not}  respond proportionately to their gyromagnetic ratios and the electrons then precess.  The alkali atoms acquire a sensitivity to non-magnetic interactions with the \He3 nucleus that is amplified by the factor $\gamma_{\rm S}/\gamma_{\rm He}=617$.}


%

The K-\He3 co-magnetometer was used to demonstrate a sensitive  gyroscope \citep{Kornack05},  set new limits on anomalous nuclear spin-dependent forces \cite{Vasilakis09}, and set new limits on Lorentz/CPT violation \citep{Brown10}.  These experiments exhibit a frequency sensitivity in the range of 18 pHz.

\subsection{Magnetometry}

Optically pumped metastable $^4$He magnetometers have been extensively developed \citep{Schearer85}  and
used for military, geomagnetic, and planetary magnetic field measurement applications \cite{Dunlop99}\footnote{Polatomic Inc., Polatomic, Inc.,  Richardson, TX}.
\Thad{Despite their
greater potential sensitivity, \He3 NMR-based magnetometers have not
been commercialized due to challenges of the NMR readout.   Development has been primarily for specialized precision physics measurements, especially  neutron EDM experiments.}
\citet{Borisov2000} pioneered a scheme where they measured the absolute \He3 polarization in a MEOP cell, let the gas expand into the measurement volume, applied a $\pi/2-T-\pi/2$ Ramsey sequence, then recompressed the gas into the MEOP cell where the final polarization was measured.  The magnetic field measurement precision was 24 fT.

Extensive studies of \He3-Cs magnetometers have been made for integration into the ultra-cold neutron \Thad{EDM} experiment being developed at PSI\footnote{Paul Scherrer Institute, Villigen PSI, Switzerland}.   The   basic principle is to transport \He3 from a MEOP pumping setup inside the magnetic shield where the \Thad{EDM} experiment is located.  They detect free-induction-decay of the \He3 with  lamp-pumped \citep{Kraft2014} or multiple laser-pumped \citep{Koch2015,Koch15b} Cs magnetometers.

For measurements of Tesla-scale magnetic fields, \He3 is very attractive due to its immunity to systematic errors.  \citet{Nikiel14} demonstrate astounding relative accuracy of 10$^{-12}$ at 1 T, with further advances promised from the use of essentially perfect spherical containers for the \He3 \citep{Maul16}.


\subsection {Searches for axion-like interactions}

Since \He3\ has similar spin and electromagnetic properties as a free neutron, it can be used to investigate possible exotic spin-dependent interactions with matter.  In particular, hypothetical axion-like particles would generate a \Thad{CP-violating} scalar-pseudoscalar coupling of strength $g_sg_p$ \cite{Moody84} with a Yukawa-like spatial dependence of length-scale $\Lambda$, \Thad{which is} inversely proportional to the axion-like mass.  The most straightforward approach is to search for a shift in the \He3 resonance frequency as the distance between a macroscopic mass  and the \He3 gas is modulated \cite{Tullney13}.  Another remarkable approach, proposed by \citet{Poko10}, is to use the fact that  spin relaxation of \He3 is sensitive to magnetic field gradients, \Thad{ which in this case means} 
gradients of the hypothetical Yukawa scalar-pseudoscalar coupling.  
\Thad{Measurements
of the transverse spin relaxation rates are particularly sensitive to
these non-magnetic gradients and, in particular, the long coherence
times of }
\citet{Gemmel10b} were used to set new limits \citep{Fu2010,Petukhov10}.  For $\Lambda$ smaller than the cell size, the classic transverse relaxation,  Eq.~\eqref{gradient} must be modified\citep{Petukov11}.  Measurements of \He3 transverse relaxation rate as a function of parameters such as bias magnetic field, density, and mass distribution then potentially reveal the existence of the scalar-pseudoscalar coupling.   A series of experiments have been done along these lines \citep{Chu13,Guigue2015,Yan15,Fu2011axion}.  At the shortest distances $\Lambda<20$ $\mu$m, this method is the most sensitive \citep{Guigue2015};  for $ \Lambda>50$~$\mu$m the \He3-Xe comagnetometry prevails  \cite{Tullney13}; the small intermediate regime is limited by a dual isotope Xe comagnetometer \citep{Bulatowicz13}. A new approach  for improving axion limits using a resonance method has been proposed as well  \cite{Arvanitaki14}.

\section{Future trends} \label{future}
We close with some of the scientific and technical challenges in the theory and practice of SEOP and MEOP.
The limiting polarization for SEOP is still not clear.  Although the discovery of excess relaxation that scales
with alkali-metal density has modified our understanding of this limit, the present analysis is purely
phenomenological as the origin of this cell-dependent relaxation is not fundamentally understood.
Indeed both its origin and magnetic field effects on \He3\ cells are new twists in the 
quest for understanding of wall relaxation, but perhaps they will turn out to be clues rather than simply a new source of confusion.
The recent higher polarizations from SEOP observed at the NCNR~\cite{Chen14} also suggest
that there is more to the limit than excess relaxation.   These results must be reproduced by other groups.

  Hybrid SEOP has yielded a significant gain in the polarizing rate.  A possibility for a 
further increase in rate is operation in a $\approx 50$~mT magnetic field, which has been 
shown to reduce alkali-alkali relaxation by a factor of $\approx$2~\cite{Kadlecek98,Kadlecek_thesis}.
This approach would be the most useful for the typical 1.5 bar pressure in NSFs,
for which this contribution dominates over alkali-buffer gas relaxation.
However, the detrimental effect of even such modest magnetic fields on \He3\ wall relaxation~\cite{Chen11,Jacob}
would have to be substantially reduced to employ this approach.
Employing the very low relaxation rate for sodium~\cite{Borel03} is hampered by browning
of aluminosilicate glass at the high temperatures required~\cite{Chen11,BabcockPhD}.
Although this issue could be addressed with the use of sapphire~\cite{Masuda05}, practical construction of such cells
for routine applications would be required. { Again, we see that poorly understood wall-relaxation properties limit the performance of SEOP}.

  The continuing development of laser technology has greatly benefitted SEOP.
Cost-effective laboratory systems can presently make use of chirped VHGs to narrow 100~W single
diode bars to 0.25~nm bandwidth.  However, fiber-coupled lasers are required or preferred for various in-situ systems.
If  VHG-narrowed, fiber-coupled systems \cite{Liu15} were commercialized and could reach this power level
with similar bandwidth and comparable cost, the performance of targets and in-situ NSFs could be improved.
Similarly, ultra-narrow lasers could yield benefits, but their utility has not been experimentally
investigated.   A study of the practical improvement in SEOP
for narrowing high power diode lasers from 2~nm to 0.2~nm was reported~\cite{Chann03}, {However,  no such study has been reported
for further narrowing from 0.2~nm to 0.04~nm \cite{Gourevitch:08}, even though this is now commercially available\footnote{Optigrate Corp., Oviedo, Florida}.}
Another possible approach to future lasers is alkali-metal lasers~\cite{Zhdanov13}.\
If 770 nm lasers were to become as available as 795 nm lasers for the same cost,
pure K pumping should be further investigated.  
Despite tremendous progress in the last two decades, a mature, common model
for SEOP in both Rb/K hybrid and pure alkali-metal cells remains to be developed and tested.

{The most significant new trend in MEOP is high field/high pressure operation.
To date the achievable polarization at high field has not exceeded that obtained
at low field and low pressure. Whereas operation in low field is typically easier,
operation at high field  is under development for a polarized \Het\ ion source at RHIC 
\cite{Maxwell16} and for high-accuracy magnetic field measurements \cite{Nikiel14,Maul16}.
Recent studies have revealed a light-induced relaxation mechanism 
that wastes most of the angular momentum deposited of the pumping light and limits the achievable polarization at higher pressures, but is not well understood.
Whereas the lower polarizations currently obtained at higher pressure are acceptable
for MRI applications, NSF and electron scattering applications typically demand the highest polarizations.
However, the current remote mode of operating large compressors inherently
introduces lower time-averaged polarization because of the need to replace cells
or perform local refilling operations.  If future work revealed the origin of this relaxation and a path to its elimination,
there is the potential for simplifying and shrinking large-scale compression apparatus
for these applications and thus perhaps allowing for continuous operation, as well as for more efficient use of available pumping light and larger production rates for MRI applications.}

\begin{figure}[htbp]
\begin{center}
\includegraphics[width=8.4cm]{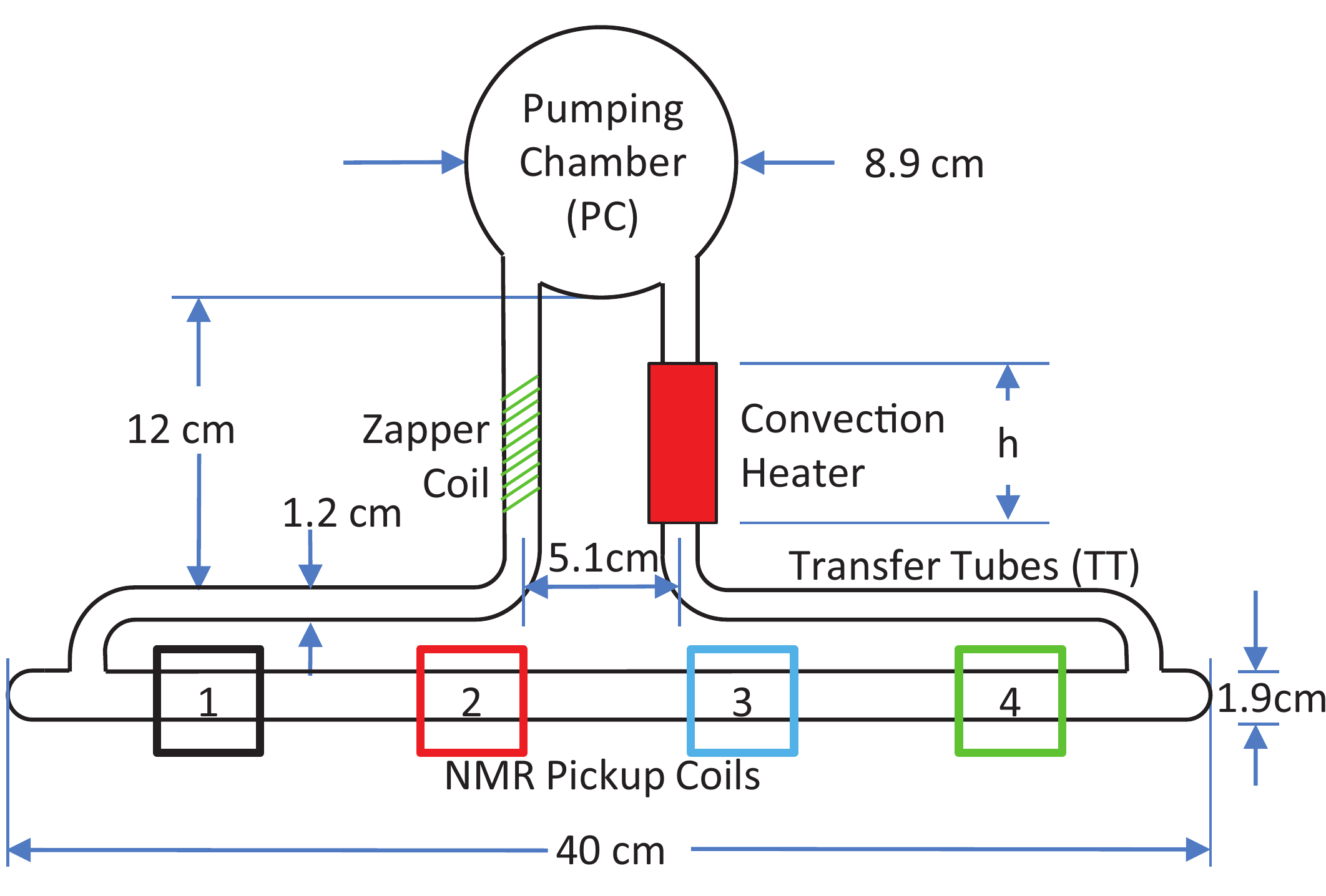}
\caption{Design of a convection cell for decreased transfer time between the two volumes
of a double cell, from \citet{Dolph11}. \Tom{ The flow of gas was monitored using an NMR tagging technique, in which the
zapper coil was used to depolarize a slug of gas and NMR signals were then detected at each of four locations
along the target chamber (labelled 1,2,3 and 4).} }
\label{fig:convection}
\end{center}
\end{figure}

  The high luminosity planned for future experiments at JLAB will put new demands on  SEOP targets
for electron scattering.  The high polarizing rates achievable with hybrid pumping make it possible to tolerate
the \He3\ relaxation produced by higher beam currents, but reducing the transfer time between the two volumes
of the double cell is desirable.  Since the \He3 polarimetry is typically done using EPR in the optical
pumping cell, more rapid transfer increases the accuracy of the determination of the polarization in the target cell.
Towards that end, targets employing convective transfer between the two volumes (see Fig.~\ref{fig:convection}) are under
development for these future experiments.\\

As NSFs are employed on an increasing range of neutron scattering instruments,
there will be new challenges for both methods.
For neutron scattering, NSFs have been typically polarized remotely and transferred to neutron beam lines,
but it is becoming of increasing interest to operate NSFs continuously or pseudo-continuously so as 
to maximize the figure of merit and decrease polarization correction uncertainties.   For SEOP,
apparatus that can fit within difficult space constraints and satisfy laser safety requirements are required.
For MEOP, local filling is likely to be used more extensively.  Increasing the range of applications will require NSFs with
large and/or complex cells and/or greater tolerance to stray magnetic fields. Finally, some NSFs may provide the
greatest impetus for achieving the highest possible \He3\ polarizations for two reasons:  
1)  some NSFs are relatively simple, single, cells at pressures near one bar, for which the 
limiting polarization in a practical application can be close to that possible under ideal conditions and
2) high neutron polarization and/or analyzing power are typically desirable for neutron scattering
experiments in which a small component of magnetic scattering is separated from a much large
component of nuclear scattering~\cite{Gentile05}.  In contrast with the typical figure of merit proportional
to $P_{\mathrm He}^2$ for the running time of an experiment, such situations yield
a stronger dependence that is closer to $P_{\mathrm He}^4$. 

\indent In the area of \He3 \ polarimetry, the high polarizations achievable
in NSFs provide both the impetus and a methodology for tests and possible improvements.
For example, if an accurate EPR apparatus were operated on an in-situ NSF, \Tom{a careful comparison
of the two methods would enable a potential improvement in the determination of $\kappa_0$ 
}
In the other direction, improved EPR, water-based NMR, and/or magnetometry could  provide
the means for a determination or better limit on the small cross section for neutron absorption
by \He3\ with parallel spins~\cite{Huber14}.  Finally if NSFs are to be used to polarize high flux beams
for fundamental neutron physics applications, double cells for neutron beams will have to be developed.

While the long-term future availability of \He3 for medical imaging remains uncertain, $^{129}$Xe is naturally abundant and relatively inexpensive. The advancements in $^{129}$Xe polarization technology and work comparing
\He3 and $^{129}$Xe support $^{129}$Xe being a viable alternative to \He3 in many, but not all, instances. { Furthermore, the unique physical properties of $^{129}$Xe  allow }for dissolved-phase imaging capable of measuring biomarkers related to gas uptake and exchange. In patient populations where the slight dissolution of $^{129}$Xe into blood is a potential issue, such as infants, \He3 may continue to be the preferred hyperpolarized gas.

Spin-polarized \He3 can be expected to be desirable for many applications in precision measurements.  The insensitivity of \He3 to non-magnetic interactions, plus its extremely small fundamental bandwidth, make it ideal for co-magnetometry applications for symmetry tests like searches for electric dipole moments.  The \He3 is relatively immune to chemical shifts, so it can be considered a primary magnetic field standard.  In a synchronous SEOP configuration \citep{Korver2013,Korver2015} there is promise to avoid systematic shifts from the alkali fields for making absolute magnetic field measurements or rotation measurements with the inherently high sensitivity of an imbedded alkali magnetometer.

\begin{acknowledgements}

The National Science Foundation (PHY-1607439) and Northrop-Grumman Corp. funding supported T.W.'s work on this paper. 

\end{acknowledgements}

\bibliography{He3RMP_BIBTEX-all}


\end{document}